\documentclass[preprint, 1p, authoryear]{elsarticle}

\let\today\relax
\makeatletter
\def\ps@pprintTitle{%
	\let\@oddhead\@empty
	\let\@evenhead\@empty
	\def\@oddfoot{\footnotesize\itshape
		{} \hfill\today}%
	\let\@evenfoot\@oddfoot
}
\makeatother

\usepackage[titletoc,toc,title]{appendix}

\usepackage{stackengine}

\usepackage[capposition=top]{floatrow}

\usepackage{epstopdf} 
\usepackage{bbm, bm}
\usepackage[inline]{enumitem}

\usepackage{makecell}

\usepackage{hvfloat}

\usepackage[hyphens]{xurl}
\usepackage[hidelinks]{hyperref}
\hypersetup{breaklinks=true}

\usepackage{cite}

\usepackage{amssymb,amsmath,amsthm,mathtools}
\usepackage{graphicx}
\usepackage{dsfont}

\usepackage{bigstrut,booktabs,varioref,float,tabularx,multirow}
\usepackage{tablefootnote}

\usepackage{pdflscape}

\usepackage{color}
\usepackage{soul}
\usepackage{adjustbox}
\usepackage{changepage}

\usepackage{afterpage}

\usepackage{tikz}
\usetikzlibrary{positioning}
\usetikzlibrary{arrows}
\usetikzlibrary{shapes.multipart}

\theoremstyle{plain}

\theoremstyle{definition}

\theoremstyle{remark}

\newcommand{\tabref}[1]{Table~\ref{#1}}

\newcommand{\figref}[1]{Figure~\ref{#1}}

\newcounter{TKcommentCounter}

\usepackage{cancel}

\title{{Insurance pricing for breast cancer under different multiple state models}}

\author[1] { Ay\c{s}e Ar{\i}k \corref{cor1}%
}
\ead{A.ARIK@hw.ac.uk}

\author[1] { Andrew J.G. Cairns }

\author[2] {Ereng{u}l Dodd}

\author[1] { Angus S. Macdonald  }

\author[3] { Adam Shao }

\author[1] { George Streftaris }

\address[1] {Department of Actuarial Mathematics and Statistics, 
Heriot-Watt University, and the Maxwell Institute for Mathematical Sciences, 
UK}
\address[2] {  Mathematical Sciences, S3RI, University of Southampton, 
Southampton, United Kingdom}
\address[3] {  Biometric Risk Modelling Chapter, SCOR, Singapore }

\cortext[cor1]{Corresponding author}

\begin{document}

\begin{abstract}
	
In this paper we consider pricing of insurance contracts for breast cancer risk based on three multiple state models. 
Using population data in England and data from the medical literature, we calibrate a collection of semi-Markov and Markov models.
Considering an industry-based Markov model as a baseline model, we demonstrate the strengths of a more detailed model while showing the importance of accounting for duration dependence in transition rates. 
We quantify age-specific cancer incidence and cancer survival by stage along with type-specific mortality rates based on the semi-Markov model which accounts for unobserved breast cancer cases and progression through breast cancer stages. 
Using the developed models, we obtain actuarial net premiums for a specialised critical illness and life insurance product.
Our analysis shows that the semi-Markov model leads to results aligned with empirical evidence. 
Our findings point out the importance of accounting for the time spent with diagnosed or undiagnosed pre-metastatic breast cancer in actuarial applications.

\end{abstract}

\maketitle

\section{Introduction}

Critical illness insurance (CII) is one of the most significant insurance products and provides coverage against a wide range of diseases, including cancer. 
Cancer remains one of the most common causes of morbidity and mortality worldwide with an estimated 19.3 million new cases and almost 10 million deaths occurred in 2020 \citep{Sungetal2021}. 
Meanwhile, in the aftermath of COVID-19 pandemic,
cancer has been considered to be one of the diseases to be impacted the most due to the dramatic changes in relevant healthcare pathways introduced as a response to the COVID-19 pandemic \citep{Laietal2020, Sudetal2020, Maringeetal2020, Alagozetal2021}.


In this study our focus is on breast cancer (BC), as this is one of the most common cancers diagnosed in women, in addition to being one of the leading causes of death for women in several countries \citep{ACS2021, McDonaldetal2008}. 
BC is also one of the most common conditions amongst CII claims, e.g. accounting 44\% of female CII claims in 2014 \citep{CMI2011, Aviva2015}. 
Besides, due to medical, scientific, and technological progress, 
BC survival has significantly improved over time. 
This, combined with a better understanding of BC risk, has led to changes in insurance practice, {such as `right to be forgotten' initiative in Europe}, with more and more BC survivors being currently considered to be insurable and with new insurance options being available to women with medical history \citep{IS2023, iam2023, SCOR2023, InsuranceEurope2021}.

It is also important to note that BC is one of the key cancer types for which cancer screening is often available. 
The availability of screening is crucial for early diagnosis of cancer, which is also a main determinant of cancer survival.  
Unfortunately, national-level lockdowns, introduced as a response to the COVID-19 pandemic, had significant consequences leading, for instance, to suspension of cancer screening programmes and treatments between March and June 2020 in the UK \citep{CRUKCIT2021}. 

In this study, we consider two multiple state models in a continuous time framework for modelling BC \citep{Soeteweyetal2022, BaioneandLevantesi2018, Doddetal2014}.
Specifically, we consider an industry-based Markov model, also discussed in \citet{BaioneandLevantesi2018}, as our baseline model. 
This is a compact model, accounting for all BC incidences in a single state. 
As an alternative to this model, we use a semi-Markov model developed by \citet{Ariketal2022} that differentiates between life histories on the basis of cancer stage and whether or not cancer diagnosis is made.
The \citet{Ariketal2022} model is further developed here in the following ways: (i) we consider a collection of semi-Markov models for a wider age range starting from age 30, and (ii) we apply generalised additive models to define transition intensities to account for changes in rates of transition over time for a given cohort. 
Particularly, we are interested in the impact of combining certain events, such as BC registrations across different cancer stages at the time of diagnosis as assumed by the industry-based model, in the absence of sufficient data, on insurance cash flows.
Thus, we show how both modelling frameworks can be implemented to quantify net insurance premiums of two related insurance contracts, specifically a specialised accelerated CII contract and a life insurance contract for women aged 30 to 60 years at the time of purchase. 
At the same time, we also explicitly address differences raised by different model assumptions, e.g. duration dependence for the transition rates of the semi-Markov model.

The remainder of this paper is organised as follows. 
In Section \ref{sec:Models} we introduce the multiple state models used in this study, 
while in Section \ref{sec:Pricing} we describe the considered insurance contracts.
In Section \ref{sec:Data} we  provide available data, and in Section \ref{sec:Numerical_Illustration}, we explain how to calibrate and implement the models under consideration. Following this, in Sections~\ref{sec:Numerical_Illustration}--\ref{sec:NetSinglePremiums}, age-specific occupancy probabilities, net survival rates from different causes, and net single premiums are calculated. 
In Section \ref{sec:Sensitivity} we provide sensitivity analyses for net insurance premiums under different parametrisations applied using the semi-Markov model. 
In Section \ref{sec:Discussion} we discuss the main findings and their implications for the insurance industry.

\section{Multiple state models}\label{sec:Models}

This section introduces two multiple state models used to define the life history of an individual exposed to BC risk. 
Specifically, we first introduce an industry-based 4-state Markov model and then an extension of this model.

\subsection{An industry-based multiple state model}

We consider a multiple state model with four states as shown in \figref{fig:CI_model}.
This model is used as a CII industry model in \citet{ReynoldsFaye2016} and is similar to the multiple state models discussed in \citet{CMI1991} and \citet{BaioneandLevantesi2018}. 

\begin{figure}[H]
	\begin{center}
		\begin{tikzpicture}[scale=.7]
			
			\draw[rounded corners, thick] (0,-0.5) rectangle (4,1.5); 
			\node at (2,1) {\textbf{ State 0}};
			\node[align=center] at (2,0) {No BC};
			
			\draw[rounded corners, thick] (12,-0.5) rectangle (16,1.5); 
			\node at (14,1) {\textbf{ State 1}};
			\node[align=center] at (14,0.2) {BC\\ Observed};
			
			\draw[rounded corners, thick] (0,-4) rectangle (4,-2); 
			\node at (2,-2.3) {\textbf{State 2}};
			\node[align=center] at (2,-3.3) {Dead,\\ Other Causes};
			
			\draw[rounded corners, thick] (12,-4) rectangle (16,-2); 
			\node at (14,-2.3) {\textbf{ State 3}};
			\node[align=center] at (14,-3.3) {Dead,\\ BC};
			
			\draw[->, thick] (4,0.5) -- (12,0.5); 
			\node[right] at (8,0.8) {$\mu^{01}_{x}$};	
			\draw[->, thick] (2,-0.5) -- (2,-2); 
			\node[right] at (2,-1.3) {$\mu^{02}_{x}$};	
			\draw[->, thick] (14,-0.5) -- (14,-2); 
			\node[right] at (14,-1.3) {$\mu^{13}_{x}$};	
			\draw[->, thick] (12,-0.5) -- (4,-2); 
			\node[right] at (8,-1.5) {$\mu^{12}_{x}$};	
			
		\end{tikzpicture}
	\end{center}
	\caption{Industry-based 4-state Markov model. Intensities $\mu$ are functions of age $x$.
		\label{fig:CI_model}}
\end{figure}

\figref{fig:CI_model} demonstrates the life history of an insured individual in a continuous time setting.
Transition intensities from state $i$ to state $j$ at age $x$ are denoted by $\mu^{ij}_x$.

This model does not differentiate between different stages of BC, or observed and unobserved BC cases. 
The model only accounts for observed BC cases 
and is characterised by a set of Kolmogorov equations provided in Appendix \ref{sec:AppendixKolmogorovCI}. 

\subsection{Semi-Markov model} \label{sec:semiMM}

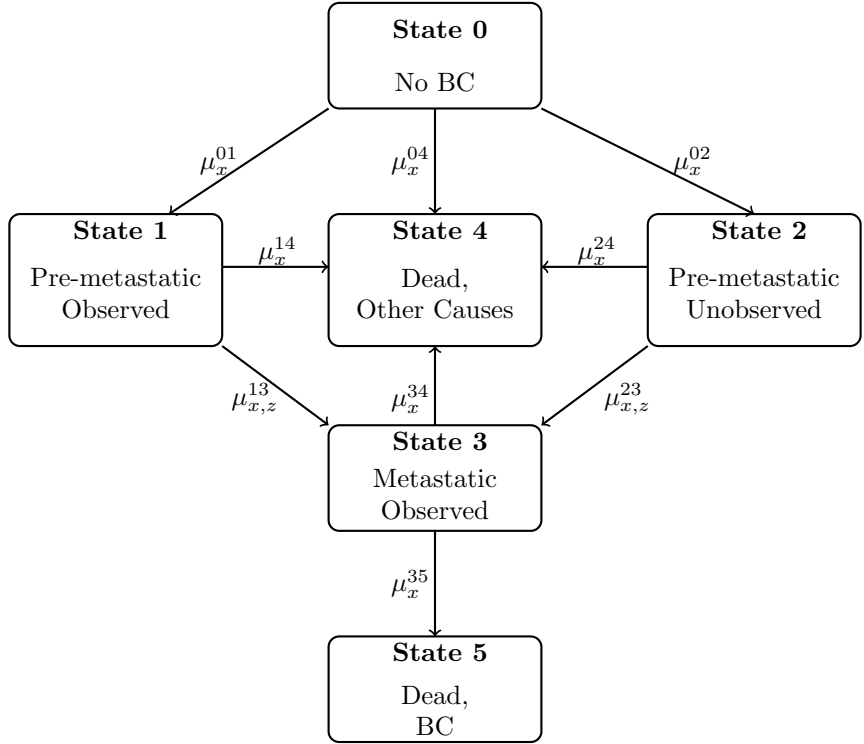
\begin{figure}[H]
	\begin{center}
		\begin{tikzpicture}[scale=.7]
			\draw[rounded corners, thick] (6,4) rectangle (10,6); 
			\node at (8,5.5) {\textbf{ State 0}};
			\node at (8,4.5) {No BC};
			
			\draw[rounded corners, thick] (0,-0.5) rectangle (4,2); 
			\node at (2,1.7) {\textbf{ State 1}};
			\node[align=center] at (2,0.5) {Pre-metastatic\\ Observed};
			
			\draw[rounded corners, thick] (12,-0.5) rectangle (16,2); 
			\node at (14,1.7) {\textbf{ State 2}};
			\node[align=center] at (14,0.5) {Pre-metastatic\\ Unobserved};
			
			\draw[rounded corners, thick] (6,-0.5) rectangle (10,2); 
			\node at (8,1.7) {\textbf{ State 4}};
			\node[align=center] at (8,0.5) {Dead,\\ Other Causes};
			
			\draw[rounded corners, thick] (6,-4) rectangle (10,-2); 
			\node at (8,-2.3) {\textbf{ State 3}};
			\node[align=center] at (8,-3.3) {Metastatic\\ Observed};

			\draw[rounded corners, thick] (6,-8) rectangle (10,-6); 
			\node at (8,-6.3) {\textbf{ State 5}};
			\node[align=center] at (8,-7.4) {Dead,\\ BC};
			
			\draw[->, thick] (6,4) -- (3,2); 
			\node[right] at (3.4,3) {$\mu^{01}_{x}$};
			\draw[->, thick] (10,4) -- (14,2); 
			\node[right] at (12.3,3) {$\mu^{02}_{x}$};
			\draw[->, thick] (8,4) -- (8,2); 
			\node[right] at (7,3) {$\mu^{04}_{x}$};
			\draw[->, thick] (4,1) -- (6,1); 
			\node[right] at (4.5,1.3) {$\mu^{14}_{x}$};	
			\draw[->, thick] (12,1) -- (10,1); 
			\node[right] at (10.5,1.3) {$\mu^{24}_{x}$};	
			\draw[->, thick] (8,-2) -- (8,-0.5); 
			\node[right] at (7,-1.5) {$\mu^{34}_{x}$};	
			\draw[->, thick] (8,-4) -- (8,-6); 
			\node[right] at (7,-5) {$\mu^{35}_{x}$};	
			\draw[->, thick] (4,-0.5) -- (6,-2); 
			\node[right] at (4,-1.5) {$\mu^{13}_{x,z}$};	
			\draw[->, thick] (12,-0.5) -- (10,-2); 
			\node[right] at (11,-1.5) {$\mu^{23}_{x,z}$};	
			
		\end{tikzpicture}
	\end{center}
	\caption{A breast cancer semi-Markov model. Intensities $\mu$ are functions of age $x$ and/or duration $z$.
		\label{fig:BC_model}}
\end{figure}

\figref{fig:BC_model} displays a continuous time semi-Markov model with 6 states for the life history of a policyholder. 
It describes the development of a single policy depending on age-specific transition intensities from state $i$ to state $j$ at age $x$, denoted by $\mu^{ij}_x$, and age- and duration-dependent transition intensities at age $x$ and duration $z$ from state $i$ to state $j$, denoted by  $\mu^{ij}_{x,z}$. 

In this model, State 0 represents individuals free of BC.
\figref{fig:BC_model} distinguish between observed and unobserved BC cases.
In particular, State 1 and State 3 show observed BC cases whereas State 2 show unobserved BC cases.
States 4 and 5 correspond to death from other causes and BC, respectively.

We note that the progression of BC is a well-defined staging process, described based on tumour (T) size, and whether or not cancer cells found in lymph nodes (N) or spread to the distant part of the body (M). This process is known as TNM Stages 1--4 \citep{ONS2017CancerSurvival}. 
In this staging, a higher stage number indicates a more progressed BC, where TNM Stage 4 (known as `metastasis') is the last stage of BC with tumours spread to the distant parts of the body. 
This latter stage, Stage 4 BC, is known to be more lethal and distinctively different than earlier BC stages \citep{Zhaoetal2020}. 
In the absence of comprehensive relevant data, and provided that one-year survival probabilities from Stages 1--3 BC are similar \citep{ONS2016CancerSurvival}, 
we model BC progression in two states, pre-metastatic and metastatic BC cases, respectively. 

We assume that the onset of BC, that is, new cancer cases at a given age $x$, can be determined by the sum of the rates of transition from State 0 to State 1 and to State 2 such that  

\begin{equation}
	\mu_x^{01} + \mu_x^{02} = \mu_x^*,
	\label{eq:BCDiagUnchnaged}
\end{equation}
where $ \mu_x^*$ corresponds to all new BC cases. \eqref{eq:BCDiagUnchnaged} leads to a convenient parametrisation such that 

\begin{equation}
	\mu_x^{01} = \alpha_x \, \mu_x^*, \qquad \mu_x^{02} = (1 - \alpha_x) \, \mu_x^*,  
	\label{eq:Parametrisation}
\end{equation}
\noindent where $0<\alpha_x<1$ quantifies a proportional relationship between $\mu^{01}_x$ and $\mu^{02}_x$.

We consider duration dependent transition intensities from pre-metastatic BC (State 1 and State 2) to metastatic BC (State 3). The reason is that in the medical literature it is stated that the risk of developing a metastatic BC after having pre-metastatic BC changes over time. This will be discussed in detail in Section~\ref{sec:Numerical_Illustration}. 
We also assume that individuals in State 1 may have treatments for BC, whereas this would not be possible 
for individuals in State 2. Thus, we assume a lower rate of transition to metastatic from State 1, compared to 
State 2, such that 
	
\begin{equation}
 \mu_{x,z}^{13} = \beta_{x,z} \, \mu_{x, z}^{23},
	\label{eq:Parametrisation2}
\end{equation}
\noindent where $\beta_{x,z} < 1$.
	
This model in \figref{fig:BC_model} is characterised by a set of modified Kolmogorov equations where a system of integral-differential equations is involved due to the existence of duration dependence from States 1 and 2 to State 3. 
These equations are explicitly provided in Appendix \ref{sec:AppendixKolmogorov}.
A fourth-order Runge-Kutta scheme is applied to numerically solve the (modified) Kolmogorov equations under consideration.  

We also consider a Markov model as a special case of the semi-Markov model in \figref{fig:BC_model}, choosing $\mu^{13}_{x}=0.01954$. This value is aligned with first distant metastasis rates based on Table 1 in \citet{Colzanietal2014}. See Section~\ref{sec:Numerical_Illustration} for further discussion regarding different modelling assumptions.

\section{Net single premiums of different insurance contracts}\label{sec:Pricing}

CII is a popular insurance contract that covers cancer as one of the core diseases, also including heart attack, stroke, and so on. 
Alongside, BC life insurance has recently been attracting more attention in the life insurance industry.
This is possibly linked to increasing demands from people with existing conditions or, perhaps, with a related-medical history. 
However, premium rates for the latter contract are noted to vary significantly from one insurance company to another, as a result of several factors impacting the underwriting process \citep{iam2023}. 
For instance, age, a common risk factor, can be a strong determinant of a BC premium, due to the distinctive age-specific curve in BC incidence resulting from very large changes in oestrogen level of women after age 50 \citep{Hendersonetal1988, Brayetal2004}. 
At the same time, provided high survival rates from BC, increasing chances of preventing recurrence with more advanced medical technology and a better understanding of the disease over time, 
it would be possible to tailor a contract for an individual surviving BC. 
In that case, other factors, such as cancer stage and time since the end of treatment, could become the main determinants of the insurance premium \citep{IS2023}. 

We consider two different insurance contracts here, noting that our focus is on the impact of different modelling assumptions.
Following the studies of \citet{Yueetal2017} and \citet{Soeteweyetal2022}, the first is a special accelerated CII contract, where a single benefit is paid when the insured
\begin{enumerate}[label=(\roman*)]
	\item is either diagnosed with BC for the first time; or
	\item dies from other causes before being diagnosed with BC.
\end{enumerate}
 
The net single premium for this contract is an accelerated death benefit along with a benefit paid at the time of diagnosis, and can be calculated based on the industry-based model in \figref{fig:CI_model} as

\begin{eqnarray}	\label{eq:AxCIMarkov2}
	{_ {\text{CI}, 1}}\bar{A}_{x}  = 	\int_{0}^{\infty}{ e^{-\delta t} 	{_ {t}} p^{00}_x \bigg( \mu^{01}_{x+t} +  \mu^{02}_{x+t} \bigg) dt}.
\end{eqnarray}

\noindent The net single premium for the same contract, based on the semi-Markov model, can be determined as

\begin{eqnarray}	\label{eq:AxSemiMarkov2}
	{_ {\text{CI}, 2}}\bar{A}_{x}  &= &	\int_{0}^{\infty}{ e^{-\delta t} 	{_ {t}} p^{00}_x \bigg( \mu^{01}_{x+t} +  \mu^{04}_{x+t} \bigg) dt} \\ \nonumber
	& + & \int_{0}^{\infty}{   e^{-\delta u} 	{_ {u}} p^{00}_x  \mu^{02}_{x+u} 	\int_{0}^{\infty}{ e^{-\delta t}  {_ {t}} p^{22}_{[x+u]} \bigg( \mu^{23}_{[x+u]+t} +  \mu^{24}_{[x+u]+t}  \bigg) dt} \,\,du  }.
\end{eqnarray}
Here, $\delta$ is the instantaneous constant force of interest rate, and we assume that 
\begin{enumerate}[label=(\roman*)]
	\item the insured purchases the product before being diagnosed with BC, i.e. in State 0 `No BC'; and 
	\item there is no waiting time between cancer diagnosis and the insurance payment.
\end{enumerate}
Note that, in order to make the formulae clearer, we have modified slightly earlier notation and have used actuarial selection notation. 
For instance, $\mu^{23}_{x,z}$ is presented based on select attained age $[x]$ with duration $z$, such that $\mu^{23}_{x,z} = \mu^{23}_{[x]+z}$. 

The second contract under consideration is a life insurance contract, that can also be purchased with an existing BC condition, and provides a single death benefit at the time of death. 
The net single premium of this contract, that is a death benefit from any cause for an insured person with no BC at the time of purchase, can be expressed based on the industry-based model as 

\begin{eqnarray}	\label{eq:AxIndustry_death1}
	{_ {\text{LI}, 1}}\bar{A}_{x}  =  \int_{0}^{\infty}{ e^{-\delta t}  	\bigg( 	{_ {t}} p^{00}_x  \mu^{02}_{x+t} +  	{_ {t}} p^{01}_x  \big(  \mu^{12}_{x+t} +  \mu^{13}_{x+t}  \big)  \bigg) dt }, 
\end{eqnarray}
 
\noindent while under the semi-Markov model it can be written as

\begin{eqnarray}	\label{eq:AxSemiMarkov_death1}
	{_ {\text{LI}, 2}}\bar{A}_{x} & = & \int_{0}^{\infty}{ e^{-\delta t}  	\bigg( 	{_ {t}} p^{00}_x  \mu^{04}_{x+t} +  	{_ {t}} p^{01}_x  \mu^{14}_{x+t} +  {_ {t}} p^{02}_x  \mu^{24}_{x+t}   \bigg) dt}  \\ \nonumber
	& + & \int_{0}^{\infty}{   e^{-\delta u} 	{_ {u}} p^{01}_x  \mu^{13}_{x+u} 	\int_{0}^{\infty}{ e^{-\delta t}  {_ {t}} p^{33}_{[x+u]} \bigg( \mu^{34}_{[x+u]+t} +  \mu^{35}_{[x+u]+t}  \bigg) dt} \,\,du  } \\ \nonumber
	& + & \int_{0}^{\infty}{   e^{-\delta u} 	{_ {u}} p^{02}_x  \mu^{23}_{x+u} 	\int_{0}^{\infty}{ e^{-\delta t}  {_ {t}} p^{33}_{[x+u]} \bigg( \mu^{34}_{[x+u]+t} +  \mu^{35}_{[x+u]+t}  \bigg) dt} \,\,du  }.
\end{eqnarray}

\noindent 
At the same time, the net single premium for the same contract for an insured individual with BC at the time of purchase, can be determined based on the industry-based model as 

\begin{eqnarray}	\label{eq:AxIndustry_death2}
	{_ {\text{LI}, 3}}\bar{A}_{x}  &= &	\int_{0}^{\infty}{ e^{-\delta t} 	{_ {t}} p^{11}_x \bigg( \mu^{12}_{x+t} +  \mu^{13}_{x+t} \bigg)dt},
\end{eqnarray}

\noindent and the premium for an insured person with pre-metastatic BC, is written based on the semi-Markov model as 

\begin{eqnarray}	\label{eq:AxSemiMarkov_death2}
	{_ {\text{LI}, 4}}\bar{A}_{x}  &= &	\int_{0}^{\infty}{ e^{-\delta t} 	{_ {t}} p^{11}_x  \mu^{14}_{x+t} dt} \\ \nonumber
	& + & \int_{0}^{\infty}{   e^{-\delta u} 	{_ {u}} p^{11}_x  \mu^{13}_{x+u} 	\int_{0}^{\infty}{ e^{-\delta t}  {_ {t}} p^{33}_{[x+u]} \bigg( \mu^{34}_{[x+u]+t} +  \mu^{35}_{[x+u]+t}  \bigg) dt} \,\,du  }.
\end{eqnarray}

Note that the net single premiums of the contracts under consideration, defined in \eqref{eq:AxCIMarkov2} -- \eqref{eq:AxSemiMarkov_death2}, are expressed for whole life insurance contracts. 
The upper bound of related integrals in each formula would change in the case of term insurance contracts. 

\section{Data}\label{sec:Data}

In this section we describe available data which consist of numbers of deaths by cause-of-death and cancer incidence registrations in England for the following age groups: 30-49, 50-54, 55-59, \ldots, 85-89. 
The cause-specific number of deaths data is available up to age 89 from 2001 to 2021 whereas the cancer registrations data is available from 2001 to 2020.

Cancer registrations are stratified by five-year age-at-diagnosis groups, type of tumour, single calendar year, and gender, for the years between 2001 and 2020. 
This data is provided by the Office for National Statistics (ONS) up to 2017 and 
later by the Health and Social Care Information Centre of the National Health Service (NHS) of England, also known as NHS Digital. 
Cause-specific death numbers, provided by the ONS, have similar granularity, where the data is split by five-year age-at-death groups, causes of death defined based on ICD 10 classification, single calendar year, and gender, between 2001 and 2021. Mid-year population estimates up to 2021 are also available from the ONS. 

\begin{figure}[H]
	\subfloat[Breast cancer incidence \label{fig:BreastIncidence}]{\includegraphics[width=0.5\textwidth]{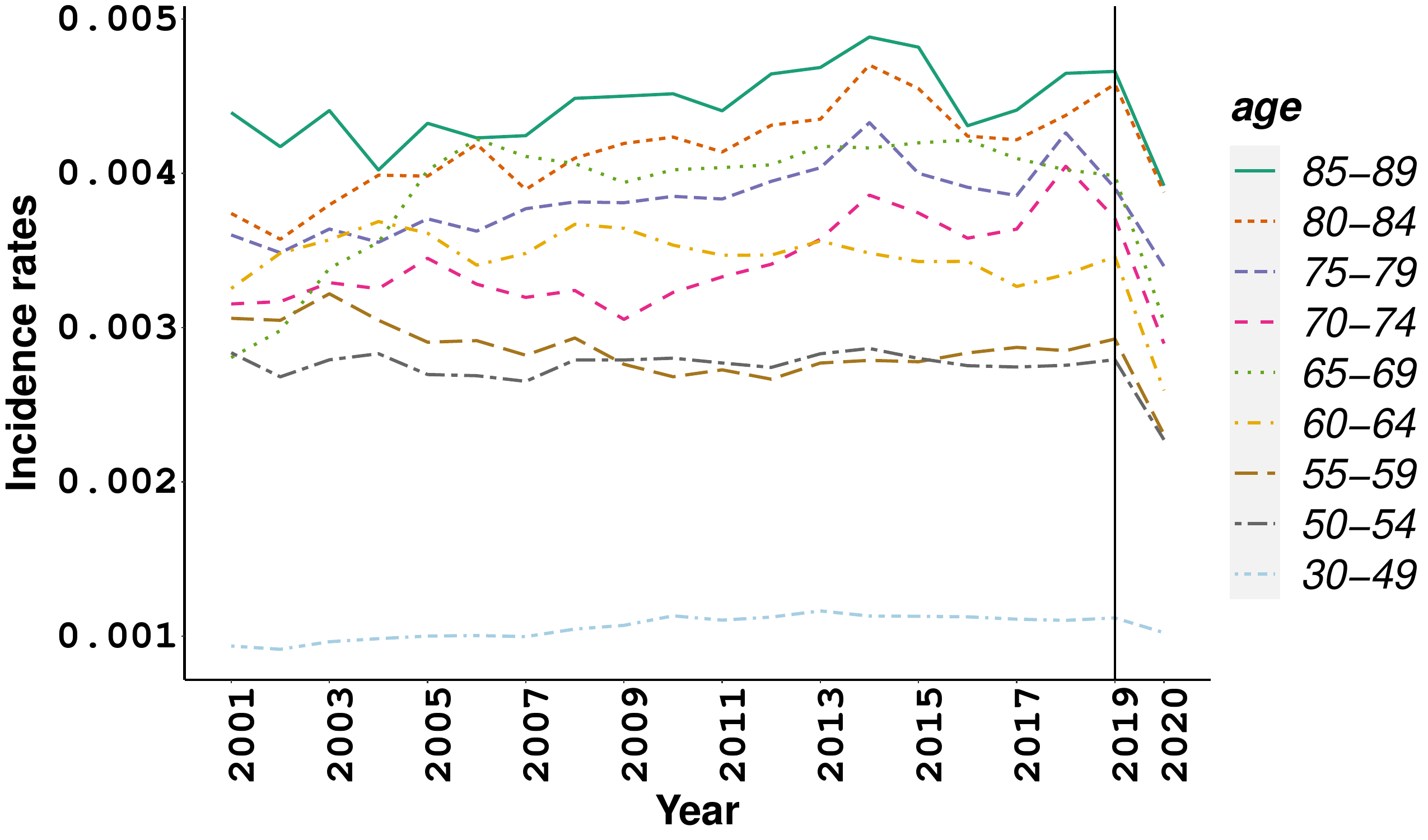}}
	\hfill
	\subfloat[Breast cancer mortality \label{fig:BreastMortality}]{\includegraphics[width=0.5\textwidth]{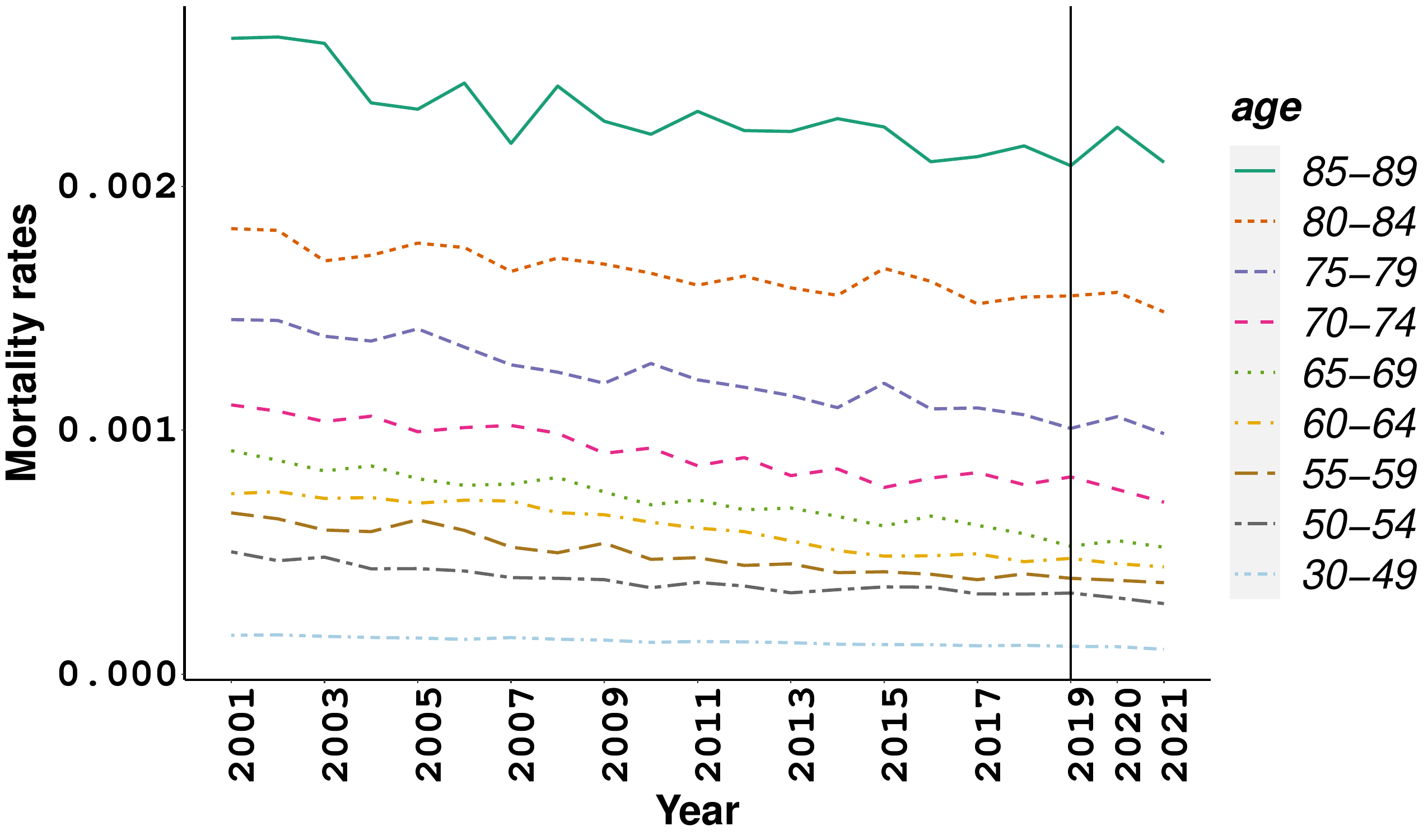}}
	\hfill
	\subfloat[Mortality from other causes  \label{fig:OtherMortality}]{\includegraphics[width=0.5\textwidth]{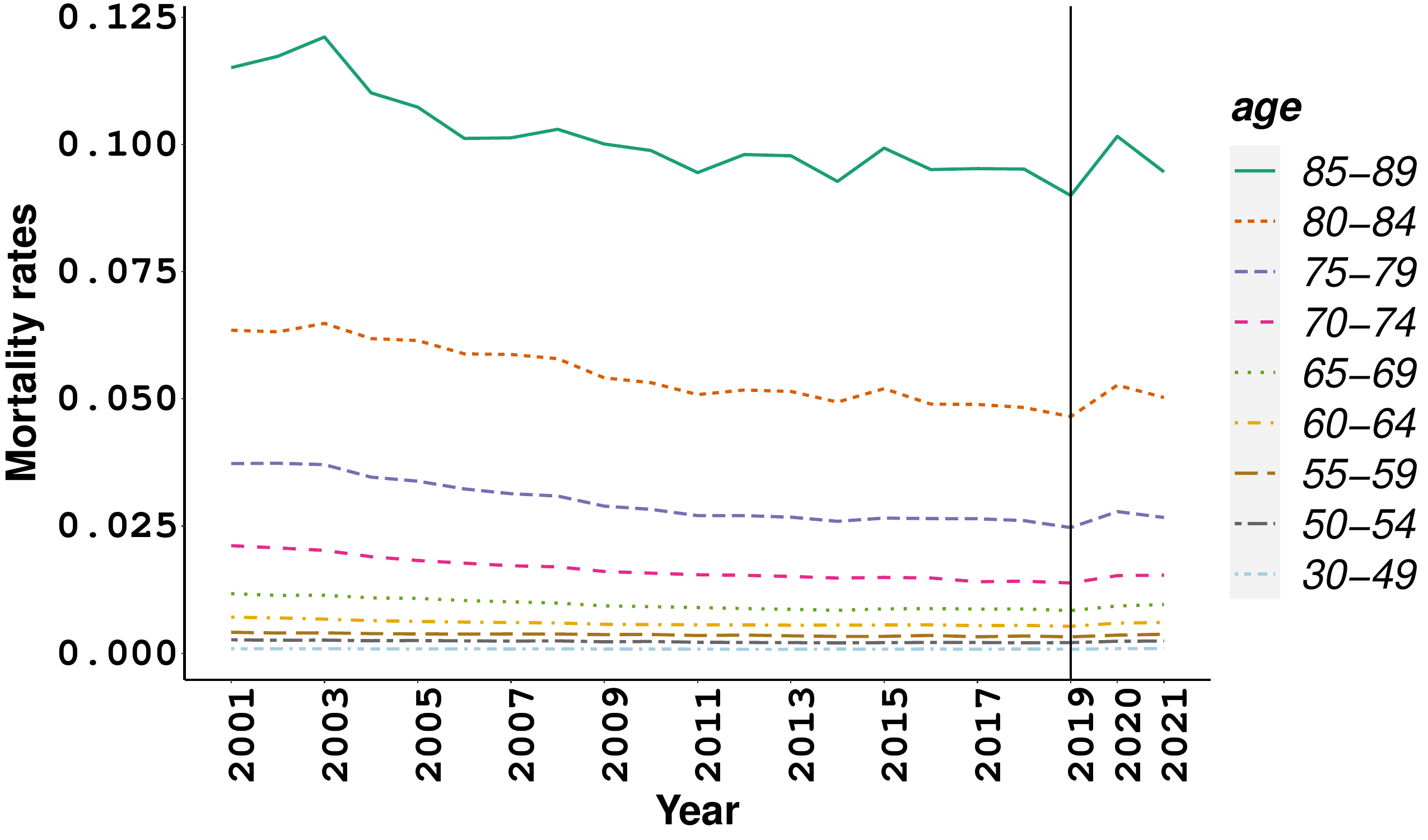}}
	\caption{Breast cancer incidence, mortality, and mortality from other causes (excluding breast cancer).}
	\label{fig:ONSData}
\end{figure}

\figref{fig:ONSData} shows age-specific BC incidence and mortality along with mortality from other causes over time. Here, BC incidence is defined as new BC registrations at a certain age-at-diagnosis group in a given year, divided by the related mid-year population estimates. Mortality from other causes is defined similarly, with deaths from BC being excluded.

\begin{figure}[H]
	\centering
	\subfloat[Breast cancer incidence  \label{fig:BreastIncidence2020}]{\includegraphics[width=0.5\textwidth]{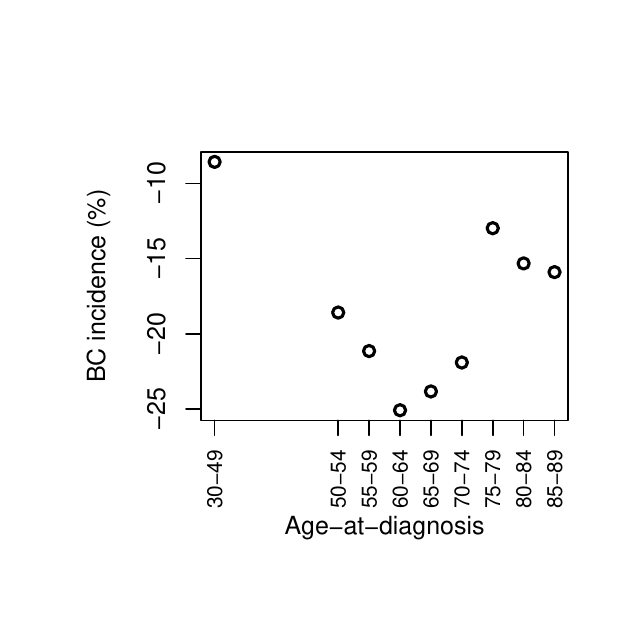}}
	\hfill
	\subfloat[Breast cancer mortality  \label{fig:BreastMortality2020}]{\includegraphics[width=0.5\textwidth]{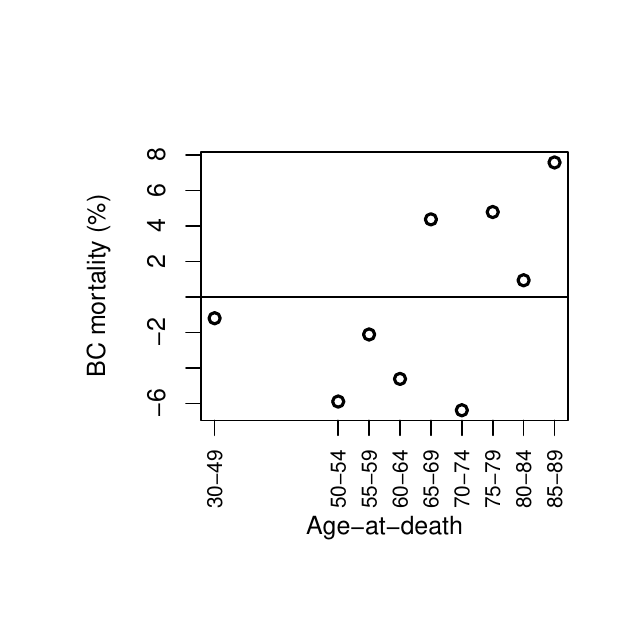}}
	\hfill
	\subfloat[Mortality from other causes (except breast cancer)  \label{fig:OtherMortality2020}]{\includegraphics[width=0.5\textwidth]{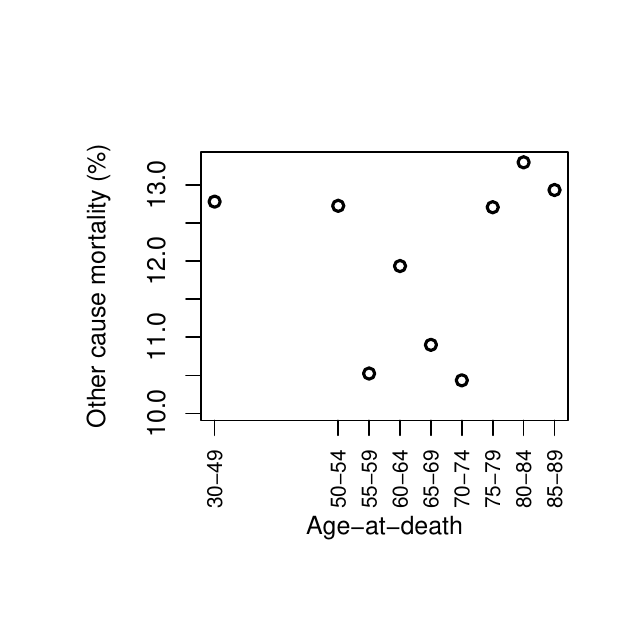}}
	\caption{Change (\%) in breast cancer incidence and mortality, and in mortality form other causes, 
	 in 2020 as compared to 2019.}
	\label{fig:BCDrop2020}
\end{figure}

\figref{fig:BreastIncidence} demonstrates that BC is a greater risk for women from age 50 onwards. Also, BC incidence mostly exhibits an increasing trend at different ages over calendar years apart from 2020, where we observe a sharp decline, as low as 25\% at ages 60--64 (see also  \figref{fig:BreastIncidence2020}). This is more evident for women older than 30--49, reflecting a reduction in cancer registrations due to the changes in availability of cancer services, e.g. a halt in cancer screening  from late March 2020 till June 2020, as a result of national lockdowns introduced as a response to the COVID-19 pandemic \citep{CRUK2021}. 

Figures \ref{fig:BreastMortality} and \ref{fig:OtherMortality} 
show a generally decreasing trend in BC and other-cause mortality.
This trend was interrupted by COVID-19 in 2020, with increasing mortality from other causes at all considered ages by 10--13\% as compared to 2019 (\figref{fig:OtherMortality2020}).
BC mortality increased in 2020, up around 7\%, for older ages, while it decreased for younger ages (\figref{fig:BreastMortality2020}).

\section{Model Results}\label{sec:Numerical_Illustration}

In this section we first present key transition intensities used to calibrate the models described in Section~\ref{sec:Models}.
We show our main findings based on different modelling assumptions, summarised in \tabref{tab:diffmodels}. 
We exclude the calendar years 2020 and 2021, while calibrating the models, as 
 the COVID-19 pandemic has had a major impact not only on cancer diagnoses and treatments, but also on health-seeking behaviour in general. 
The models are then used to estimate age-specific occupancy probabilities in future years, starting from 1 January 2020 onwards. 
The occupancy probabilities are subsequently used to derive net cancer survival at different ages of diagnosis. 

\begin{table}[H]
	\centering
	\caption{Models used in the numerical illustrations.}
	\label{tab:diffmodels}
	\begin{tabular}{cl} 
		M0  & Industry-based Markov model \\
		M1 & Semi-Markov model\\
		M2 & Markov model, as a special case of the semi-Markov model        \\
	\end{tabular}
\end{table}
	
\subsection{Transition intensities}\label{sec:TransitionRates}

Key transition intensities used in this paper are summarised in \tabref{tab:TransitionIntensities}. 
In the models tabulated in \tabref{tab:diffmodels}, we use available data 
as introduced in Section~\ref{sec:Data}, up to 2019, to avoid the impact of the COVID-19 pandemic in 2020 and 2021.
To be specific, we use average cancer incidence and type-specific mortality between 2001 and 2019 to determine age-specific rates of transition in the models under consideration, and by doing so we ignore the time trend. 
This is mainly for easiness of computation. 
We then fit generalised additive models to these rates, see Section~\ref{sec:GAMKeyTransitions}, while obtaining insurance premiums to account for changes in incidence and mortality rates at different ages. 

%

\begin{table}[H]
	\centering
	\caption{Age-specific transition intensities for the BC models M0--M2 based on available data and medical literature. 
	}
	\begin{tabular}{ccccc}
		\hline
		Age     & $\mu^{01}_x\, \text{in M0}$& $\mu^{01}_x\, \text{in M1\&M2}$& 
		\begin{tabular}[c]{@{}l@{}} $\mu^{02}_x\, \text{in M0}$\\$\mu^{04}_x\, \text{in M1\&M2}$\end{tabular}   & 	 \begin{tabular}[c]{@{}l@{}}$\mu_{x}^{13}\, \text{in M0}$\\$\mu^{35}_x\, \text{in M1\&M2}$\end{tabular} \\ \hline
		30--49  &0.00106 &  0.00086   &   0.00084  & 0.16739  \\
		50--54  & 0.00277  & 0.00224   &   0.00228  & 0.24005  \\
		55--59 &  0.00287   & 0.00233  &    0.00363   & 0.24005  \\
		60--64  & 0.00349    &  0.00282 &0.00588  & 0.28060 \\
		65--69  & 0.00393  &  0.00318  &  0.00952 & 0.28060 \\
		70--74  & 0.00345  &  0.00280  &  0.01643 & 0.36002  \\
		75--79 &  0.00384   &  0.00311  & 0.02987 & 0.40000  \\
		80--84 &  0.00417  &  0.00338  &  0.05496  & 0.49711 \\
		85--89 &  0.00447    & 0.00362  &  0.10112  & 0.50000  \\
		\hline
	\end{tabular} \label{tab:TransitionIntensities}
\end{table}

Hereby, we use average cancer incidence between 2001 and 2019 to determine age-specific rates of transition from State 0 to State 1, $\mu^{01}_x$. 
We note that BC registrations by stage in England are available between years of diagnosis 2012--2015 \citep{ONSData2016}. 
Following Table 1 in \citet{ONS2016CancerSurvival}, 81\% of new cancer registrations between 2001 and 2019 is used to represent Stages 1--3 BC, i.e. transitions to State 1 in M1 and M2.

Furthermore, in the models summarised in \tabref{tab:diffmodels}, we use average mortality from other causes between 2001 and 2019 to determine age-specific rates of transition from State 0 to death from other causes, i.e. $\mu^{02}_x$ in M0, and $\mu^{04}_x$ in M1 and M2.
For simplicity, in the absence of  comprehensive data, we assume that transitions to death due to other causes from all live states are equal to the transition to death due to other causes from State 0, i.e.

\[
\mu^{12}_x = \mu^{02}_x,
\]
in M0, and 
\[
\mu^{14}_x = \mu^{24}_x = \mu^{34}_x = \mu^{04}_x,
\]
in M1 and M2.

We determine rates of transition to death from BC denoted by $\mu^{13}_x$, in M0, and by $\mu^{35}_x$ in M1 and M2, using the information reported in \citet{Zhaoetal2020}, which is based on a cohort, representing 28\% of the population in the US, between 2010 and 2015, obtained from the National Cancer Institute Surveillance, Epidemiology and End Results (SEER) database.
Particularly, we use the  `No early death' and `Total early death' reported figures, where patients with Stage 4 BC that survived for 12 months are counted under `No early death'. 
In order to quantify $\mu^{35}_x$, we assume that  a `No Early Death' contributes a full year to exposure, while each `Early Death' contributes half a year on average to exposure, following the common `uniform distribution of deaths' actuarial assumption \citep{Hossain1994}.

We refer the study of \citet{Colzanietal2014} to determine rates of transition for developing metastatic BC after being diagnosed with pre-metastatic BC, denoted as $\mu^{13}_{x,z}$ in M1. 
\citet{Colzanietal2014} is based on women diagnosed with BC in Stockholm and Gotland Swedish counties between 1990 and 2016. 
Rates of transition from State 1 to State 3 in M1 can be determined based on the data in \tabref{tab:ColzaniData}.

\begin{table}[H]
	\centering
	\caption{Rates of transition from State 1 to State 3 for different {durations (years) in M1}.
	}\label{tab:ColzaniData}
	\stackunder{
		\begin{tabular}{lcccccccccc}
			\hline
			{duration (z)}             & 0 & 1    & 2    & 3    & 4     & 5     & 6    & 8      & 10    & 60 \\
			{$\mu^{13}_{x,z}$} & 0 & 0.03 & 0.04 & 0.03 & 0.024 & 0.021 & 0.02 & 0.0194 & 0.0194  & 0.0194 \\ \hline
		\end{tabular}
	}
	{\parbox{4in}{
			\footnotesize Note: The values are determined based on Figure 1 in \citet{Colzanietal2014}.
		}
	}
\end{table}

{Note that we assume $\mu^{13}_x=0.0194$ in M2, which is a special case of M1. 
	This value is consistent with first distant metastasis rates based on Table 1 in \citet{Colzanietal2014}. 
	Rates of transition from State 2 to State 3, $\mu^{23}_{x,z}$, are determined based on $\mu^{13}_{x,z}$ (see \eqref{eq:Parametrisation2}).
}

Last, in the absence of data to support other assumptions, 
we assume $\alpha_x = 0.6 $ and $\beta_{x,z}=1/7$ in M1 and M2, following \citet{Ariketal2022}. 

\subsection{Generalised additive models for key transition intensities}\label{sec:GAMKeyTransitions}

We consider Markov (M0 and M2), and semi-Markov (M1) models, and we assume smooth functions of the logarithm of key transition intensities, presented in \tabref{tab:TransitionIntensities}.
Hereby, we apply generalised additive models of the following form to the observed transition intensities in \tabref{tab:TransitionIntensities}:

\begin{align}
	g(E(y_i)) = \alpha + \sum_{p}{ s_p( x_{ip}) },
	\label{eq:GeneralAdditiveModels}
\end{align}
where $\alpha $ is the intercept; 
$g(.)$ is a smooth monotonic link function used to transform the expectation of the outcome $y$; and $y$ is 
modelled as the sum of smooth functions, $s(.)$, of covariates $x$ \citep{Wood2017}.
We use cubic splines as basis functions. 

\figref{fig:KeyIntensitiesGAM} displays observed, provided in \tabref{tab:TransitionIntensities} and \tabref{tab:ColzaniData}, and fitted values of key transition intensities, 
as functions of attained age based on the additive models in \eqref{eq:GeneralAdditiveModels}.
The plots in \figref{fig:KeyIntensitiesGAM} demonstrate smooth fitted rates across different ages. 
\figref{fig:mu01OurGAM} shows a significant difference in BC morbidity before and after age 50, where the increase in BC incidence seems to slow down from age 65 onwards. 
\figref{fig:mu04GAM} and \figref{fig:mu35GAM} exhibit increasing mortality rates from other causes and BC, respectively, for higher ages. 
\figref{fig:mu13GAM} suggests a decrease in developing a higher stage of BC, i.e. a more progressed BC, after being diagnosed with BC over time. 

\afterpage{%
	\begin{landscape}
\begin{figure}[!ht]
	\centering
	\subfloat[$\mu^{01}_{30+t}$ in M1\&M2 \label{fig:mu01OurGAM}]{\includegraphics[width=0.37\textwidth]{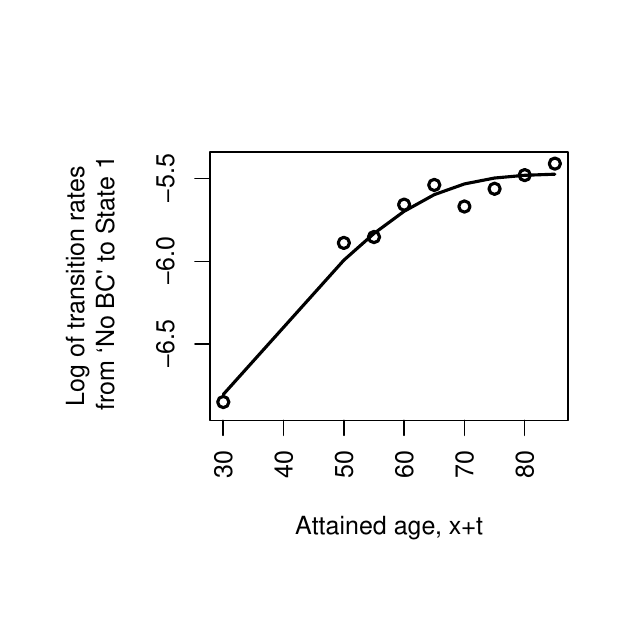}}
	\hfill
	\subfloat[ $\mu^{02}_{30+t}$ in M0 and $\mu^{04}_{30+t}$ in M1\&M2 \label{fig:mu04GAM}]{\includegraphics[width=0.37\textwidth]{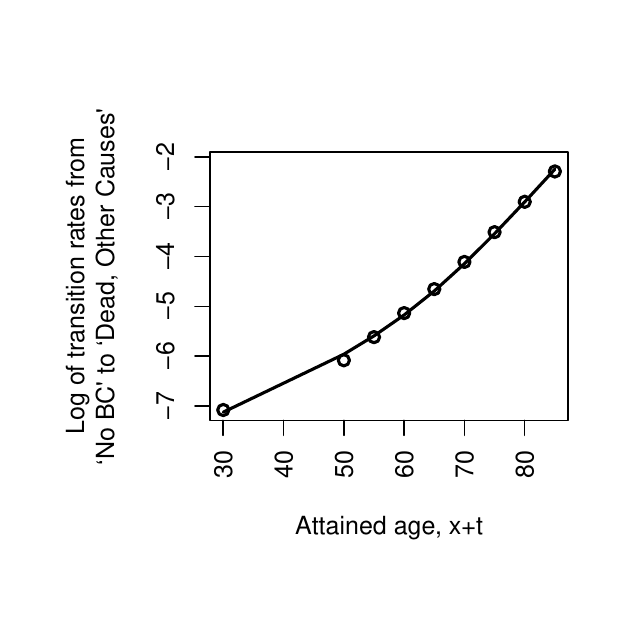}}
	\hfill
	\subfloat[ $\mu^{13}_{30+t}$ in M0 and $\mu^{35}_{30+t}$ in M1\&M2 \label{fig:mu35GAM}]{\includegraphics[width=0.37\textwidth]{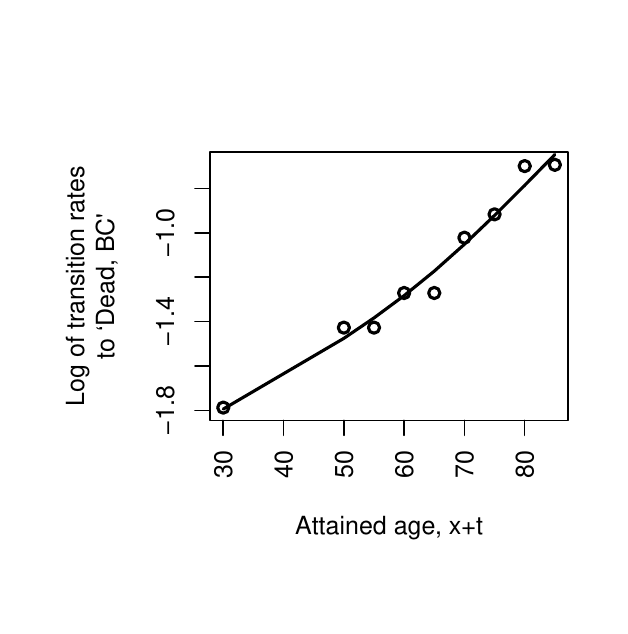}}
	\hfill
	\subfloat[ $\mu^{13}_{x,t}$ in M1 \label{fig:mu13GAM}]{\includegraphics[width=0.37\textwidth]{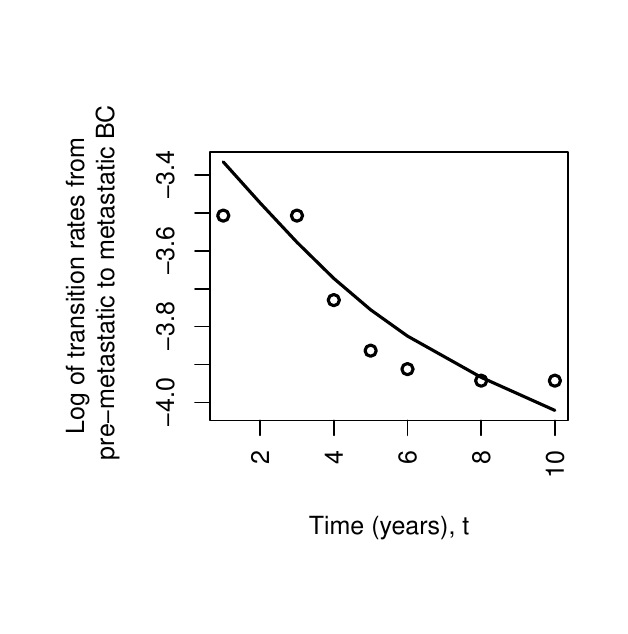}}
	\caption{Key transition rates, in the models listed in \tabref{tab:diffmodels}, as functions of attained age, $x+t$, where circles show observed values taken from \tabref{tab:TransitionIntensities}, and lines show fitted values from the relevant generalised additive models.} 	
	\label{fig:KeyIntensitiesGAM}
\end{figure}
\end{landscape}

}

\newpage
\subsection{Occupancy probabilities for different insured ages}

In Markov-type models, the occupancy probability, ${_ {t}} p^{ij}_x$, is defined to be the probability that an individual in state $i$ at age $x$ will be in state $j$ at age $x+t$ \citep{Macdonaldetal2018}.

\begin{figure}[H]
	\centering
	{\includegraphics[width=0.8\textwidth]{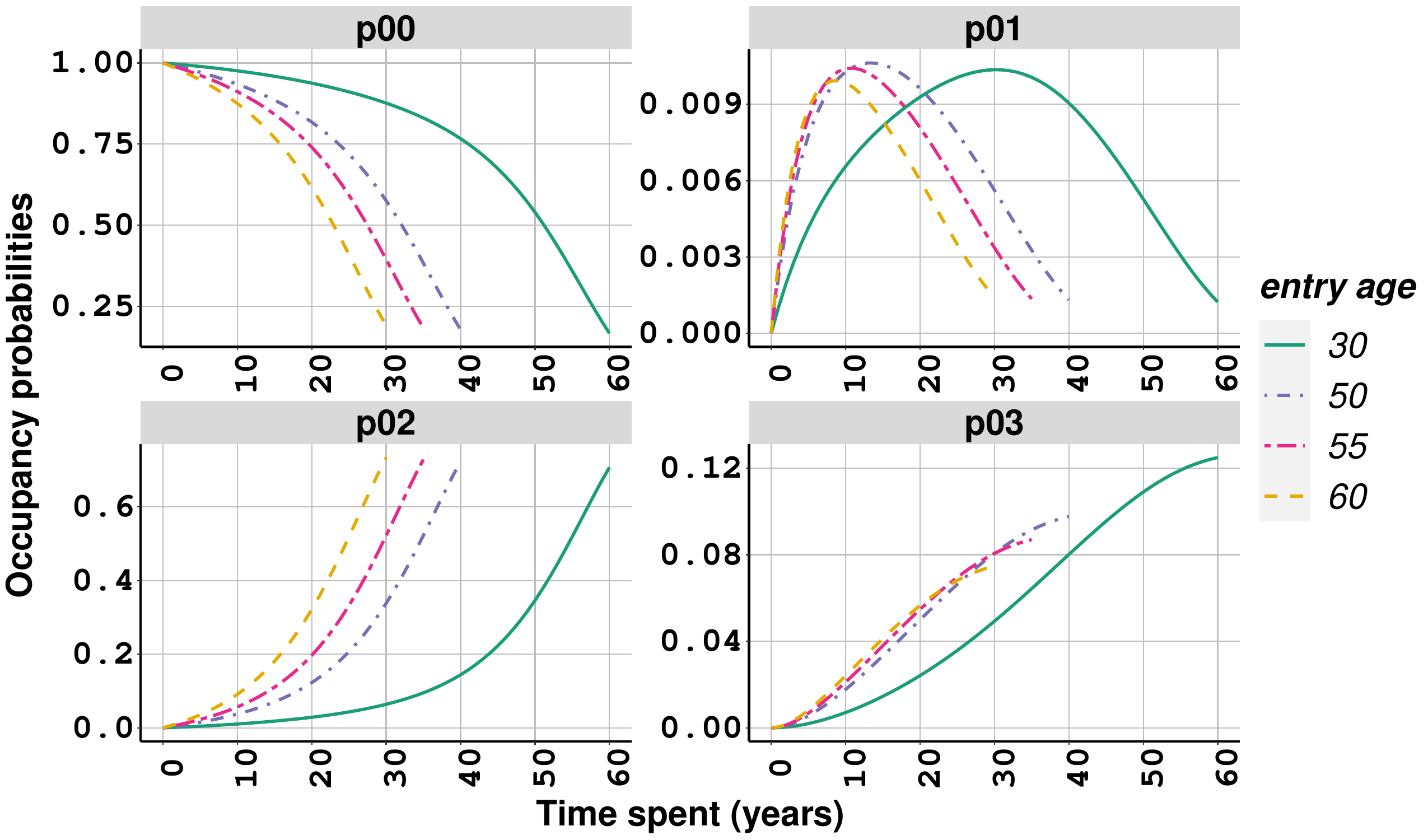}}
	\caption{Occupancy probabilities for a policyholder with no breast cancer, at different contract entry ages, based on M0.} 	
	\label{fig:KeyProbabGAMCI}
\end{figure}

\figref{fig:KeyProbabGAMCI} presents occupancy probabilities for an individual aged between 30 to 60 years, with no BC, at the beginning of the insurance contract based on M0. 
Note that these probabilities can be associated with a policyholder at these ages at the time of purchase of a particular insurance product. 
Hereby, in the figure, we assume that time spent on the contract depends on the policyholder's age at purchase.
We also assume that the maximum age for these insurance contracts is 90. This means that the policy purchased by a person who is aged 50 could be in-force for at most 40 years, whereas policy purchased by a 60 year old could be in-force for at most 30 years.
Accordingly, in each subplot, we see shorter time periods for higher ages on $x$ axis.

\begin{figure}[H]
	\centering
	{\includegraphics[width=1\textwidth]{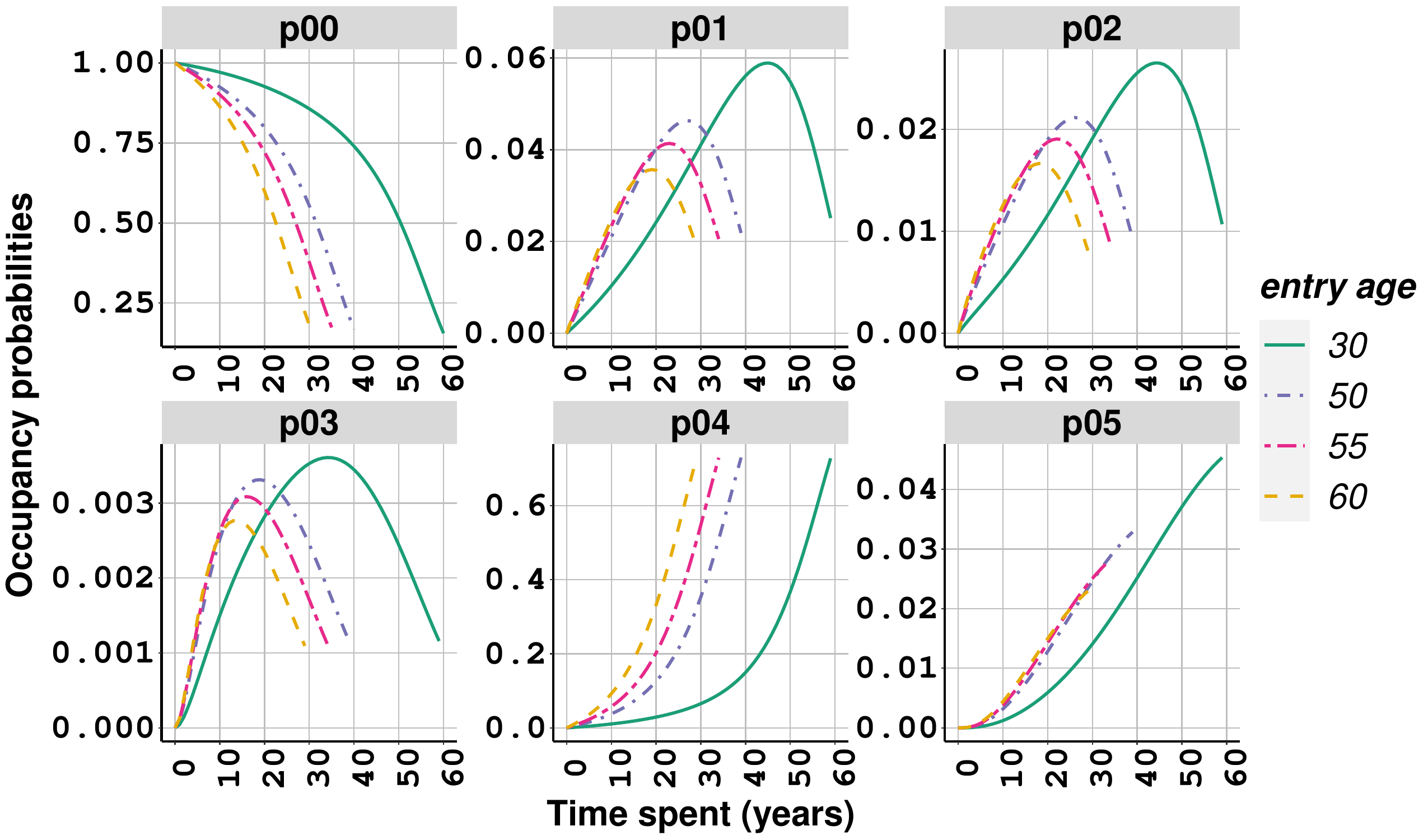}}
	\caption{Occupancy probabilities for a policyholder with no breast cancer, at different contract entry ages, based on M1.} 	
	\label{fig:KeyProbabGAMSM}
\end{figure}

Similarly, \figref{fig:KeyProbabGAMSM} displays occupancy probabilities for women with different entry ages ranging from 30 to 60, with no BC at time zero, based on M1. 

We note that the death probabilities from BC for a healthy woman under M0, ${_{t}} p^{03}_x$, are estimated as significantly higher than those under the semi-Markov model, M1, ${_{t}} p^{05}_x$, for the same $t$. 
This is related to the assumption about the risk of dying from BC after receiving a BC diagnosis, that is $\mu^{13}_x$. 
In the absence of relevant data, we use a similar assumption for $\mu^{35}_x$ under M1, leading to high deaths from BC (\tabref{tab:TransitionIntensities}).
See \ref{sec:Impactmu13} for further discussion related to this assumption.

\subsection{Model validation}

We calculate age-specific net cancer survival, following the ONS definition  \citep{ONSSurv2019}.
This measures survival from a given cancer type after receiving a diagnosis, assuming that this cancer can be the only cause of death \citep{Mariottoetal2014, SwaminathanandBrenner2011}. 
As an example, we can consider a woman diagnosed with pre-metastatic BC at age $x$. 
BC survival of this woman in $t$ years can be obtained, based on M1, as follows:

\begin{equation}
	\frac{1 -  {_{t}} p^{14}_x  \, - {_{t}} p^{15}_x \, }{1 - {_{t}} p^{14}_x  },
	\label{eq:CancerSurvival}
\end{equation}
where ${_{t}} p^{14}_x$  represents mortality from other causes, and ${_{t}} p^{15}_x$ represents mortality from BC at age $x$ in $t$ years time. 

\begin{figure}[H]
	\subfloat[Net survival from pre-metastatic BC, M1 \label{fig:NetCancerSurvivalPreMet}]{\includegraphics[width=0.5\textwidth]{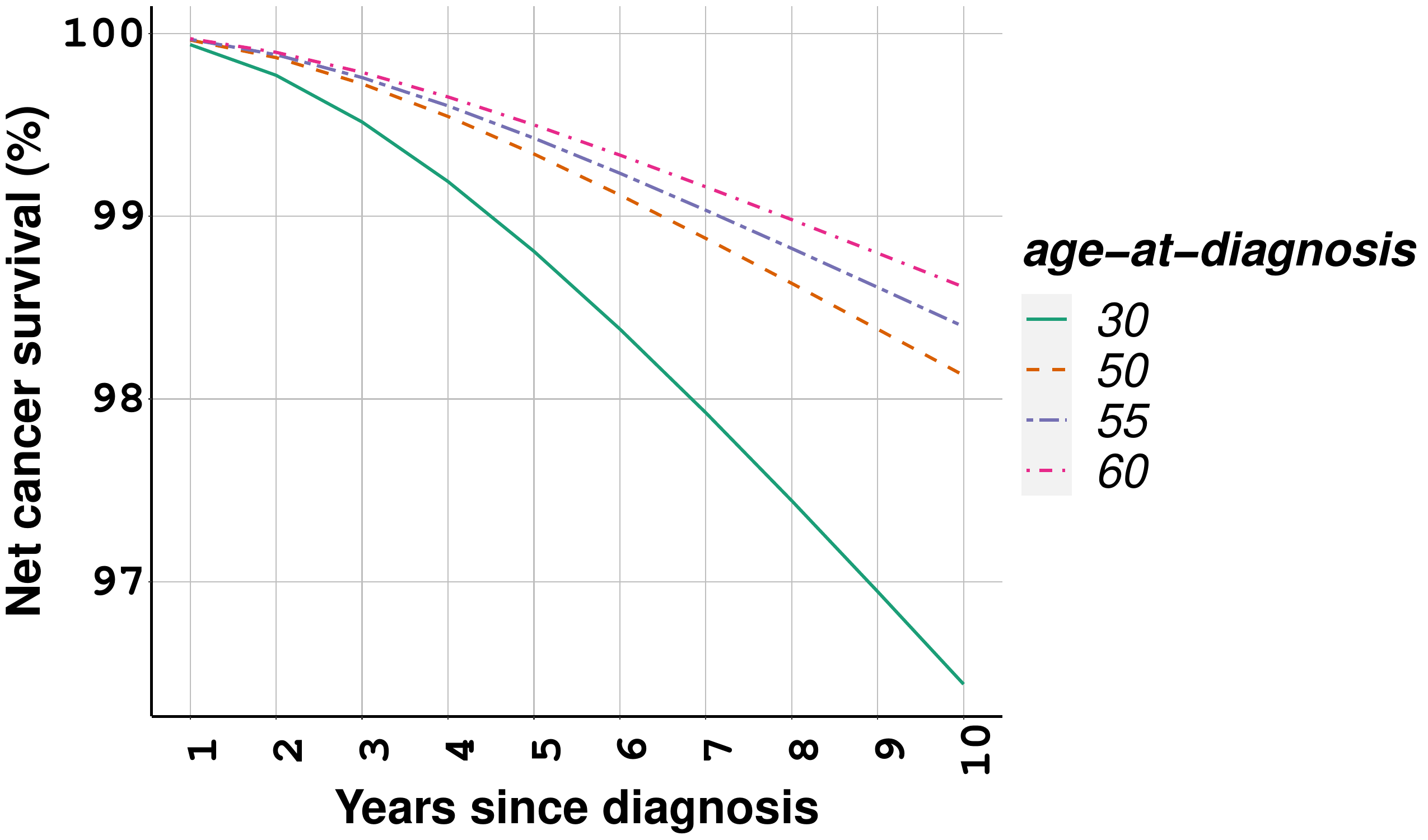}}
	\hfill
	\subfloat[Net survival from metastatic BC, M1 \label{fig:NetCancerSurvivalMet}]{\includegraphics[width=0.5\textwidth]{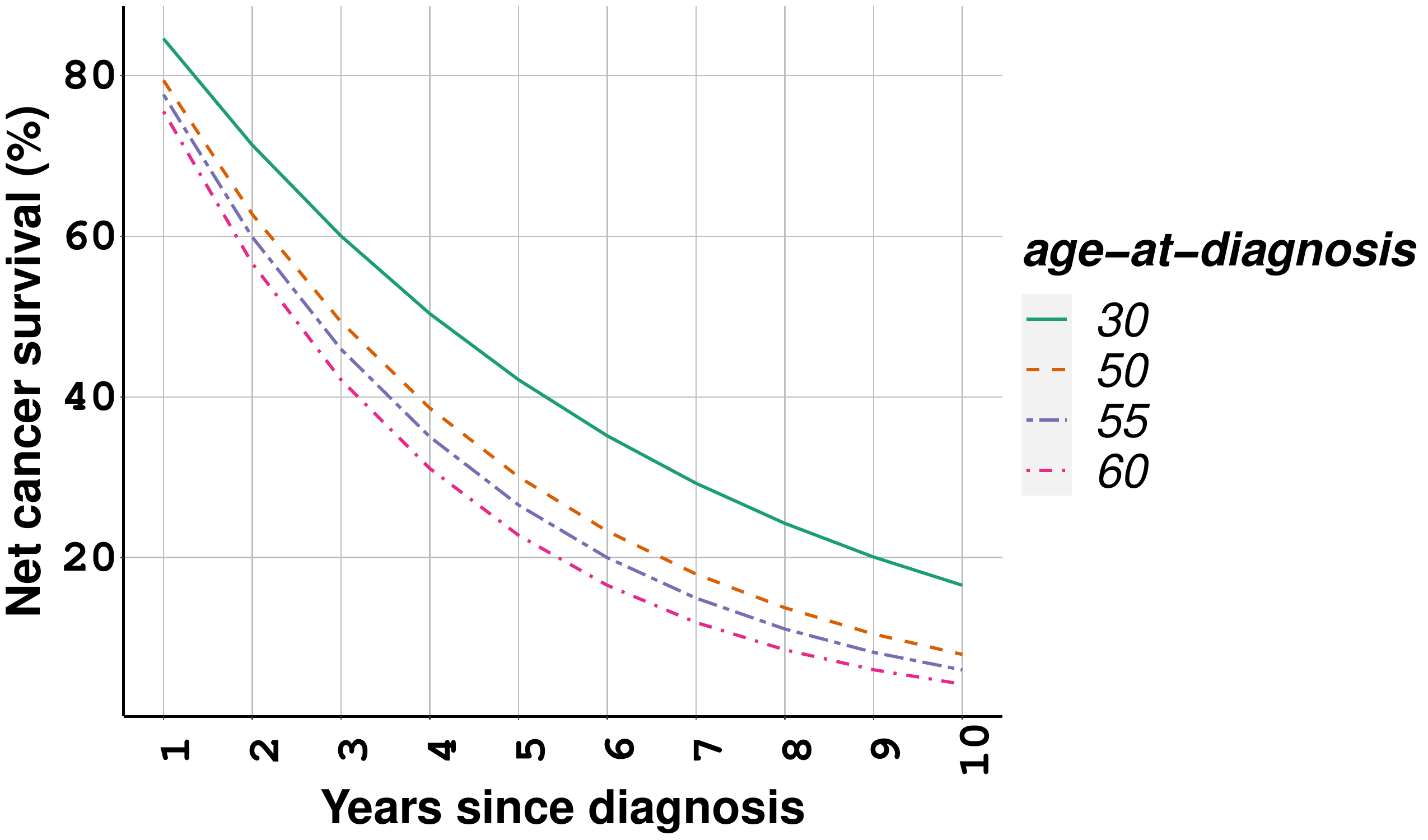}}
	\caption{Estimated net cancer survival at different ages under M1, for a woman diagnosed with: (a) pre-metastatic breast cancer; and (b) metastatic breast cancer.}
	\label{fig:NetCancerSurvival}
\end{figure}

\figref{fig:NetCancerSurvival} shows net cancer survival from pre-metastatic and metastatic BC for a woman diagnosed with the related BC at different ages, based on M1. 
The relationship between BC survival with a pre-metastatic BC and age, \figref{fig:NetCancerSurvivalPreMet}, demonstrates an inverse pattern, where survival seems to increase 
from age 30 to 60. 
These estimates are aligned with BC statistics in England. 
For instance, 5-year cancer survival for women aged 15 to 39 years at diagnosis between 2010 and 2014, followed up to 2015, is reported to be 85.5\%,
whereas the same survival for women aged 40 to 69 years 
is between 90.7\% to 93.0\%.
Higher survival at older ages may be linked to the availability of BC screening for women aged 50 to 70 at the time \citep{ONS2016CancerSurvival2}. 
Another reason could be the type of BC, which is more likely to be a more aggressive cancer for younger women as compared to older women \citep{ONS2019CancerSurvival}.

At the same time, \figref{fig:NetCancerSurvivalMet} points out significantly lower cancer survival estimates for women with metastatic BC, as compared to those with pre-metastatic BC (in \figref{fig:NetCancerSurvivalPreMet}), where higher age is associated with lower cancer survival. 
One- and 5-year age standardised survival rates (aged 15 to 99 years) for women diagnosed with Stage 4 BC in 2012 to 2016, followed up to 2017, are reported to be 66\% and 27.9\%, respectively, 
whilst even 5-year survival estimates remain very high for women diagnosed with Stages 1--3 BC, that is around and above 96.5\% 
\citep{ONS2019CancerSurvival}.

Note that survival estimates obtained under M0 are identical to the ones in \figref{fig:NetCancerSurvivalMet} due to the assumptions about rates of transition to death from BC and other causes, respectively (see \tabref{tab:TransitionIntensities}).

\subsection{Deriving the proportion of breast cancer deaths: $k_x$-method}

The industry-based model described in \figref{fig:CI_model} is widely applied in CII by the insurance industry. 
Also, in the absence of reliable cause of death data, a particular approach, known as $k_x$-method, is employed based on this model \citep{DashGramshaw1990, SASMWG2011, IFoA2014, ReynoldsFaye2016}.
Specifically, this approach aims to identify the proportion of deaths from a certain CI condition at age $x$, denoted by $k_x$, with respect to all deaths, in a given year. This is also aimed to be used to determine the proportion of deaths from other causes in the same year. 

In this section we clarify the definition of $k_x$ and  explain how  models M0 and M1
can be used to calculate values of $k_x$. 
This also allows to comment on the performance of these models, based on a comparison between model-calculated and observed $k_x$ values. 

We consider the pricing of a CII product, see \eqref{eq:AxCIMarkov2}, based on M0, 
where two rates of transition are essential: ${\mu}^{01}_x$ and ${\mu}^{02}_x$. 
In the absence of a reliable cause of death data, 
it is challenging to differentiate between CI causes and other causes. 
Provided that all-cause mortality is available, along with an estimate for the proportion of deaths due to CI at a given age $x$, ${k}_x$, 
it is possible to estimate ${\mu}^{02}_x$ as follows:

\begin{align}
	\hat{\mu}^{02}_x  = (1 - k_x) \hat {\mu}_x, 
\end{align}
where $k_x = D_{x}^{13}/D_x$ 
and $D_x$ quantifies the deaths from all causes such that 
\[
D_x = D^{02}_x + D^{12}_x  + D^{13}_x,
\]
with $D^{ij}_x $ being the number of transfers from state $i$ to state $j$ at age $x$ (see Appendix \ref{sec:DetailedKx} and Section~\ref{sec:TransitionRates}). 
Here, estimated all-cause mortality, denoted by $\hat {\mu}_x$, can be derived as 

\begin{align*}
	\hat {\mu}_x =  \frac{D_x }{E^{0}_x + E^{1}_x },
	\label{eq:OverallMort}
\end{align*} 
 where $E^i_x$ is the relevant person-year exposure in the departed state $i = 0, 1$. 
 We also assume that $\mu^{12}_x = \mu^{02}_x$. 

In this paper, we define the proportion of BC deaths at atdtained age $x$ as 

\begin{align*}
	\hat{k}_{x} = \frac{  { {_{x}} p^{01}_0} \mu^{13}_x}{    {_{x}} p^{00}_0 \mu^{02}_x + {_{x}} p^{01}_0  \mu^{12}_x +  { {_{x}} p^{01}_0} \mu^{13}_x  },
\end{align*}
based on M0; and 

\begin{align*}
	\hat{k}_{x} = \frac{  { {_{x}} p^{03}_0 \mu_{x}^{35} } }{{ {_{x}} p^{00}_0 \mu_{x}^{04} + {_{x}} p^{01}_0 \mu_{x}^{14}  + {_{x}} p^{02}_0 \mu_{x}^{24} + {_{x}} p^{03}_0 \mu_{x}^{34}  +{_{x}} p^{03}_0 \mu_{x}^{35} } },
\end{align*}
based on M1 and M2. 
Note that, for simplicity, the formulas above are defined for a cohort starting in State 0 at age 0.

\begin{figure}[H]
	\centering
		\subfloat[Based on M0--M2 \label{fig:KxAge30}]{\includegraphics[width=0.5\textwidth]{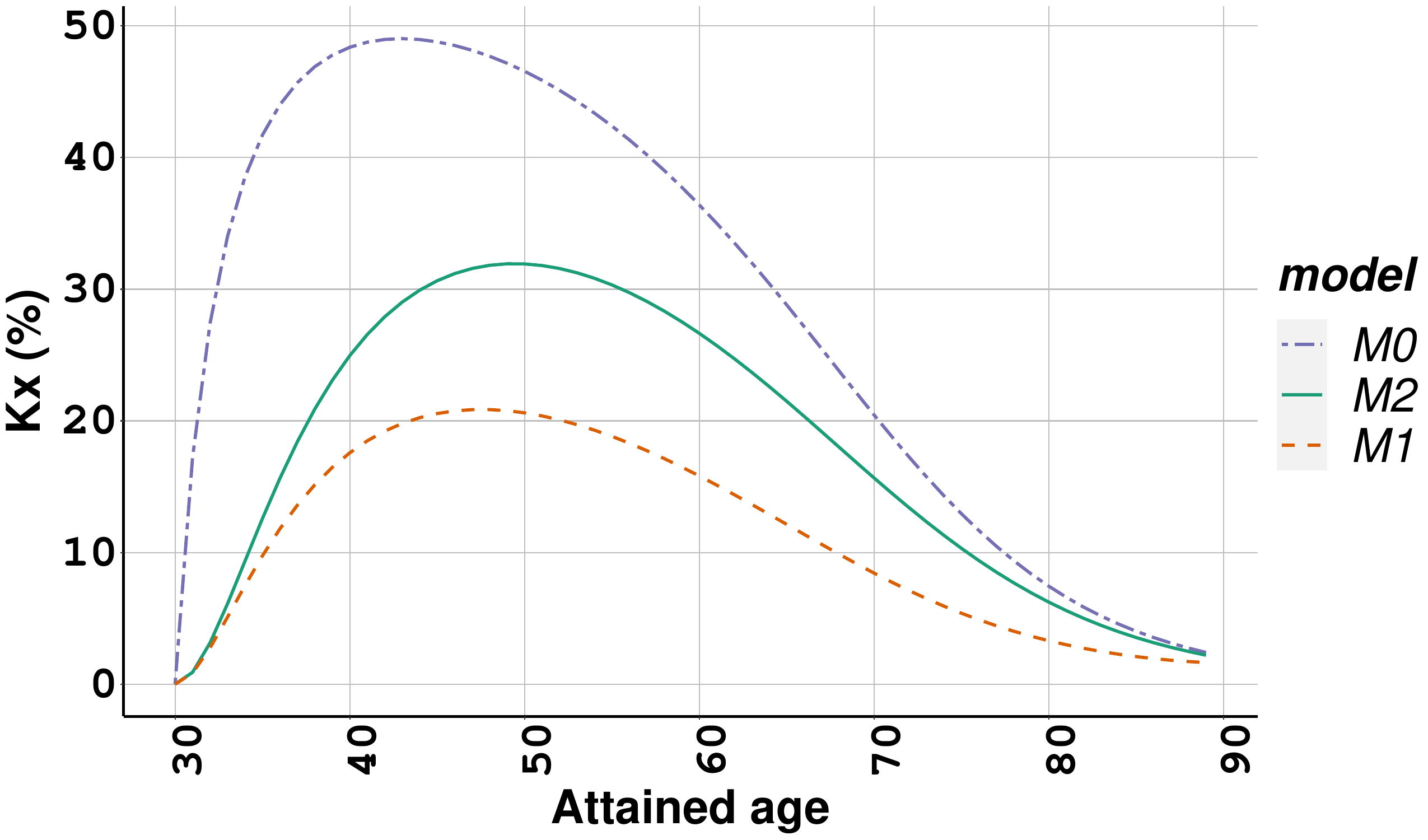}}
	\hfill
	\subfloat[Based on the ONS data \label{fig:KxONSData}]{\includegraphics[width=0.5\textwidth]{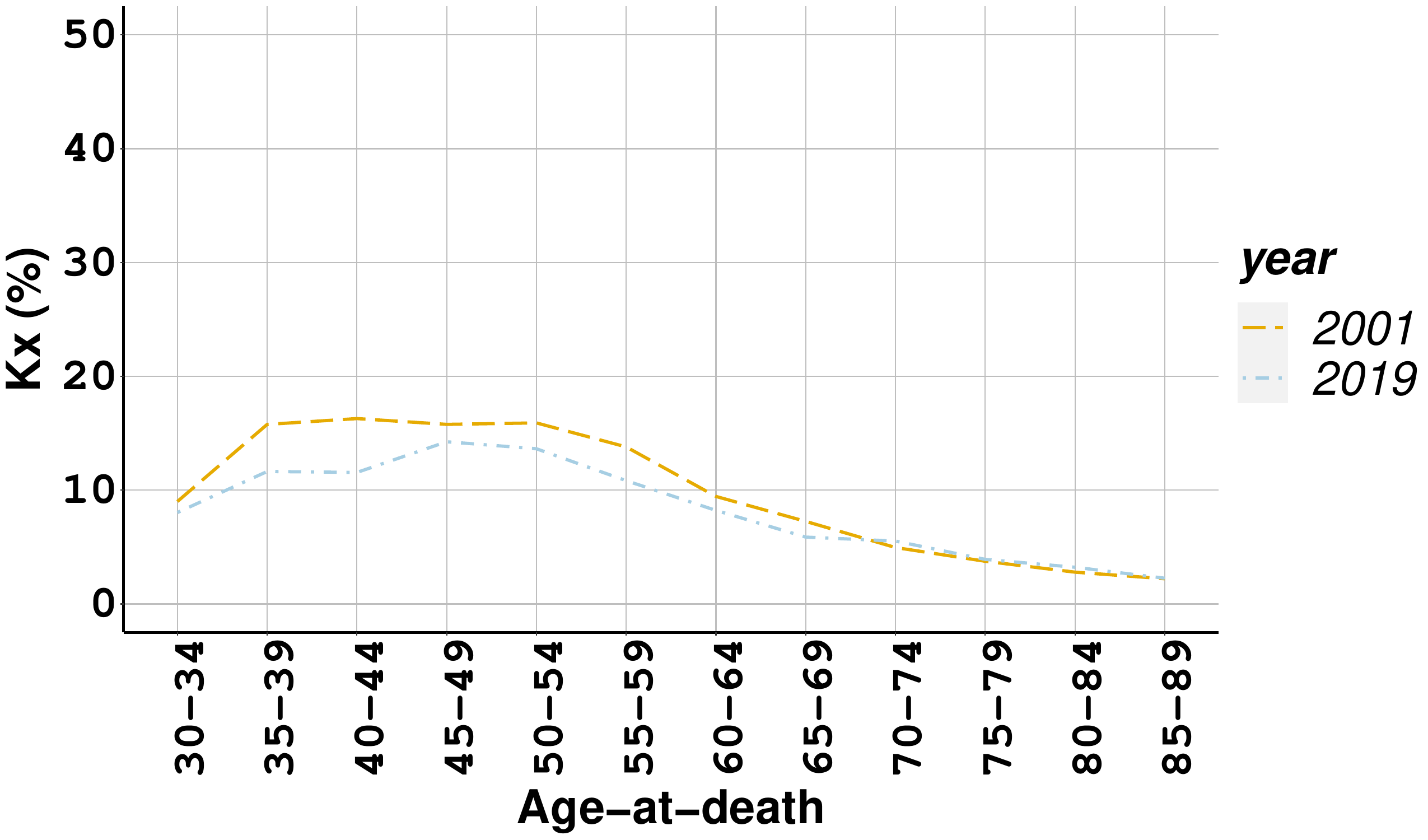}}
	\caption{$\hat{k}_{x}$ values (a) based on different models, and observed $k_x$ values (b) based on the ONS data.}
\label{fig:OccupancyProb_Kx_DiffModels}
\floatfoot{Note: $\text{Attained age} = \text{Age-at-entry} + \text{Time}$
}
\end{figure}

\figref{fig:OccupancyProb_Kx_DiffModels} exhibits estimated and observed values of $k_x$.
\figref{fig:KxAge30} shows values of $\hat{k}_{x}$ for policyholders aged 30, with no BC, at time zero based on different models, where the models are calibrated as described in Sections \ref{sec:TransitionRates}--\ref{sec:GAMKeyTransitions}. 
We see significantly higher estimates based on M0 as compared to the observed values, such as around 48\%, for a woman aged 30, after staying for 10 years in State 0. The same value is estimated to be 18\% and 25\% based on M1 and M2, respectively, which are more aligned with the observed value at ages 40--44.
The main reason for this difference is that the rates of transition to death from BC after a diagnosis under M0, have been defined using the mortality of women with metastatic BC. 
It is important to note that this assumption, made in the absence of relevant data, results to higher than expected BC deaths in that modelling setting.
\figref{fig:KxONSData} displays the proportion of BC deaths over all deaths, based on the population data of England in 2001 and 2019. 
Provided that we rely on average rates of transition in 2001--2019, 
we see that the estimates based on M1 and M2 are broadly aligned with the empirical observation. 

\section{Net premium rates} \label{sec:NetSinglePremiums}

In this section we present net insurance premiums for women at different ages at the time of purchase for the CII contract described in \eqref{eq:AxCIMarkov2} and \eqref{eq:AxSemiMarkov2}, and the life insurance contract described in \eqref{eq:AxIndustry_death1}--\eqref{eq:AxSemiMarkov_death2}.

\subsection{Net single premium rates for the CII contract}

\begin{figure}[H]
	\subfloat[Whole life accelerated CII  \label{fig:WLIupto2019}]{\includegraphics[width=0.5\textwidth]{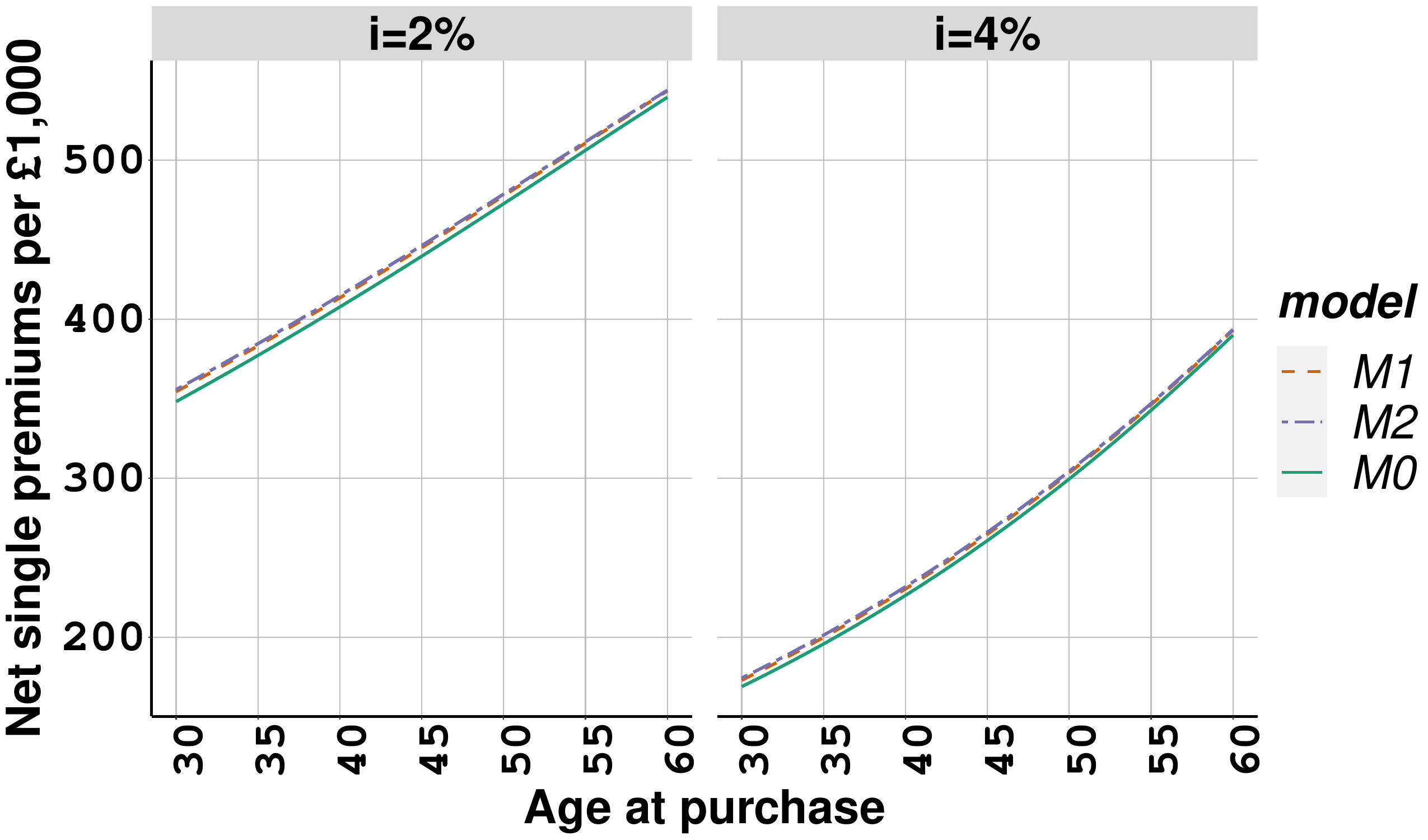}}
	\hfill
	\subfloat[25-year accelerated CII  \label{fig:25TIupto2019}]{\includegraphics[width=0.5\textwidth]{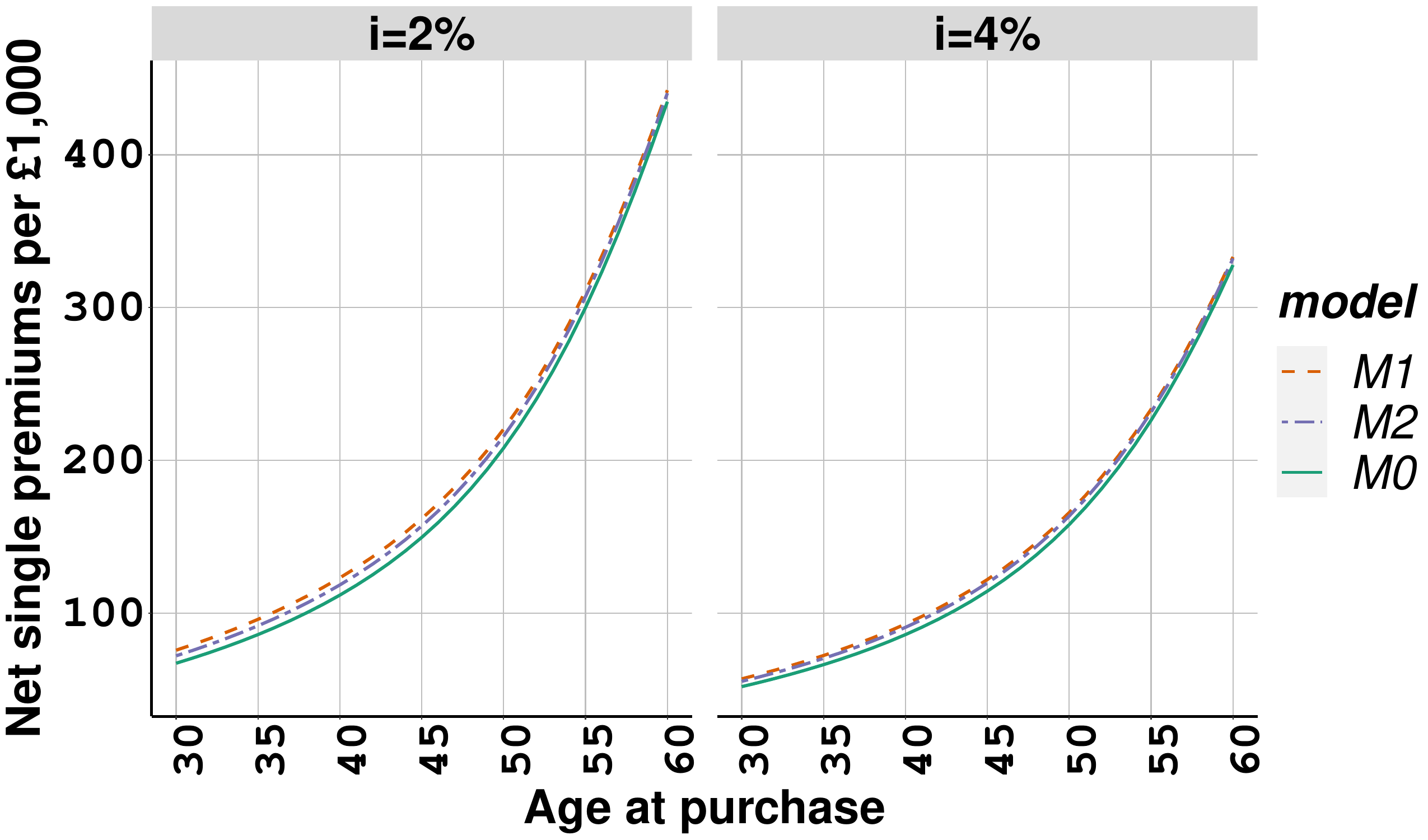}}
	\hfill
	\subfloat[10-year accelerated CII  \label{fig:10TIupto2019}]{\includegraphics[width=0.5\textwidth]{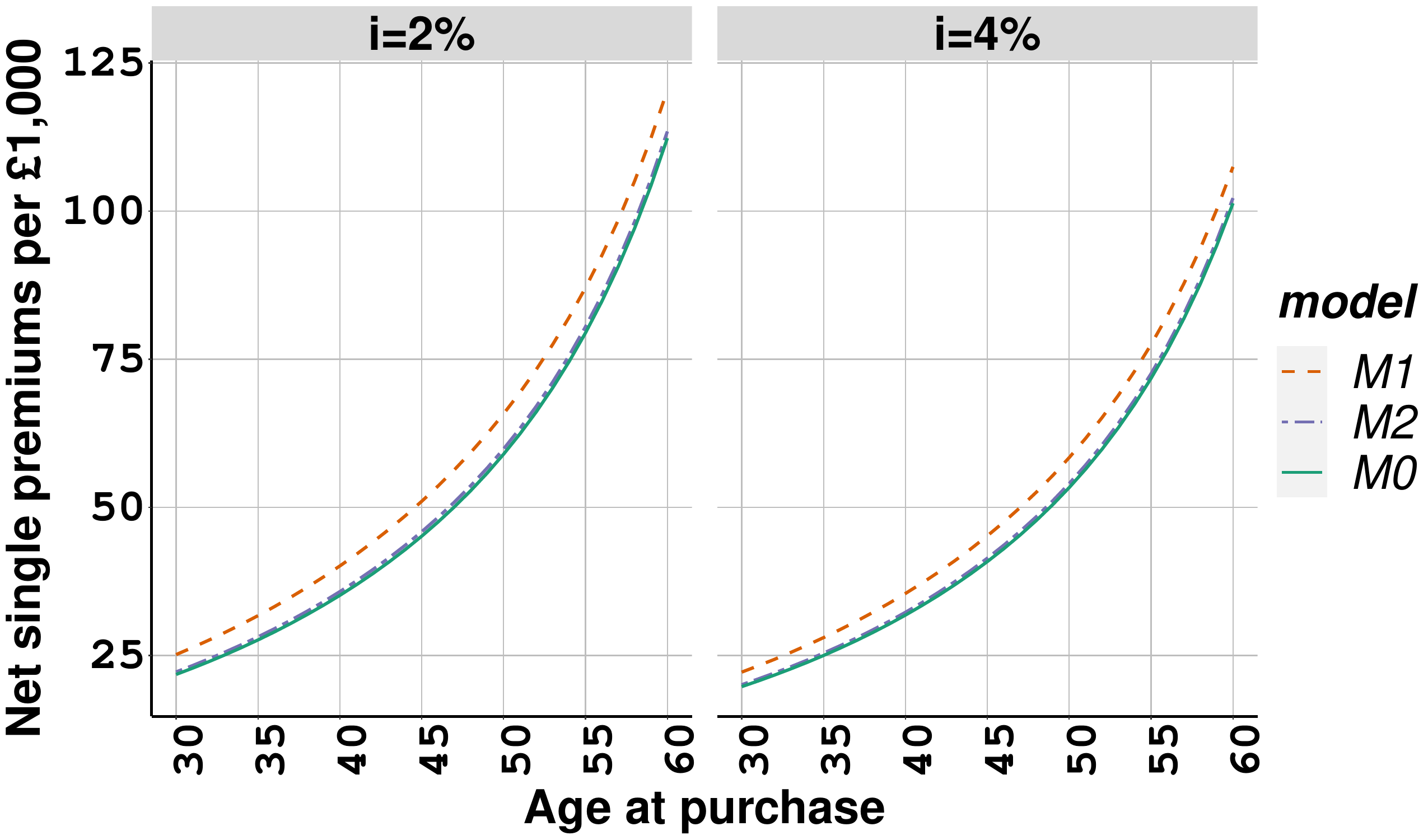}}
	\caption{Net single premium rates of a specialised CII contract, \eqref{eq:AxCIMarkov2} and \eqref{eq:AxSemiMarkov2}, for policyholders with no breast cancer at the time of purchase, \pounds1,000 benefit, {payable at the time of diagnosis of BC or at the time of death}, based on different model assumptions.}
	\label{fig:SinglePremiums_DiffModels}
\end{figure}

\figref{fig:SinglePremiums_DiffModels} shows the net single premiums for the accelerated CII contracts, described in \eqref{eq:AxCIMarkov2} and \eqref{eq:AxSemiMarkov2}, based on M0--M2 at different maturities, for  policyholders at age-at-entry between 30--60. We have used 2\% and 4\% effective rates of interest.
As expected, a higher age-at-purchase, a longer maturity or a lower rate of interest, lead to higher net single premiums.

The highest premiums have been obtained based on the semi-Markov model, M1, 
whereas the lowest premiums are obtained under the industry-based model, M0. 
Time to maturity seems to be a significant factor impacting the differences between estimated premiums under different models.
The differences across these models are more evident with a shorter time to maturity.

Pricing differences across different models may be justified with respect to the following points: 
\begin{enumerate*}[label=(\roman*)]
	\item number of departures from State 0; and
	\item definition of State 1.
\end{enumerate*}
Specifically, under M1 and M2, we allow three departures from State 0. 
One of these three departures, to `Pre-metastatic Unobserved', is not considered under M0.
Meanwhile, one of the other two departures, to State 1, linked to cancer registrations, is defined differently (see Section~\ref{sec:TransitionRates}). 
These have two main consequences. 
First, the occupancy probability ${_ {t}} p^{00}_x$, which is crucial for pricing purposes,  
is estimated to be lower under M1 and M2, as compared to M0, because of the higher number of departures in the former model(s). 
Second, State 1 in M1 and M2 is defined to be a state involving pre-metastatic BC cases, i.e. Stages 1--3 BC, whereas 
State 1 in M0 combines all BC registrations into a single state without accounting for cancer stage information. 
This then implies that we have lower rates of transition to State 1, $\mu^{01}_{x}$, in M1 and M2, in comparison to the rates used in M0 (\tabref{tab:TransitionIntensities}). 

To fully address differences between model pricing results, particularly between M1, M2 compared to M0, we also need to refer to the existence of unobserved BC cases, State 2, under M1 and M2. 
Unobserved BC cases are not considered to be a risk factor under M0. 
Although the occupancy probability from State 0 to State 2, ${_ {t}} p^{02}_x$, is much smaller as compared to ${_ {t}} p^{00}_x$ (\figref{fig:KeyProbabGAMSM}), 
the risk of having a BC without a diagnosis, together with the time spent in this state, seems to become an important factor, especially, to price a 10-year CII contract. 
This is considered to be associated with the increased risk of developing a metastatic BC due to the lack of treatment in the absence of BC diagnosis (see, for instance, \eqref{eq:Parametrisation2}).


Furthermore, we note the differences in the estimated premiums based on M1 and M2. 
We account for the time spent after BC diagnosis and the time spent with BC without having a diagnosis in State 1 and State 2, respectively, under M1, 
whereas we assume a constant rate of transition by ignoring the impact of duration in any of these states under M2. 
Provided that there exists an increased risk of having a more progressed BC in the first 2 years of BC diagnosis, (\tabref{tab:ColzaniData}), 
the estimates under M1 are found to be higher. This is more evident for a 10-year CII contract.
In later years following diagnosis, the fact that this risk gradually decreases, and being stable after 10 years, leads to almost identical results across M1 and M2 for a whole life CII contract. 


\subsection{Net single premium rates for the life insurance contract} \label{sec:NetSinglePremiumsLifeInsur}

\figref{fig:SinglePremiums_LI_DiffModels} displays the estimated net single premiums for the life insurance contracts, described in \eqref{eq:AxIndustry_death1}--\eqref{eq:AxSemiMarkov_death2}, based on M0--M2 at different maturities, for different policyholders aged between 35--60.  Again, we use 2\% and 4\% effective rates of interest.
The figure mainly compares the estimated net single premiums for a woman with no BC, with those for a woman with pre-metastatic BC diagnosis at the time of purchase or 5 years before the time of purchase.

First, we can see smaller net single premiums in \figref{fig:SinglePremiums_LI_DiffModels} for the life insurance contracts for a woman with no BC, as compared to those calculated for the CII contracts (\figref{fig:SinglePremiums_DiffModels}). 
Note that the differences between related premiums are much smaller under M0 due to the definition of State 1 and the assumption linked to BC deaths. 
The main reason for observing smaller life insurance premiums in general, is that BC morbidity is a bigger risk than BC mortality for a healthy woman, and both BC diagnosis or death from other causes lead to a benefit payment under the CII contract. 
Meanwhile, only death from any cause leads to a benefit payment under the life insurance contract.
Besides, in the context of our calculations, for instance under M1, the occupancy probabilities relating to moving to States 1--2 from State 0, are much smaller as compared to 
staying in State 0 (see \eqref{eq:AxSemiMarkov2}, \eqref{eq:AxSemiMarkov_death1}, and \figref{fig:KeyProbabGAMSM}).

\afterpage{%
\begin{landscape}
\begin{figure}[!ht]
	\subfloat[Whole life insurance, $i=2\%$ \label{fig:WLIupto2019_LI_delta1}]{\includegraphics[width=0.50\textwidth]{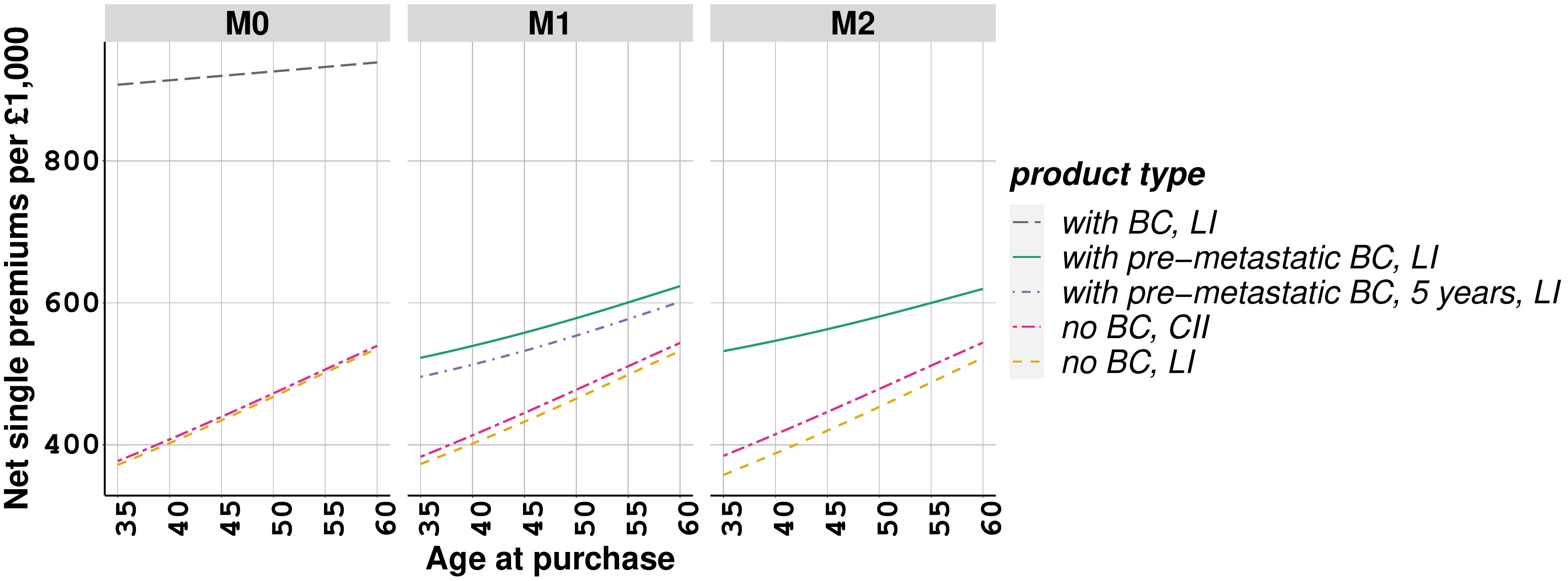}}
	\hfill
			\subfloat[Whole life insurance, $i=4\%$ \label{fig:WLIupto2019_LI_delta2}]{\includegraphics[width=0.50\textwidth]{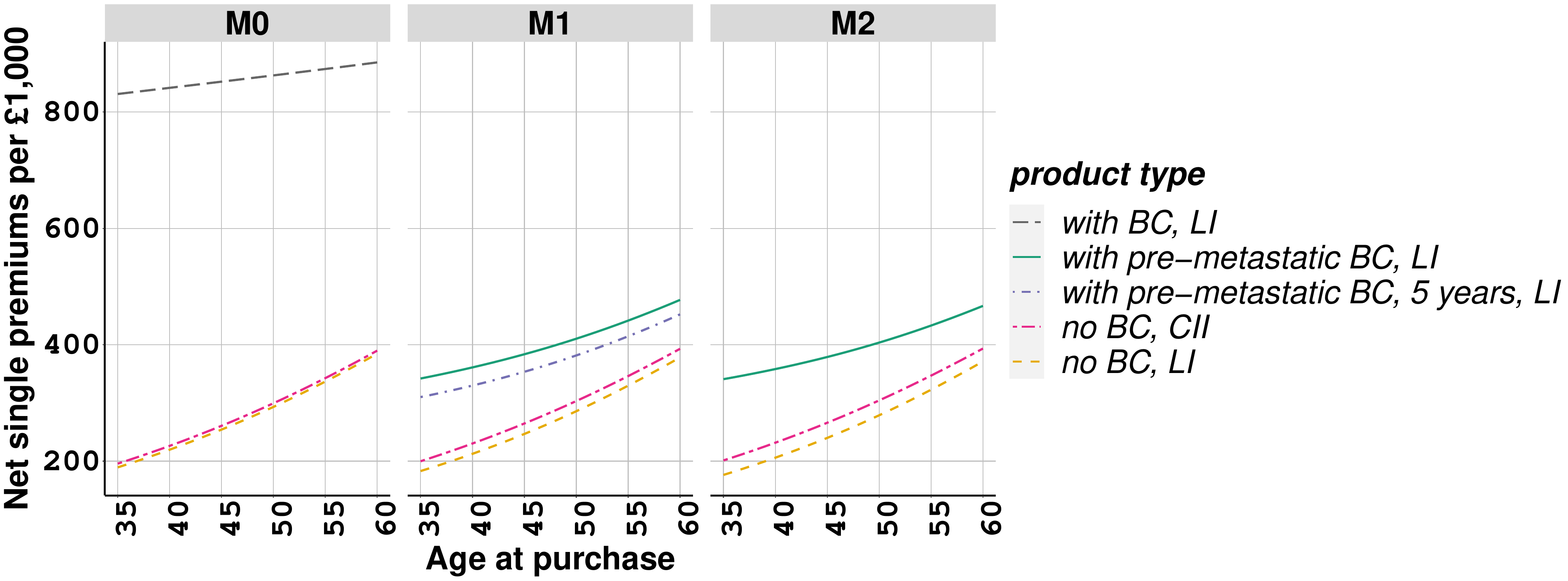}}
	\hfill
	\subfloat[25-year life insurance, $i=2\%$ \label{fig:25TIupto2019_LI_delta1}]{\includegraphics[width=0.50\textwidth]{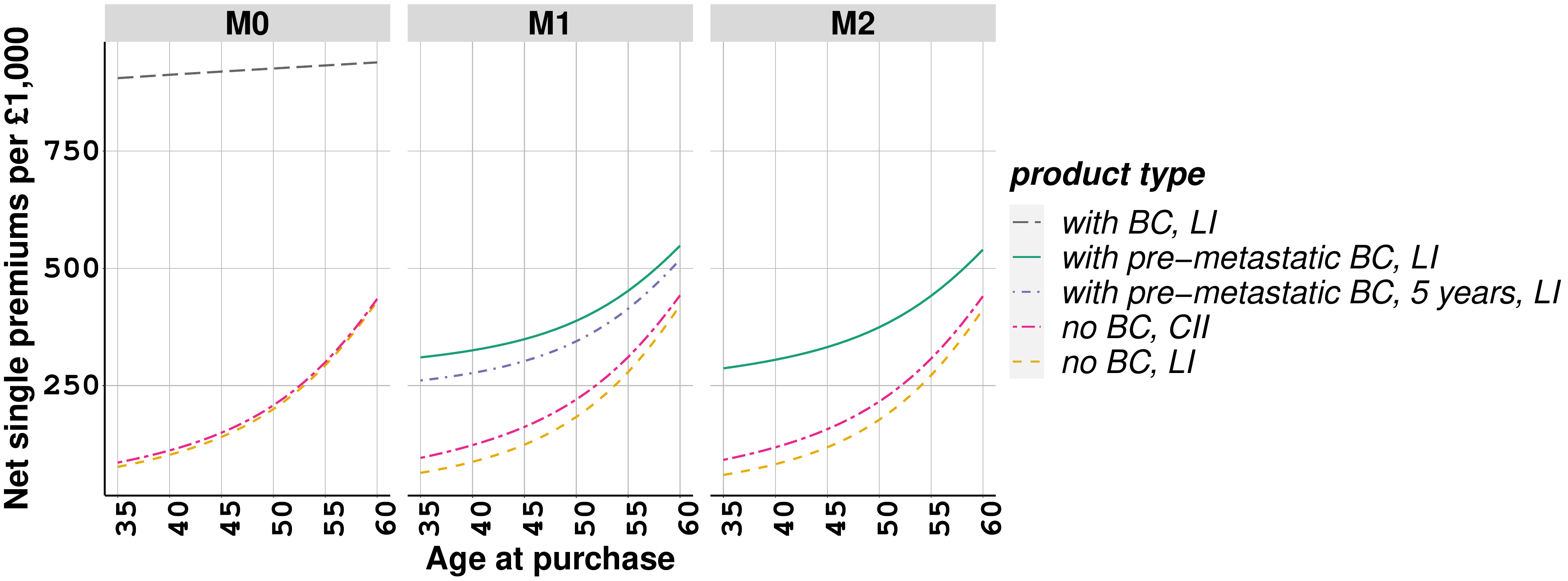}}
	\hfill
	\subfloat[25-year life insurance, $i=4\%$ \label{fig:25TIupto2019_LI_delta2}]{\includegraphics[width=0.50\textwidth]{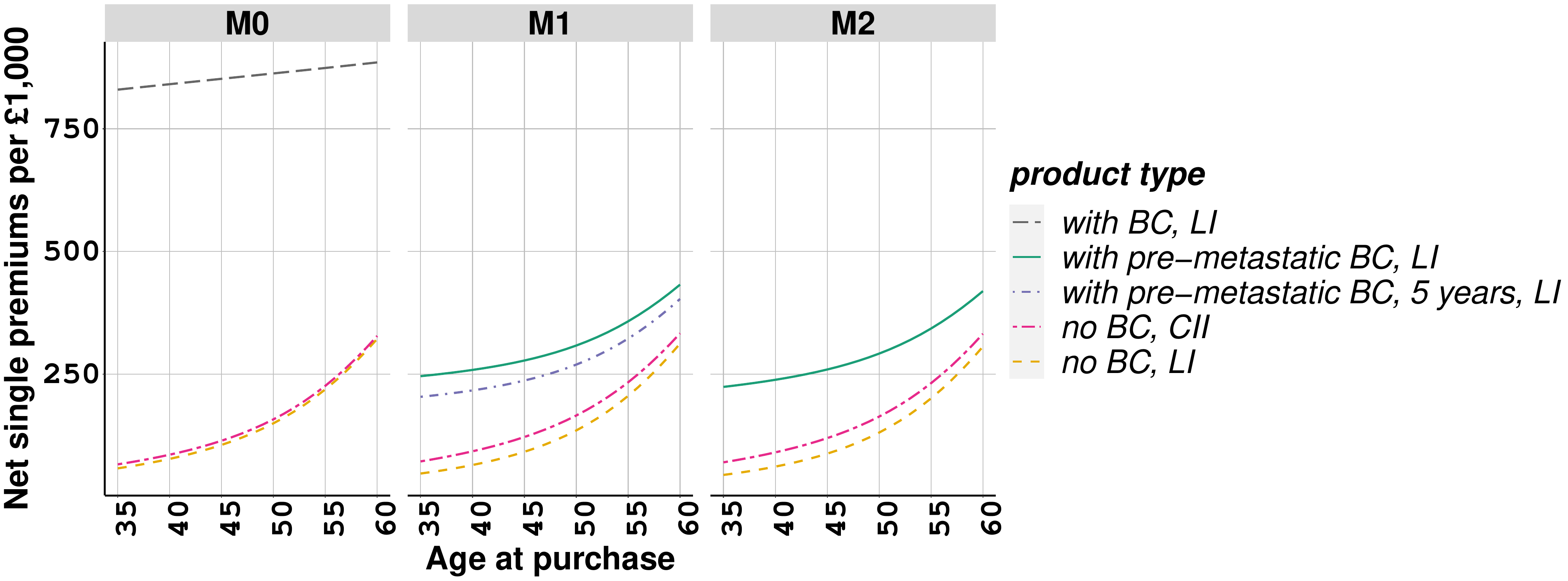}}
	\hfill
		\subfloat[10-year life insurance, $i=2\%$ \label{fig:10TIupto2019_LI_delta1}]{\includegraphics[width=0.50\textwidth]{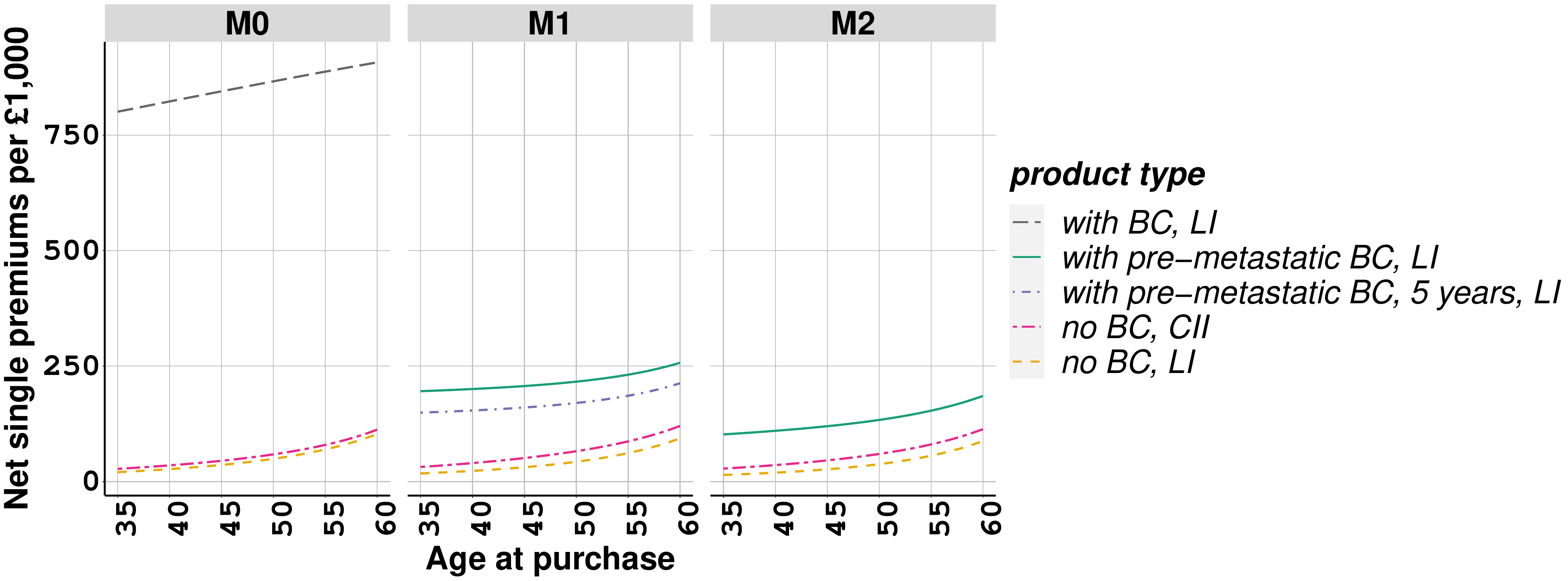}}
	\hfill
	\subfloat[10-year life insurance, $i=4\%$ \label{fig:10TIupto2019_LI_delta2}]{\includegraphics[width=0.50\textwidth]{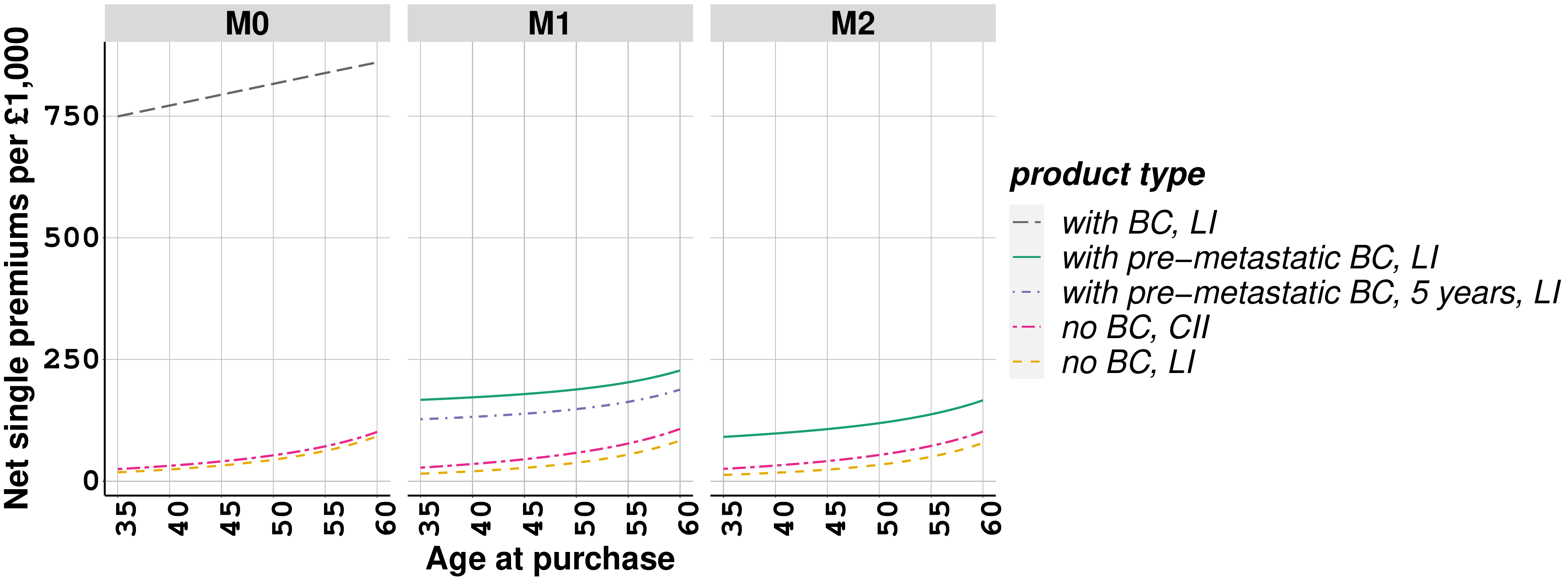}}
	\caption{Net single premium rates of a specialised life insurance contract, \eqref{eq:AxIndustry_death1}--\eqref{eq:AxSemiMarkov_death2}, and \eqref{eq:AxSemiMarkov_death2_v2}, for
		policyholders with or without breast cancer at the time of purchase, \pounds1,000 benefit, {payable at the time of death}, based on M1 and M2, when $\alpha = 0.6$ and $\beta = 1/7$.}
	\label{fig:SinglePremiums_LI_DiffModels}
\end{figure}	
\end{landscape}

}

Furthermore, we observe considerably higher premium estimates for a woman with BC diagnosis at the time of purchase under M0, in comparison to the estimates for a woman with pre-metastatic BC diagnosis under M1--M2. This highlights the importance of our more detailed modelling in M1 and M2.

Once again, our findings for women with pre-metastatic BC under M1 and M2 point towards the significance of modelling for the time spent with observed and unobserved pre-metastatic BC. 
The combined impact on the insurance premiums leads to higher premiums under M1, as compared to M2. This is more evident for a contract with shorter maturity. 
We can see, e.g. in a whole life insurance contract under M1, that the differences in the estimated premiums for women with no BC or pre-metastatic BC diagnosis at the time of purchase, get smaller with higher age-at-purchase or age-at-diagnosis and a longer time to maturity (\figref{fig:WLIupto2019_LI_delta1} and \figref{fig:WLIupto2019_LI_delta2}). 
This can be because the risk of developing a metastatic cancer, after being diagnosed with pre-metastatic BC, becomes fairly stable after about 5 years with a peak at about the first 2 years (see \tabref{tab:ColzaniData}, and also see Figure 1 in \citet{Colzanietal2014}), where BC risk becomes a relatively lower risk as opposed to other risk factors.   


We have also estimated net single premiums for a life insurance contract for a woman diagnosed with pre-metastatic BC 5 years ago at time of purchase, under M1. 
This aims to determining the impact of eliminating the high risk years of developing metastatic BC on the net insurance premiums, where we would expect lower insurance premiums by allowing the insurance contract to be tailored in a way to be more inclusive. 
This calculation can be carried out by modifying \eqref{eq:AxSemiMarkov_death2} as follows:

\begin{eqnarray}	\label{eq:AxSemiMarkov_death2_v2}
	{_ {\text{LI}, 4}}\bar{A}_{x} &= &	\int_{0}^{\infty}{ e^{-\delta t} {_ {t}} p^{11}_{[x-5]+5}  \mu^{14}_{x+t} dt} \\ \nonumber
	& + & \int_{0}^{\infty}{   e^{-\delta u}  {_ {u}} p^{11}_{[x-5]+5} \mu^{13}_{[x-5]+5+u} 	\int_{0}^{\infty}{ e^{-\delta t}  {_ {t}} p^{33}_{x+u} \bigg( \mu^{34}_{x+u+t} +  \mu^{35}_{x+u+t}  \bigg) dt} \,\,du  }, 
\end{eqnarray}
where the upper boundary in the integrals would be set differently for a term life insurance contract. 

Our findings suggest lower premiums for a woman after 5 years of BC diagnosis, where smaller differences are observed with an increasing time to maturity.
Our results demonstrate intuitive outcomes, aligned with medical observations.
{Yet, these premiums can be higher than expected by considering, for instance, the premiums for a woman after 5 years of BC diagnosis under `right to be forgotten' initiative \citep{InsuranceEurope2021}.
The difference between the estimated premiums under M1 and the premiums under the `right to be forgotten' initiative could be linked to two assumptions that we maintained during our calculations: (i) the woman with pre-metastatic BC is assumed to be in State 1 at the time of purchase, where the implicit assumption is that the woman is not free of BC, even after 5 years, but the risk of developing metastatic BC has been considerably reduced, and (ii) we haven't accounted for time trend in any of the transition rates in our modelling, including the trend in the risk of developing metastatic BC, which may have improved significantly as a result of medical advances \citep{Colzanietal2014}.
 }

We note that a similar calculation as \eqref{eq:AxSemiMarkov_death2_v2} under M2 would lead to the same results for a woman with pre-metastatic BC diagnosis due to the lack of duration dependence assumption in M2.
Also, such calculation would not be possible under M0, as this model does not distinguish between different cancer stages and it does not account for duration dependence in transition intensities after BC diagnosis.
%

\section{Sensitivity Analysis}\label{sec:Sensitivity}

Our analyses in Section \ref{sec:Numerical_Illustration} assume that $\alpha = 0.6$ and $\beta = 1/7$ under the modelling framework described in \figref{fig:BC_model}, M1 and M2. 
Furthermore, we assume that the rates of transition  from BC to death in M0 can be determined using the rates based on the risk of death from BC for women with metastatic BC (Section \ref{sec:TransitionRates}). 
In this section we examine the sensitivity of estimated net insurance premium values, net cancer survival rates, and proportion of BC deaths out of total deaths, to different values of $\alpha$ and $\beta$ parameters in M1 and M2, and to a different set of rates of transition to death from BC in M0, where relevant. 

\subsection{Impact of  parameter $\alpha$} \label{sec:ImpactAlpha}

The value of the $\alpha$ parameter is associated with the onset of BC, 
allowing us to distinguish between diagnosed and undiagnosed BC cases, 
by considering the proportion of BC diagnoses out of all BC cases. 
We consider $\alpha =0.8$ and $\alpha=0.4$, suggesting a higher and lower proportion of diagnosed BC cases, respectively.
We change the level of $\alpha$ while all other model quantities are calibrated in the same way.   

Our analysis suggests that a change in the value of $\alpha$ has no impact on life insurance premiums, described in \eqref{eq:AxIndustry_death2}, \eqref{eq:AxSemiMarkov_death2} and \eqref{eq:AxSemiMarkov_death2_v2}, for women with (pre-metastatic) BC (see \eqref{eq:Parametrisation2} and \tabref{tab:TransitionIntensities} for the definitions of related transition intensities). 
However, any change in this parameter affects the insurance premiums for a healthy woman (see \eqref{eq:AxSemiMarkov2} and \eqref{eq:AxSemiMarkov_death1}). 
\figref{fig:OccupancyProb_Kx_DiffModels_Alpha1} demonstrates that a lower diagnosis proportion leads to higher number of BC deaths by causing an increase in the insurance premiums across different models,
as compared to 
the baseline calibration when $\alpha=0.6$. 
Meanwhile, a decrease in the premiums has been shown when a higher diagnosis proportion is assumed. 
In particular, the net single premiums for a healthy woman
have changed between 4--30\% under M1 and 2--22\% under M2 across different ages based on the two considered interest rates and different values of $\alpha$.
When $\alpha= 0.4$ we observe bigger changes for women at younger ages, for instance an increase as high as 30\% in the premiums for a 25-year life insurance contract for a woman aged 30 under M1 and around 22\% under M2.

In addition, a lower proportion of BC diagnoses also leads to higher premiums for the CII contract,
as compared to those when $\alpha = 0.6$. 
Once more the differences are bigger for younger age groups and the contracts with shorter time to maturity, especially under the semi-Markov model. 
Specifically, estimated premiums are up to 28\% higher for a 10-year CII contract purchased by a woman aged 30 under M1, when $\alpha=0.4$, whereas the premium under M2 is estimated to be almost 15\% higher. 
The differences in the premiums for a higher proportion of BC diagnoses 
is less dramatic, being estimated as low as 14\% under M1 and 7\% under M2 (as compared to when $\alpha = 0.6$). 

\subsection{Impact of parameter $\beta$} \label{sec:ImpactBeta}

The value of  parameter $\beta$ is associated with the availability of cancer treatment, 
with this parametrisation being used to quantify the difference between rates of transition to metastatic BC from observed or unobserved pre-metastatic BC. 
We consider a higher- or lower-level access to BC treatment 
by changing the value of $\beta$ to be $1/5$ and $1/10$, respectively. 

A change in the value of $\beta$ parameter has an impact on the insurance premiums for a healthy woman, described in \eqref{eq:AxSemiMarkov2} and \eqref{eq:AxSemiMarkov_death1}. 
\figref{fig:OccupancyProb_Kx_DiffModels_Beta1} demonstrates that lower-level access to BC treatment, with $\beta=1/10$, leads to higher number of BC deaths. As a result, there is an increase in insurance premiums, as well. This is as high as 8\%, with larger changes in the youngest age group, across different maturity terms under M1 and 2\% under M2 in comparison to the baseline calibration when $\beta=1/7$.
Meanwhile, higher-level access to BC treatment has resulted in lower premiums, as high as 5\% across different maturity terms under M1 and 2\% under M2, with higher changes at the youngest age group.  

Furthermore, a lower value of $\beta$ also causes higher insurance premiums for the CII contract.
Under both models, changes in the premiums are around or less than 2\% with bigger changes in younger age groups and shorter time to maturity.

\subsection{Impact of assumed rates of transition to death from breast cancer}  \label{sec:Impactmu13}

State 1 in M0 involves all BC registrations, where rates of transition from State 1 to death from BC, $\mu^{13}_x$, at a given age $x$, should be determined by considering the risk of death from BC for women with any type of BC. However, in the absence of relevant data, these have been determined by considering the risk of death from BC for women with metastatic BC \citep{Zhaoetal2020}. 
Although we can assume that death from BC without metastatic BC is rare enough to ignore \citep{RedigandMcAllister2014}, 
this assumption suggests that everyone in State 1 has the same risk of death from BC, similar to a woman with metastatic BC.
Thus, this leads to significantly higher BC deaths, see \figref{fig:OccupancyProb_Kx_DiffModels}, as compared to M1.
We therefore consider the sensitivity of the main findings to a different set of $\mu^{13}_x$ values.

\citet{Giannakeasetal2020} estimate the risk of death from BC for women diagnosed with primary ductual carcinoma in situ (DCIS) between 1995 and 2014, based on the SEER database. 
DCIS is the appearance of cancer cells or tumours within the breast area without showing any presence beyond that area. 
Hereby, this type of cancer can be associated with early stage BC, such as Stage 1 BC. 
The risk of dying from BC has been found to be approximately 3-fold greater than that for women at the same age with no BC in the general population \citep{Giannakeasetal2020}. 
Following this study, we can determine $\mu^{13}_x$ by using the population mortality from BC in England (\figref{fig:BreastMortality}). 
In particular, we use average mortality from BC between 2001 and 2019, with this being increased by a factor of 3 for all ages (\tabref{tab:mu13M2}).
Note that the rates in \tabref{tab:mu13M2} accept that everybody in State 1 would be exposed to the same risk of death from BC similar to a woman with DCIS at the time of diagnosis. 
Although this would lead to lower estimates than expected, we could consider these rates to be, perhaps, a lower boundary to define $\mu^{13}_{x}$.

\begin{table}[H]
	\centering
	\caption{Rates of transition from State 1 to State 3 at different ages in M0, based on the ONS data and \citet{Giannakeasetal2020}.}\label{tab:mu13M2}
	\resizebox{1\columnwidth}{!}{
		\begin{tabular}{lccccccccc}
			\hline
			{Age}             & 30--49 & 50--54    & 55--59    & 60--64    & 65--69     & 70--74     & 75--79    & 80--84      & 85--89     \\
			{$\mu^{13}_{x}$} & 0.00041 & 0.00115 & 0.00150 & 0.00182 & 0.00214 & 0.00274 & 0.00369 & 0.00497 & 0.00687 \\ \hline
		\end{tabular}
	}
\end{table}
 
 Changing rates of transition to death from BC has a significant impact on $\hat{k}_x$ values. 
 As a result of using the rates of transition in \tabref{tab:mu13M2}, we have found that the proportion of estimated BC deaths has dramatically declined to a level that is lower than that obtained under M1 (\figref{fig:OccupancyProb_Kx_DiffModels_Mu13}). 
 This result is aligned with the medical literature. 
 The risk of death from BC with DCIS within 20 years is very low, such that 
 approximately 3\% of the women with DCIS would be expected to die from BC \citep{Narodetal2015}.   
 This clearly demonstrates that the rates of transition to death from BC, $\mu^{13}_{x}$, under M0 have a crucial role in identifying $k_x$ values, and they should be determined with caution. 
 
It is also important to note the impact of the change in $\mu^{13}_{x}$ values, on net cancer survival (\figref{fig:NetCancerSurvivalMu13}). 
 As expected, setting $\mu^{13}_{x}$ with respect to the women with DCIS leads to higher cancer survival rates, which is more in line with net cancer survival from a pre-metastatic BC. 
 However, the model seems not to be able to capture the adverse pattern in age-specific cancer survival observed in \figref{fig:NetCancerSurvivalPreMet}.
 
The impact on life insurance premiums is also relevant, with premiums obtained with  $\mu^{13}_{x}$ in \tabref{tab:mu13M2} under M0 being shown in \figref{fig:SinglePremiums_LI_DiffModels_Appendix}, along with the related insurance premiums under M1 and M2, when $\alpha = 0.6$ and $\beta = 1/7$. 
As a result of having considerably fewer BC deaths due to the change in $\mu^{13}_{x}$ (see \tabref{tab:mu13M2}), 
the estimated premiums for a life insurance contract, with or without BC diagnosis at the time of purchase, based on M0, are estimated to be considerably lower. For example, they are lower that the premiums for a woman after 5 years of pre-metastatic BC diagnosis under M1. 
Note that the impact on CII insurance pricing is not discussed here, since the definition of $\mu^{13}_{x}$ is not relevant to the pricing of the CII contract under M0 (see, e.g., \eqref{eq:AxCIMarkov2}).

\section{Discussion} \label{sec:Discussion}

We have examined actuarial net prices for two important insurance products, considering three related models:
an industry-based Markov model (M0), a semi-Markov model (M1), and a Markov model (M2) providing a simplified case of the semi-Markov model. 
We have obtained net single premiums of a specialised CII contract for healthy women and have compared the estimated premiums under these models.
Alongside, we have examined net single premiums for a specialised life insurance contract for women with and without (pre-metastatic) BC. 
We have also considered the overall impact of the COVID-19 pandemic years on insurance premiums, by including observations in 2020 and 2021, to describe the rates of transition from State 0 to State 1 and State 4, i.e. $\mu^{01}_x$ and $\mu^{04}_x$, under M1 and M2 (Section~\ref{sec:Data}). 
We found that net single premiums would only differ by less than 1\% in all cases. 
Our findings under the semi-Markov model give reasonable mortality estimates that are aligned with the medical literature. This is important provided since these estimates are used as an input in pricing calculations. 
The differences in premiums for a given insurance contract across different models 
seem to be reduced with increasing age and longer time to maturity.

We have provided a semi-Markov approach and calibration of the related model (using various sources of data) to show the potential of this methodology as compared to the other presented approaches. While over-interpretation of the absolute numerical results should be avoided, our findings suggest that the industry-based model should be approached with caution, since it is particularly sensitive to model assumptions, such as the risk of death from BC for women with BC diagnosis (i.e. definition of $\mu^{13}_x$).
Furthermore, the industry-based model is not able to capture the relationship between age and BC survival in general. 

The estimates based on the semi-Markov model are broadly in agreement with the empirical evidence related to net cancer survival from pre-metastatic or metastatic BC, and the proportion of BC deaths over all deaths (\figref{fig:OccupancyProb_Kx_DiffModels}, \figref{fig:OccupancyProb_Kx_DiffModels_Alpha1} and \figref{fig:OccupancyProb_Kx_DiffModels_Beta1}). 
Furthermore, our work shows that the semi-Markov model has demonstrated insightful results by combining important information, such as 
cancer stage and the availability of BC diagnostic and treatment services, in a pragmatic way. 
Our results also demonstrate the significance of assuming duration dependence in the modelling, i.e. accounting for the time spent with pre-metastatic BC. 
This model, together with our findings, can help life insurers understand the impact of different modelling assumptions on insurance cash flows. 
This is also particularly relevant when considering new insurance products developed to meet the needs of individuals with medical history of BC. 

There are ongoing efforts to develop new medical technologies, such as liquid biopsy, or increase the availability of - and access to - cancer screening programmes in order to improve early cancer diagnosis.
In this context, the semi-Markov model can offer a useful tool, for instance in quantifying the effect of early BC diagnosis on BC survival. 
This would also be relevant to medical underwriting for related insurance contracts. 
In particular, the semi-Markov model can help insurers implement a more inclusive approach for underwriting related life insurance as demonstrated in Section~\ref{sec:NetSinglePremiumsLifeInsur}.

At the same time, our analysis shows that the results under M1 and M2 can be sensitive to the choice of $\alpha$ and $\beta$ parameters, especially at the youngest age group for a short-term contract. 
We note that this can be linked to how key transition intensities at different ages, or cancer stages are specified in the models. For example, in the absence of comprehensive data, we have grouped ages 30--49 to be a single age group, and Stages 1--3 BC to represent pre-metastatic BC. These groupings may be too broad given the nature of BC. 

We also note that our modelling has not taken into account BC potential recovery from pre-metastatic BC, which could have an impact on exposure in State 0 `No BC'. This implies that the exposure in the initial state used in the modelling, may have been lower than expected. 
{Furthermore, we have not accounted for a time trend in cancer incidence, type-specific mortality, or the risk of developing metastatic BC, which may also impact the results.} 
Thus, further research needs to be carried out for incorporating  a possible time trend in these aspects of the models.

\clearpage
\Urlmuskip=0mu plus 1mu\relax
\bibliographystyle{abbrvnat}
\bibliography{reference}

\begin{thebibliography}{43}
\providecommand{\natexlab}[1]{#1}
\providecommand{\url}[1]{\texttt{#1}}
\expandafter\ifx\csname urlstyle\endcsname\relax
  \providecommand{\doi}[1]{doi: #1}\else
  \providecommand{\doi}{doi: \begingroup \urlstyle{rm}\Url}\fi

\bibitem[Alagoz et~al.({2021})Alagoz, Lowry, Kurian, Mandelblatt, and
  et~al.]{Alagozetal2021}
O.~Alagoz, K.~Lowry, A.~Kurian, J.~Mandelblatt, and et~al.
\newblock Impact of the {COVID-19} pandemic on breast cancer mortality in the
  {US}: Estimates from collaborative simulation modeling.
\newblock \emph{Journal of the National Cancer Institute}, 113(11), {2021}.

\bibitem[Ar{\i}k et~al.(2023)Ar{\i}k, Cairns, Dodd, Macdonald, and
  Streftaris]{Ariketal2022}
A.~Ar{\i}k, A.~Cairns, E.~Dodd, A.~Macdonald, and G.~Streftaris.
\newblock Estimating the impact of the {COVID-19} pandemic on breast cancer
  deaths among older women.
\newblock Living to 100 Research Symposium, Society of Actuaries, 2023.

\bibitem[Aviva(2015)]{Aviva2015}
Aviva.
\newblock {UK:}breast cancer accounts for 44\% of female critical illness
  claims, 2015.
\newblock URL
  \url{https://www.aviva.com/newsroom/news-releases/2015/09/uk-breast-cancer-accounts-for-44-of-female-critical-illness-claims-17538/}.

\bibitem[Baione and Levantesi({2018})]{BaioneandLevantesi2018}
F.~Baione and S.~Levantesi.
\newblock Pricing critical illness insurance from prevalence rates: {G}ompertz
  versus {W}eibull.
\newblock \emph{North American Actuarial Journal}, 22(2):\penalty0 270--288,
  {2018}.

\bibitem[Bray et~al.({2004})Bray, McCarron, and Parkin]{Brayetal2004}
F.~Bray, P.~McCarron, and D.~Parkin.
\newblock The changing global patterns of female breast cancer incidence and
  mortality.
\newblock \emph{Breast Cancer Research}, 6(06), {2004}.

\bibitem[CMI({1991})]{CMI1991}
CMI.
\newblock Report no. 12, the analysis of permanent health insurance data.
\newblock Technical report, Faculty of Actuaries and Institute of Actuaries,
  Edinburgh and London, {1991}.

\bibitem[CMI({2011})]{CMI2011}
CMI.
\newblock Report no. 52, cause-specific cmi critical illness diagnosis rates
  for accelerated business, 2003--2006.
\newblock Technical report, Faculty of Actuaries and Institute of Actuaries,
  Edinburgh and London, {2011}.

\bibitem[Colzani et~al.({2014})Colzani, Johansson, Liljegren, Foukakis, and
  et~al.]{Colzanietal2014}
E.~Colzani, A.~Johansson, A.~Liljegren, T.~Foukakis, and et~al.
\newblock Time-dependent risk of developing distant metastasis in breast cancer
  patients according to treatment, age and tumour characteristics.
\newblock \emph{British Journal of Cancer}, 110 (5):\penalty0 1378--84, {2014}.

\bibitem[CRUK(2021)]{CRUK2021}
CRUK.
\newblock Survival for all stages of lung cancer, 2021.
\newblock URL
  \url{https://www.cancerresearchuk.org/about-cancer/lung-cancer/survival}.

\bibitem[CRUK({2021})]{CRUKCIT2021}
CRUK.
\newblock Evidence of the impact of covid-19 across the cancer pathway: Key
  stats.
\newblock Technical report, Cancer Research UK, {2021}.

\bibitem[Dash and Grimshaw({1990})]{DashGramshaw1990}
A.~Dash and D.~Grimshaw.
\newblock Dread disease cover: An actuarial perspective.
\newblock Technical report, Continuous Mortality Investigation, {1990}.

\bibitem[Europe(2021)]{InsuranceEurope2021}
I.~Europe.
\newblock Beating cancer: Right to be forgotten must account for how insurance
  works, 2021.
\newblock URL
  \url{https://www.insuranceeurope.eu/news/2478/right-to-be-forgotten-must-account-for-how-insurance-works/}.

\bibitem[Giannakeas et~al.({2020})Giannakeas, Sopik, and
  Narod]{Giannakeasetal2020}
V.~Giannakeas, V.~Sopik, and S.~Narod.
\newblock Association of a diagnosis of ductual carcinoma in situ with death
  from breast cancer.
\newblock \emph{JAMA Network Open}, 3(9), {2020}.

\bibitem[Henderson et~al.({1988})Henderson, Ross, and
  Bernstein]{Hendersonetal1988}
B.~Henderson, R.~Ross, and L.~Bernstein.
\newblock Estrogens as a cause of human cancer: the {R}ichard and {H}inda
  {R}osenthal {F}oundation {A}ward {L}ecture.
\newblock \emph{Cancer Research}, 48):\penalty0 246--253, {1988}.

\bibitem[Hossain({1994})]{Hossain1994}
S.~Hossain.
\newblock A note on force of mortality.
\newblock Technical report, ACTUARIAL RESEARCH CLEARING HOUSE, {1994}.

\bibitem[iam(2023)]{iam2023}
iam.
\newblock Breast cancer life insurance, 2023.
\newblock URL
  \url{https://iaminsured.co.uk/conditions/breast-cancer-life-insurance/#link-7}.

\bibitem[IFoA(2014)]{IFoA2014}
IFoA.
\newblock Extending the critical path, 2014.
\newblock URL
  \url{https://www.actuaries.org.uk/system/files/documents/pdf/20140217-ifoa-meeting-extending-critical-path-v4.pdf}.

\bibitem[Lai et~al.({2020})Lai, Pasea, Banerjee, and et~al.]{Laietal2020}
A.~Lai, L.~Pasea, A.~Banerjee, and et~al.
\newblock Estimated impact of the covid-19 pandemic on cancer services and
  excess 1-year mortality in people with cancer and multimorbidity: near
  real-time data on cancer care, cancer deaths and a population-based cohort
  study.
\newblock \emph{BMJ Open}, {2020}.

\bibitem[Macdonald et~al.(2018)Macdonald, Richards, and
  Currie]{Macdonaldetal2018}
A.~Macdonald, S.~Richards, and I.~Currie.
\newblock \emph{Modelling Mortality with Actuarial Applications}.
\newblock Cambridge, 2018.

\bibitem[Maringe et~al.({2020})Maringe, Spicer, Morris, Purushotham, Nolte, and
  Sullivan]{Maringeetal2020}
C.~Maringe, J.~Spicer, M.~Morris, A.~Purushotham, E.~Nolte, and R.~e.~a.
  Sullivan.
\newblock The impact of the {COVID}-19 pandemic on cancer deaths due to delays
  in diagnosis in {E}ngland, {UK}: a national, population-based, modelling
  study.
\newblock \emph{The LANCET Oncology}, 21(8):\penalty0 1023--1034, {2020}.

\bibitem[Mariotto et~al.({2014})Mariotto, Noone, Howlader, and
  Cho]{Mariottoetal2014}
A.~Mariotto, A.~Noone, N.~Howlader, and H.~e.~a. Cho.
\newblock Cancer survival: An overview of measures, uses, and interpretation.
\newblock \emph{Journal of the National Cancer Institute. Monographs},
  49:\penalty0 145--186, {2014}.

\bibitem[McDonald et~al.(2008)McDonald, Hertz, and
  Pitman~Lowenthal]{McDonaldetal2008}
M.~McDonald, R.~Hertz, and S.~Pitman~Lowenthal.
\newblock The burden of cancer in {A}sia, 2008.
\newblock URL
  \url{https://cdn.pfizer.com/pfizercom/products/cancer_in_asia.pdf}.

\bibitem[Narod et~al.({2015})Narod, Giannakeas, Sopik, and Sun]{Narodetal2015}
J.~Narod, S.A. and﻿~Iqbal, V.~Giannakeas, V.~Sopik, and P.~Sun.
\newblock Breast cancer mortality after a diagnosis of ductal carcinoma in
  situ.
\newblock \emph{JAMA Oncology}, 1(7):\penalty0 888--896, {2015}.

\bibitem[ONS({2016})]{ONS2016CancerSurvival2}
ONS.
\newblock Cancer survival in {E}ngland: Patients diagnosed between 2010 and
  2014 and followed up to 2015.
\newblock Technical report, Office for National Statistics, {2016}.

\bibitem[ONS(2016)]{ONSData2016}
ONS.
\newblock One-year net cancer survival for bladder, breast, colorectal, kidney,
  lung, melanoma, ovary, prostate and uterus, by stage at diagnosis, 2016.
\newblock URL
  \url{https://www.ons.gov.uk/peoplepopulationandcommunity/healthandsocialcare/conditionsanddiseases/datasets/oneyearnetcancersurvivalforbladderbreastcolorectalkidneylungmelanomaovaryprostateanduterusbystageatdiagnosis}.

\bibitem[ONS({2016b})]{ONS2016CancerSurvival}
ONS.
\newblock Cancer survival by stage at diagnosis for {E}ngland (experimental
  statistics): Adults diagnosed 2012, 2013 and 2014 and followed up to 2015.
\newblock Technical report, Office for National Statistics, {2016b}.

\bibitem[ONS({2017})]{ONS2017CancerSurvival}
ONS.
\newblock Cancer survival in {E}ngland: Adult, stage at diagnosis and childhood
  – patients followed up to 2016.
\newblock Technical report, Office for National Statistics, {2017}.

\bibitem[ONS({2019}{\natexlab{a}})]{ONS2019CancerSurvival}
ONS.
\newblock Cancer survival in {E}ngland: national estimates for patients
  followed up to 2017.
\newblock Technical report, Office for National Statistics,
  {2019}{\natexlab{a}}.

\bibitem[ONS({2019}{\natexlab{b}})]{ONSSurv2019}
ONS.
\newblock Cancer survival statistical bulletins {QMI}.
\newblock Technical report, Office for National Statistics,
  {2019}{\natexlab{b}}.

\bibitem[Ozkok~Dodd et~al.({2014})Ozkok~Dodd, Streftaris, Waters, and
  Stott]{Doddetal2014}
E.~Ozkok~Dodd, G.~Streftaris, H.~Waters, and A.~Stott.
\newblock The effect of model uncertainty on the pricing of critical illness
  insurance.
\newblock \emph{Annals of Actuarial Science}, 9(1):\penalty0 108--133, {2014}.

\bibitem[Redig and McAllister({2013})]{RedigandMcAllister2014}
A.~Redig and S.~McAllister.
\newblock Breast cancer as a systemic disease: a view of metastasis.
\newblock \emph{Journal of Internal Medicine}, 274(2):\penalty0 113--126,
  {2013}.

\bibitem[Reynolds and Faye({2016})]{ReynoldsFaye2016}
C.~Reynolds and M.~Faye.
\newblock Ci pricing detectives.
\newblock Technical report, Partner Reviews, {2016}.

\bibitem[SAS(2011)]{SASMWG2011}
SAS.
\newblock Singapore insured lives: Mortality investigation 2004--2008, 2011.
\newblock URL
  \url{https://www.actuaries.org.sg/files/library/other/Other%20Reports/MortalityInvestigationReport%2020111103v18.pdf}.

\bibitem[SCOR(2020)]{SCOR2023}
SCOR.
\newblock Insuring more cancer survivors through inclusive underwriting, 2020.
\newblock URL
  \url{https://www.scor.com/en/expert-views/insuring-more-cancer-survivors-through-inclusive-underwriting}.

\bibitem[Society(2021)]{ACS2021}
A.~C. Society.
\newblock Cancer facts for women, 2021.
\newblock URL
  \url{https://www.cancer.org/healthy/cancer-facts/cancer-facts-for-women.html}.

\bibitem[Soetewey et~al.({2022})Soetewey, Legrand, Denuit, and
  Silversmit]{Soeteweyetal2022}
A.~Soetewey, C.~Legrand, M.~Denuit, and G.~Silversmit.
\newblock Semi‐markov modeling for cancer insurance.
\newblock \emph{European Actuarial Journal}, {2022}.
\newblock \doi{https://doi.org/10.1007/s13385-022-00308-2}.

\bibitem[Sud et~al.({2020})Sud, Torr, Jones, and et~al.]{Sudetal2020}
A.~Sud, B.~Torr, M.~Jones, and et~al.
\newblock Effect of delays in the 2-week-wait cancer referral pathway during
  the {COVID}-19 pandemic on cancer survival in the {UK}: a modelling study.
\newblock \emph{The LANCET: Oncology}, {2020}.

\bibitem[Sung et~al.({2021})Sung, Ferlay, Siegel, Laversanne, Soerjomataram,
  Jemal, and Bray]{Sungetal2021}
H.~Sung, J.~Ferlay, R.~Siegel, M.~Laversanne, I.~Soerjomataram, A.~Jemal, and
  F.~Bray.
\newblock Global cancer statistics 2020:{ GLOBOCAN} estimates of incidence and
  mortality worldwide for 36 cancers in 185 countries.
\newblock \emph{CA: A Cancer Journal for Clinicians}, 71 (3):\penalty0
  209--249, {2021}.

\bibitem[Surgery(2023)]{IS2023}
T.~I. Surgery.
\newblock Breast cancer life insurance, 2023.
\newblock URL
  \url{https://www.the-insurance-surgery.co.uk/medical-conditions-life-insurance/breast-cancer-life-insurance/}.

\bibitem[Swaminathan and Brenner(2011)]{SwaminathanandBrenner2011}
R.~Swaminathan and H.~Brenner.
\newblock Stastistical methods for cancer survival analysis, 2011.
\newblock URL \url{https://survcan.iarc.fr/survival/chap2.pdf}.

\bibitem[Wood(2017)]{Wood2017}
S.~Wood.
\newblock \emph{Generalized Additive Models: An Introduction with R}.
\newblock Chapman and Hall/CRC, 2017.

\bibitem[Yue et~al.({2018})Yue, Wang, Leong, and Su]{Yueetal2017}
J.~Yue, H.~Wang, Y.~Leong, and W.~Su.
\newblock Using {T}aiwan {N}ational {H}ealth {I}nsurance {D}atabase to model
  cancer incidence and mortality rates.
\newblock \emph{Insurance: Mathematics and Economics}, 78:\penalty0 316--324,
  {2018}.

\bibitem[Zhao et~al.({2020})Zhao, Xu, Guo, and et~al.]{Zhaoetal2020}
Y.~Zhao, G.~Xu, X.~Guo, and et~al.
\newblock Early death incidence and prediction in stage iv breast cancer.
\newblock \emph{Medical Science Monitor}, 26, {2020}.

\end{thebibliography}

\clearpage
\begin{appendices}
	\renewcommand\thetable{\thesection\arabic{table}}
	\renewcommand\thefigure{\thesection\arabic{figure}}
	
 \section{Kolmogorov equations in the alternative model}\label{sec:AppendixKolmogorovCI}
	
	Kolmogorov equations for the industry-based Markov model are given as follows:
	
	\begin{subequations}
		\label{eq:Kolmogorov_Markov}
		\begin{align*}
			\frac{d}{dt}	{_ {t}} p^{00}_x  &= - {_ t} p^{00}_x \,    \bigg(   \mu^{01}_{x+t}   + \mu^{02}_{x+t}  \bigg)  \\
			\frac{d}{dt}	{_ {t}} p^{01}_{x}  &=  {_ {t}} p^{00}_x \mu^{01}_{x+t} - {_ t} p^{01}_{x} \, \, \bigg(  \mu^{12}_{x+t} + \mu^{13}_{x+t}  \bigg) \\
			\frac{d}{dt}	{_ {t}} p^{02}_{x}  &= {_ {t}} p^{00}_x \mu^{02}_{x+t} + {_ {t}} p^{01}_x \mu^{12}_{x+t} \\
			\frac{d}{dt}	{_ {t}} p^{03}_{x}  &=  {_ {t}} p^{01}_x \mu^{13}_{x+t} \\
			\frac{d}{dt}	{_ {t}} p^{11}_{x}  &= - {_ t} p^{11}_x \,    \bigg(   \mu^{12}_{x+t}   + \mu^{13}_{x+t}  \bigg)  \\
			\frac{d}{dt}	{_ {t}} p^{12}_{x}  &=  {_ t} p^{11}_x \mu^{12}_{x+t} \\
			\frac{d}{dt}	{_ {t}} p^{13}_{x}  &=  {_ t} p^{11}_x \mu^{13}_{x+t}
		\end{align*}
	\end{subequations}
	
	\section{Modified Kolmogorov equations with duration dependence in the Semi-Markov model}\label{sec:AppendixKolmogorov}

Modified Kolmogorov equations for the semi-Markov BC model are given as below. 
Note that more details can be found in \citet{CMI1991}, based on a 3-state multiple model, allowing recovery from the disease under inspection along with duration dependence.
Here, in order to make integrals clearer, we introduce actuarial selection notation. 
For instance, $\mu^{13}_{x,t}$ is shown based on select attained age $[x]$ with duration $t$, specifically $\mu^{13}_{x,t} = \mu^{13}_{[x]+t}$.

\begin{subequations}
	\label{eq:KolmogorovSemi_markov}
	\begin{align*}
		\frac{d}{dt}	{_ {t}} p^{00}_x  &= - {_ t} p^{00}_x \,    \bigg(   \mu^{01}_{x+t}   + \mu^{02}_{x+t} +\mu^{04}_{x+t}   \bigg)  \\
		\frac{d}{dt}	{_ {t}} p^{01}_{x}  &= {_ {t}} p^{00}_x \mu^{01}_{x+t} - {_ t} p^{01}_{x} \, \,  \mu^{14}_{x+t} 
		- \int_{u=0}^t{  {_ {u}} p^{00}_x \,\, \mu^{01}_{x+u} \,\,  {_ {t-u}} p^{11}_{[x+u]} \,\,  \mu^{13}_{[x+u]+t-u}\,\, du  }  \\
		\frac{d}{dt}	{_ {t}} p^{02}_x  &=  {_ {t}} p^{00}_x \mu^{02}_{x+t}  - {_ {t}} p^{02}_x \mu^{24}_{x+t}   - 
		\int_{u=0}^{t} {   {_ {u}} p^{00}_x \,\, \mu^{02}_{x+u} \,\,  {_ {t-u}} p^{22}_{[x+u]} \,\,  \mu^{23}_{[x+u]+t-u}\,\, du  } \\
		\frac{d}{dt}	{_ {t}} p^{03}_x  &=  	\int_{u=0}^{t} {   {_ {u}} p^{00}_x \,\, \mu^{01}_{x+u} \,\,  {_ {t-u}} p^{11}_{[x+u]} \,\,  \mu^{13}_{[x+u]+t-u}\,\, du  } +\\\nonumber
		& \int_{u=0}^{t} {   {_ {u}} p^{00}_x \,\, \mu^{02}_{x+u} \,\,  {_ {t-u}} p^{22}_{[x+u]} \,\,  \mu^{23}_{[x+u]+t-u}\,\, du  } 
		- {_ t} p^{03}_x \, \bigg(    \mu^{34}_{x+t} + \mu^{35}_{x+t}  \bigg) \\
		\frac{d}{dt}	{_ {t}} p^{04}_x  &= {_ {t}} p^{00}_x \mu^{04}_{x+t} +  {_ {t}} p^{01}_x \mu^{14}_{x+t}  +
		{_ {t}} p^{02}_x \mu^{24}_{x+t}  + {_ {t}} p^{03}_x \mu^{34}_{x+t}  \\
		\frac{d}{dt}	{_ {t}} p^{05}_x  &=  {_ {t}} p^{03}_x \mu^{35}_{x+t} 
	\end{align*}
\end{subequations}

{We note that the select notation on age $[x]$ is kept in the equations below, where this is based on the assumption of being in the relevant initial state. }

\begin{subequations}
	\label{eq:Kolmogorov_markov}
	\begin{align*}
		\frac{d}{dt}	{_ {t}} p^{11}_{[x]}  &= - {_ t} p^{11}_{[x]} \, \bigg(  \mu^{13}_{[x]+t}   +  \mu^{14}_{[x]+t} \bigg)  \\
		\frac{d}{dt}	{_ {t}} p^{13}_{[x] } &=   {_ {t}} p^{11}_{[x]} \,\, \mu^{13}_{[x]+t}   
		-    {_ {t}} p^{13}_{[x]} \mu^{34}_{[x]+t} 
		-    {_ {t}} p^{13}_{[x]} \mu^{35}_{[x]+t}  \\
		\frac{d}{dt}	{_ {t}} p^{14}_{[x] } &= {_ {t}} p^{11}_{[x]}\,\, \mu^{14}_{[x]+t}  
		+  {_ {t}} p^{13}_{[x]} \mu^{34}_{[x]+t}   \\ 
		\frac{d}{dt}	{_ {t}} p^{15}_{[x]}  &=  	 {_ {t}} p^{13}_{[x]} \mu^{35}_{[x]+t} \\
		\frac{d}{dt}	{_ {t}} p^{22}_{[x]}  &= - {_ t} p^{22}_{[x]} \, \bigg(  \mu^{23}_{[x]+t}   +  \mu^{24}_{[x]+t} \bigg)  \\
		\frac{d}{dt}	{_ {t}} p^{23}_{[x]}  &=  {_ {t}} p^{22}_{[x]} \,\, \mu^{23}_{[x]+t} - {_ {t}} p^{23}_{[x]}\,  \bigg(   \mu^{34}_{[x]+t} +   \mu^{35}_{[x]+t}  \bigg) \\ 
		\frac{d}{dt}	{_ {t}} p^{24}_{[x]}  &=   {_ {t}} p^{22}_{[x]} \,\, \mu^{24}_{[x]+t} +  {_ {t}} p^{23}_{[x]} \,\, \mu^{34}_{[x]+t}  \\
		\frac{d}{dt}	{_ {t}} p^{25}_{[x]}  &=   {_ {t}} p^{23}_{[x]} \,\, \mu^{35}_{[x]+t} \\
		\frac{d}{dt}	{_ {t}} p^{33}_{[x]}  &= - {_ t} p^{33}_{[x]} \, \bigg(  \mu^{34}_{[x]+t}   +  \mu^{35}_{[x]+t} \bigg)  \\
		\frac{d}{dt}	{_ {t}} p^{34}_{[x]}  &=  {_ t} p^{33}_{[x]} \,\, \mu^{34}_{[x]+t}   \\
		\frac{d}{dt}	{_ {t}} p^{35}_{[x]}  &=  {_ t} p^{33}_{[x]} \,\, \mu^{35}_{[x]+t}  
	\end{align*} 
\end{subequations}

\section{Computation of insurance premiums in the Semi-Markov model}\label{sec:AppendixPricing}

We take the net single premium of the CII contract, without a death rider, described in \eqref{eq:AxSemiMarkov2}, as an example, that is

\begin{eqnarray*}
	{_ {\text{CI}, 2}}\bar{A}_{x}  &= &	\int_{0}^{\infty}{ e^{-\delta t} 	{_ {t}} p^{00}_x  \mu^{01}_{x+t}  dt} \\ \nonumber
	& + & \int_{0}^{\infty}{   e^{-\delta u} 	{_ {u}} p^{00}_x  \mu^{02}_{x+u} 	\int_{0}^{\infty}{ e^{-\delta t}  {_ {t}} p^{22}_{[x+u]}  \mu^{23}_{[x+u]+t} dt} \,\,du  }.
\end{eqnarray*}

\noindent The first component in the equation above is straightforward to carry out. 
That is why we focus on the second component between time zero and $n h$ which is 

\[
\int_{u=0}^{ T= n h}{   e^{-\delta u} 	{_ {u}} p^{00}_x  \mu^{02}_{x+u} 	\int_{ t=0}^{nh - u }{ e^{-\delta t}  {_ {t}} p^{22}_{[x+u]} \mu^{23}_{[x+u]+t} dt} \,\,du  },
\]
where we can simplify this expression by denoting the inner integral by $g(x+u, T-u)$ as 
\[
\int_{u=0}^{ T= n h}{   e^{-\delta u} 	{_ {u}} p^{00}_x  \mu^{02}_{x+u} \,\, g(x+u, T-u) du  },
\]
and we can approximately calculate this part of the premium by choosing each time step to be equal to $h$, e.g. based on trapezoidal rule, as follows:
\begin{eqnarray*}
\int_{u=0}^{ T= n h}{   e^{-\delta u} 	{_ {u}} p^{00}_x  \mu^{02}_{x+u}  \,\,  g(x+u, T-u) du  } & \approx &
\frac{h}{2} \Bigg(   e^{-\delta 0}   {_ {0}} p^{00}_x  \,\, \mu^{02}_{x}  \,\,  g(x , nh)   \\  \nonumber 
&+ &2  e^{-\delta h}   {_ {h}} p^{00}_x  \,\,  \mu^{02}_{x+h}  \,\,  g(x + h , nh - h)   \\ \nonumber
&+ &2  e^{-\delta 2h}   {_ {2h}} p^{00}_x \,\,  \mu^{02}_{x+2h}  \,\,  g( x +2 h , nh - 2h)   
+ \ldots \\ \nonumber
&+  &e^{-\delta nh}   {_ {nh}} p^{00}_x  \,\,  \mu^{02}_{x + nh}  \,\,  g( x+nh , nh-nh)  
 \Bigg).
\end{eqnarray*}
Here, the important thing to notice is that $g(x+u, T-u)$ shows different net single premiums to be paid at the time of diagnosis for policyholders with unobserved BC at selected ages $[x+u]$.

\section{Net single premiums of different insurance contracts under M2}

Section \ref{sec:Pricing} explains how to calculate the net single premiums of a special CII contract and a couple life insurance contracts based on M0 and M1. 
This section further explains how to calculate net single premiums of these contracts under M2. 

The net single premium for the CII contract, formulated in \eqref{eq:AxSemiMarkov2}, can be re-defined under M2 as

\begin{eqnarray}	\label{eq:AxMarkov2}
	{_ {\text{CI}, 2}}\bar{A}_{x}  &= &	\int_{0}^{\infty}{ e^{-\delta t} 	{_ {t}} p^{00}_x \bigg( \mu^{01}_{x+t} +  \mu^{04}_{x+t} \bigg) dt} \\ \nonumber
	& + & \int_{0}^{\infty}{   e^{-\delta t} 	{_ {t}} p^{02}_x 	\bigg( \mu^{23}_{x+t} +  \mu^{24}_{x+t}  \bigg) dt }.
\end{eqnarray}

The net single premiums of the life insurance contracts for a woman without BC, \eqref{eq:AxSemiMarkov_death1}, and with pre-metastatic BC, \eqref{eq:AxSemiMarkov_death2}, 
can be re-defined under M2,  as well, where \eqref{eq:AxMarkov_death1} shows the premium for a healthy woman at the time of purchase as

\begin{eqnarray}	\label{eq:AxMarkov_death1}
	{_ {\text{LI}, 2}}\bar{A}_{x} & = & \int_{0}^{\infty}{ e^{-\delta t}  	{_ {t}} p^{00}_x  \mu^{04}_{x+t} +  	{_ {t}} p^{01}_x  \mu^{14}_{x+t} +  {_ {t}} p^{02}_x  \mu^{24}_{x+t}  + 
	{_ {t}} p^{03}_x \big(  \mu^{34}_{x+t} +\mu^{35}_{x+t} \big )  dt},  \\ \nonumber
\end{eqnarray}

\noindent whereas \eqref{eq:AxMarkov_death2} shows the premium for a woman with a pre-metastatic BC diagnosis at the time of purchase as

\begin{eqnarray}	\label{eq:AxMarkov_death2}
	{_ {\text{LI}, 4}}\bar{A}_{x}  &= &	\int_{0}^{\infty}{ e^{-\delta t} 	{_ {t}} p^{11}_x  \mu^{14}_{x+t} dt} \\ \nonumber
	& + & \int_{0}^{\infty}{   e^{-\delta t} 	{_ {t}} p^{13}_x   \bigg( \mu^{34}_{x+t} +  \mu^{35}_{x+t}  \bigg) dt}.
\end{eqnarray}

\section{Deriving an expression for $\mu^{02}_x$ in the absence of cause of death data} \label{sec:DetailedKx}

All-cause mortality, denoted by $\hat {\mu}_x$, can be estimated as follows:

\begin{align*}
	\hat {\mu}_x =  \frac{D_x }{E^{0}_x + E^{1}_x }, 
\end{align*} 
with $D_x = D^{02}_x + D^{12}_x  + D^{13}_x$ based on M0. 

\noindent If we accept that deaths from BC at age $x$, $D^{13}_x$, are a percentage of all deaths, such that $k_x$, and $\mu^{12}_x = \mu^{02}_x$, we could then write
\[
D^{02}_x + D^{12}_x = D_x - k_x D_x.
\]
Here, we can deduce 
\begin{align*}
	\mu^{12}_x &= \mu^{02}_x \Rightarrow  \frac{D^{12}_x }{E_x^1} =  \frac{D^{02}_x }{E_x^0} \Rightarrow D^{12}_x  =  D^{02}_x \frac{ E_x^1 }{E_x^0}, 
\end{align*}
and 

\begin{align*}
	D^{02}_x + D^{12}_x &= D_x - k_x D_x \Rightarrow 	D^{02}_x + D^{02}_x \frac{ E_x^1 }{E_x^0}  = ( 1 - k_x) D_x \\
	\frac{D^{02}_x }{E_x^0 }& = (1 - k_x)  \frac{D_x}{E_x^0 + E_x^1},
\end{align*}
that leads to 
\begin{align*}
	\hat{\mu}^{02}_x  = (1 - k_x) \hat {\mu}_x,
\end{align*}
where rates of transition to death from other causes, $	\hat{\mu}^{02}_x$, are indirectly determined by using all-cause mortality, $\hat {\mu}_x$, and ${k}_x$. 


\section{Main findings in Section \ref{sec:ImpactAlpha}}

\begin{figure}[H]
	\subfloat[Whole life accelerated CII \label{fig:WLIupto2019v2}]{\includegraphics[width=0.5\textwidth]{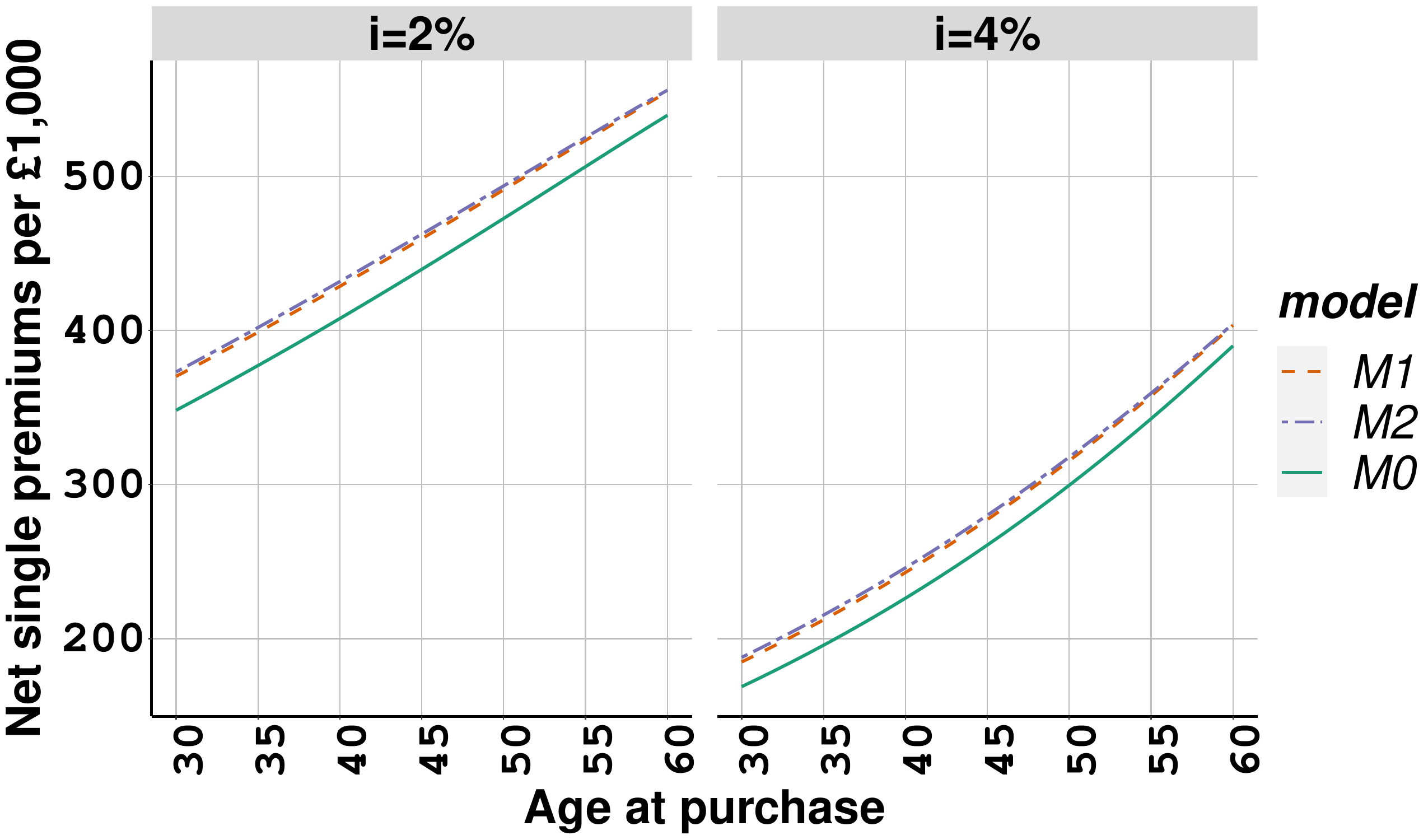}}
	\hfill
	\subfloat[25-year accelerated CII \label{fig:25TIupto2019v2}]{\includegraphics[width=0.5\textwidth]{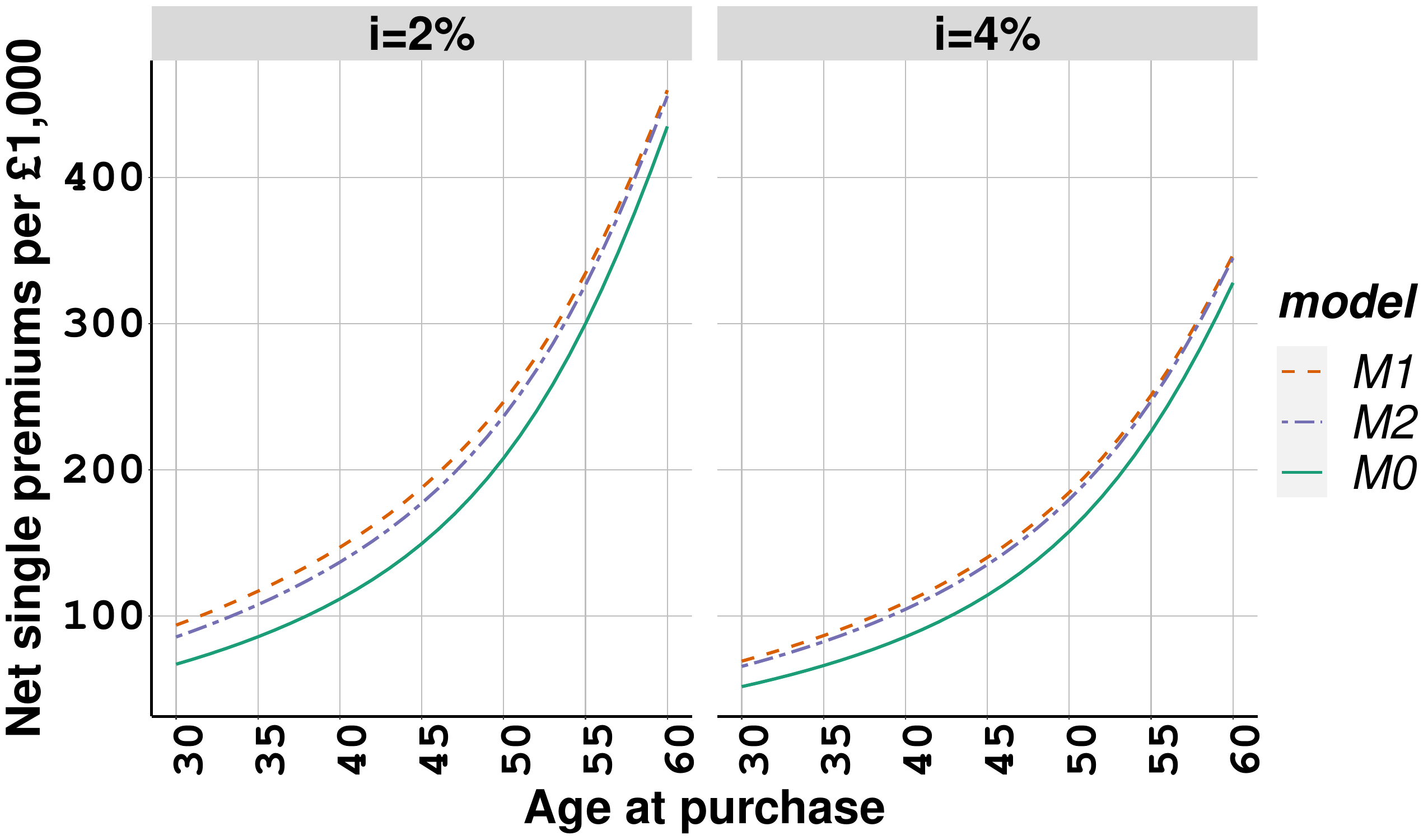}}
	\hfill
	\subfloat[10-year accelerated CII \label{fig:10TIupto2019v2}]{\includegraphics[width=0.5\textwidth]{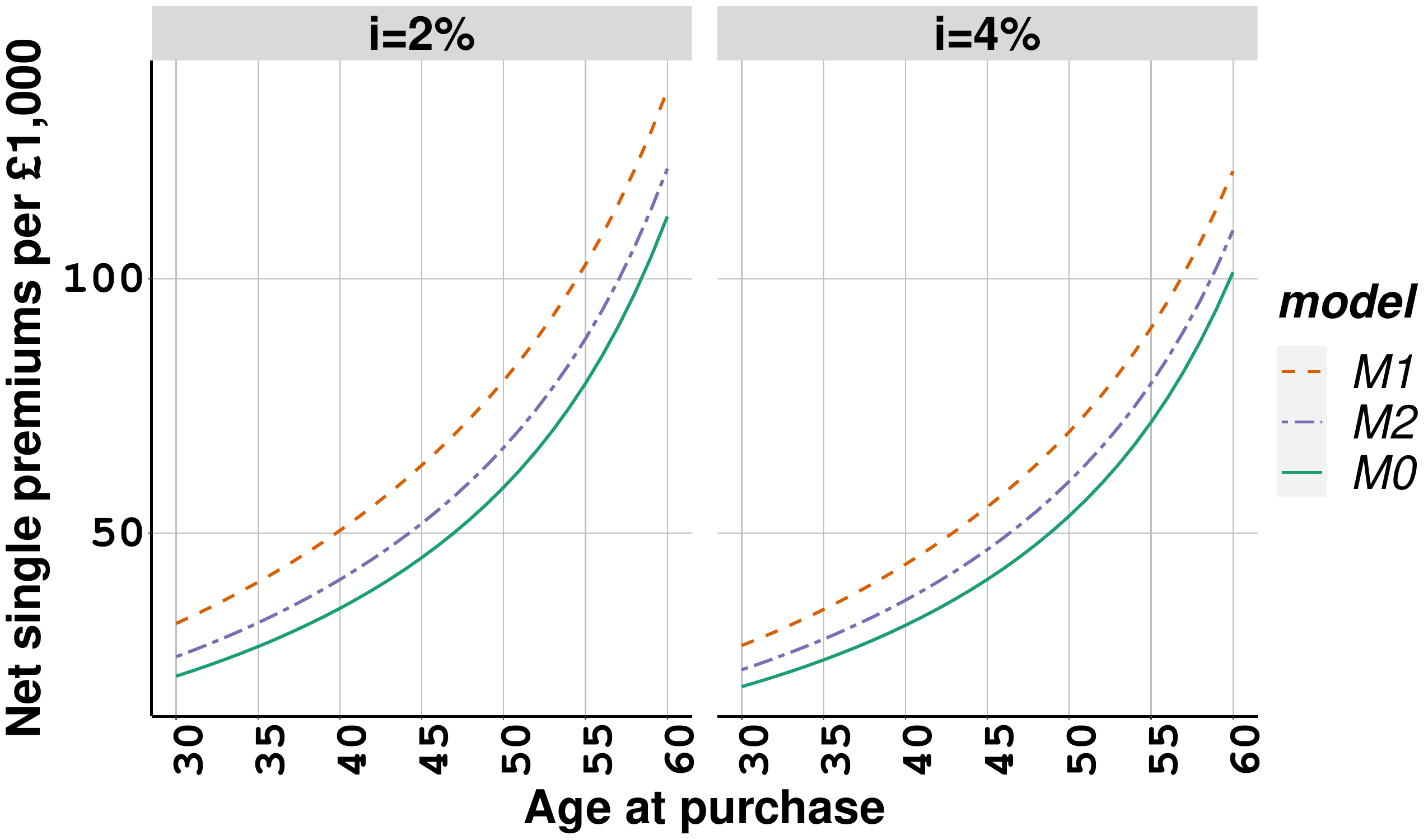}}
	\caption{Net single premium rates of a specialised CII contract, \eqref{eq:AxCIMarkov2} and \eqref{eq:AxSemiMarkov2}, for policyholders between ages 30--60 with \pounds1,000 benefit, {payable at the time of diagnosis of BC or at the time of death}, based on different model assumptions for $\alpha = 0.4$ and $\beta = 1/7$.}
	\label{fig:SinglePremiums_DiffModelsv2}
\end{figure}

\begin{figure}[H]
	\subfloat[Whole life accelerated CII \label{fig:WLIupto2019v3}]{\includegraphics[width=0.5\textwidth]{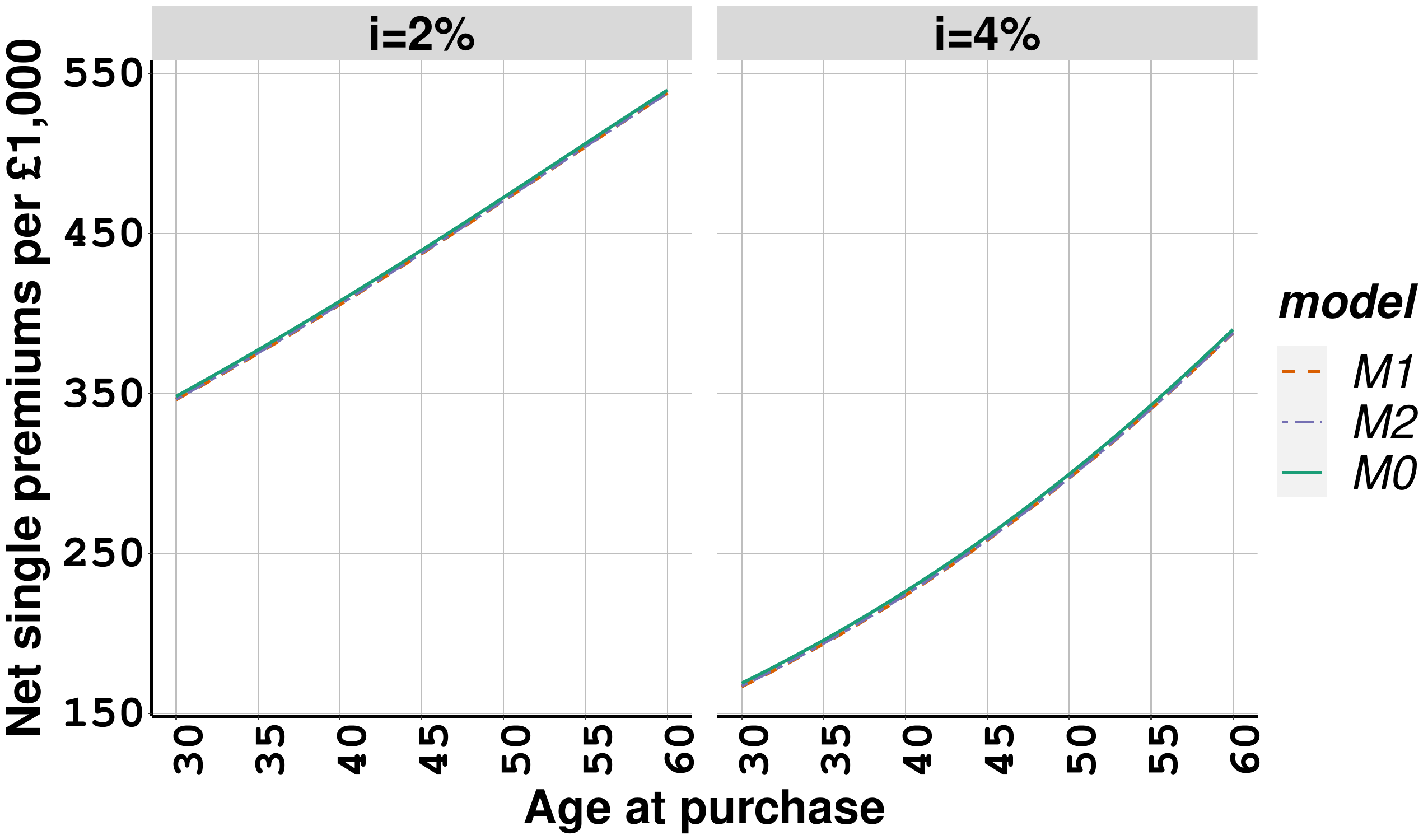}}
	\hfill
	\subfloat[25-year life accelerated CII \label{fig:25TIupto2019v3}]{\includegraphics[width=0.5\textwidth]{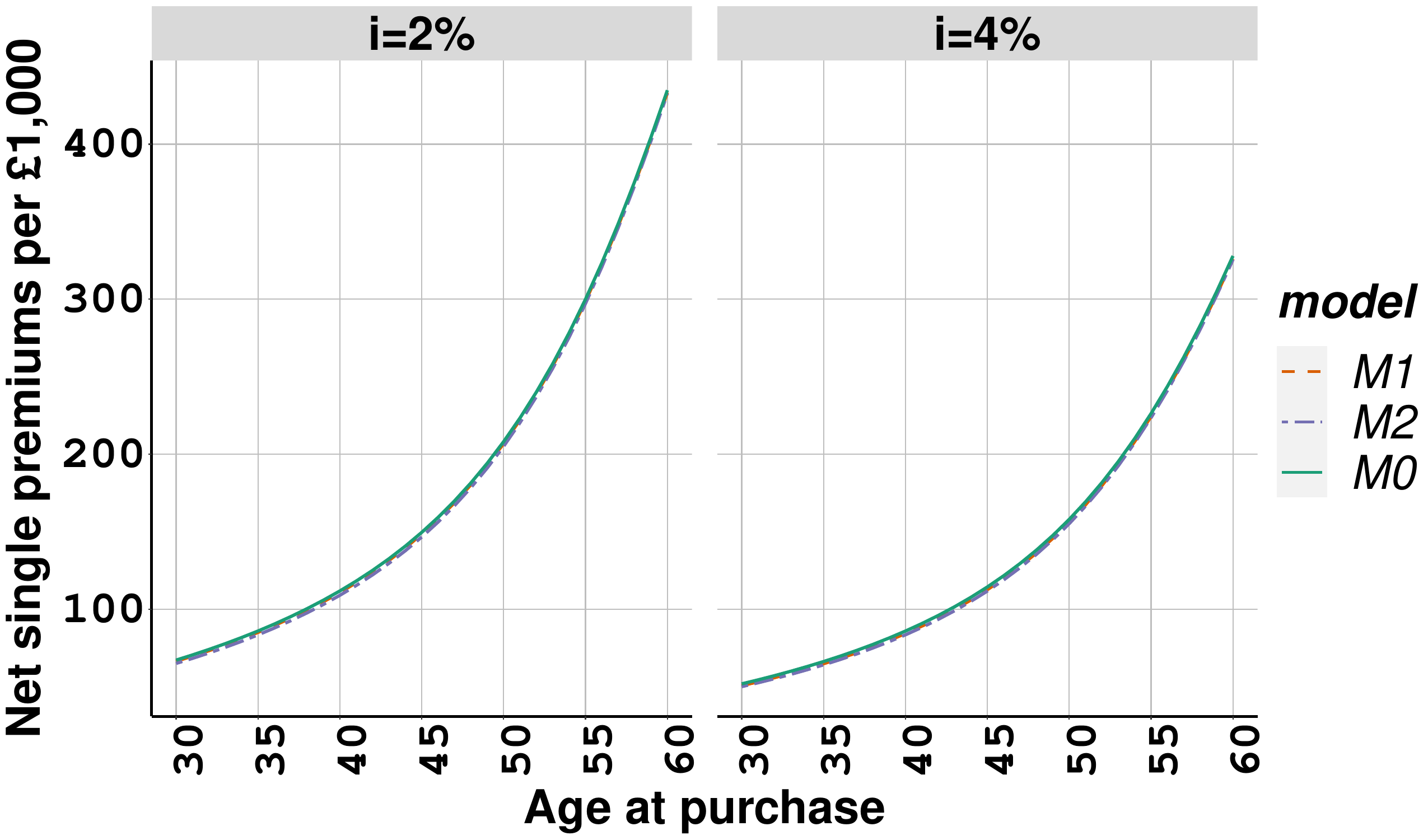}}
	\hfill
	\subfloat[10-year life accelerated CII \label{fig:10TIupto2019v3}]{\includegraphics[width=0.5\textwidth]{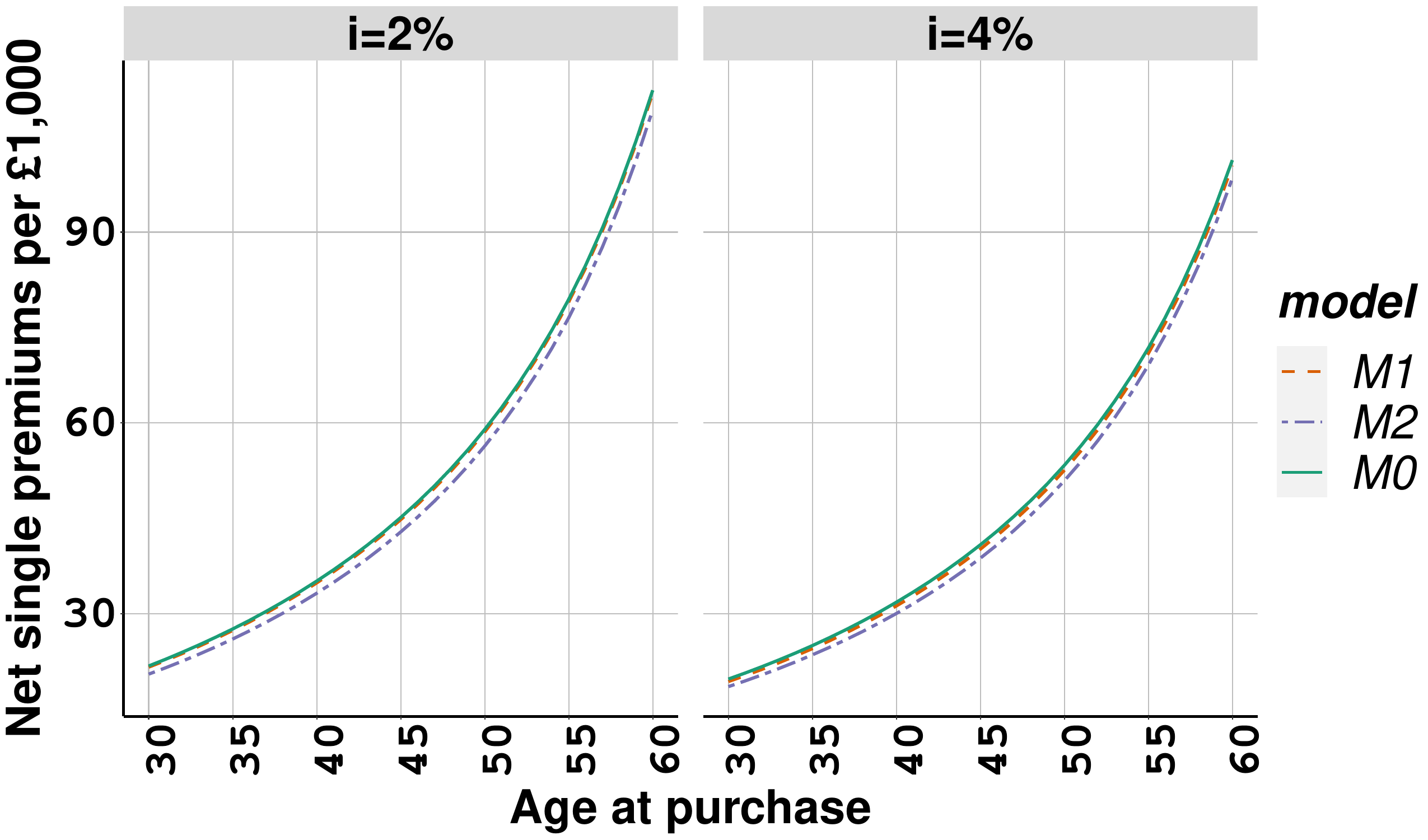}}
	\caption{Net single premium rates a specialised CII contract, \eqref{eq:AxCIMarkov2} and \eqref{eq:AxSemiMarkov2}, for policyholders between ages 30--60 with \pounds1,000 benefit, {payable at the time of diagnosis of BC or at the time of death}, based on different model assumptions for $\alpha = 0.8$ and $\beta = 1/7$.}
	\label{fig:SinglePremiums_DiffModelsv3}
\end{figure}

{%
		\begin{figure}[H]
			\subfloat[Whole life insurance, $i=2\%$ \label{fig:WLIupto2019_LIv2_delta1}]{\includegraphics[width=0.50\textwidth]{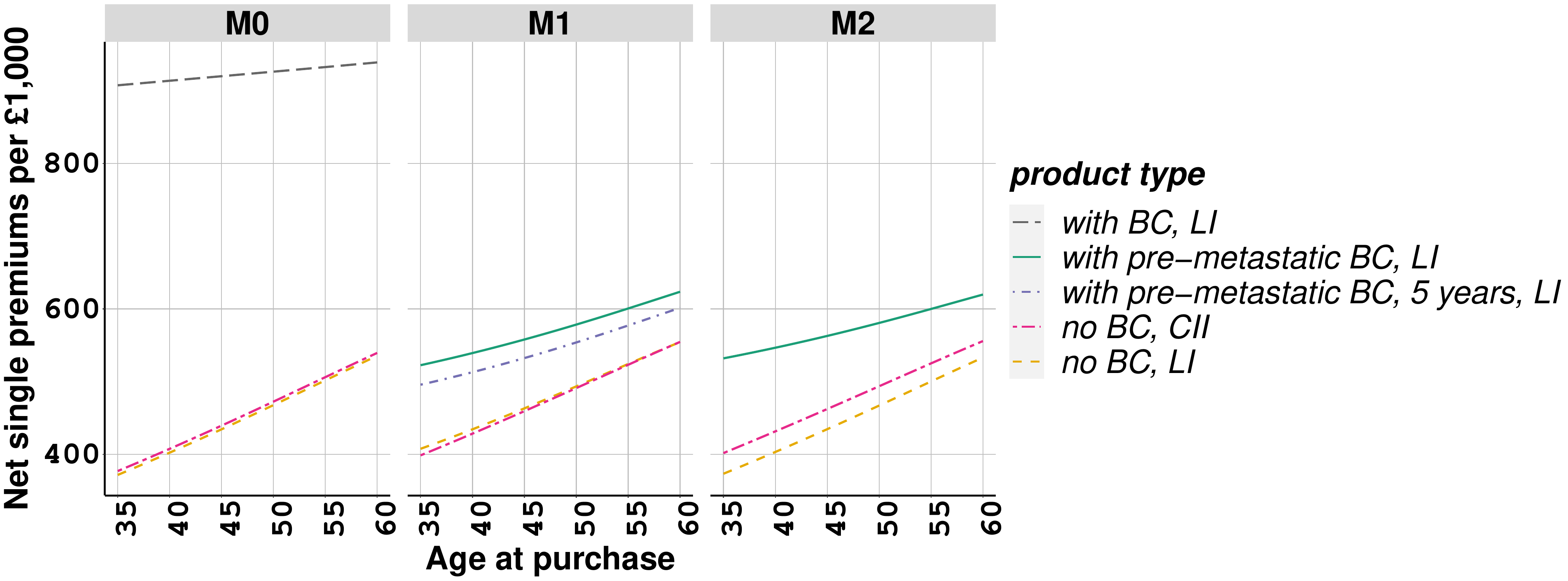}}
			\hfill
			\subfloat[Whole life insurance, $i=4\%$ \label{fig:WLIupto2019_LIv2_delta2}]{\includegraphics[width=0.50\textwidth]{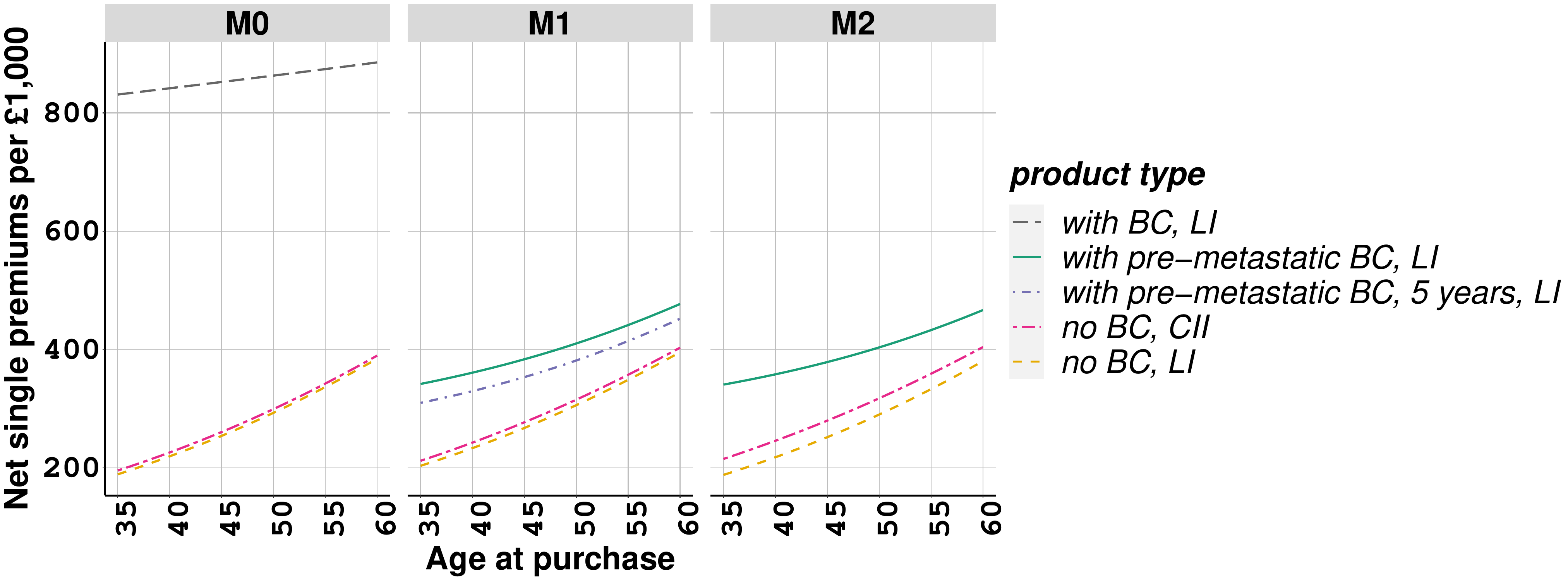}}
			\hfill
			\subfloat[25-year life insurance, $i=2\%$ \label{fig:25TIupto2019_LIv2_delta1}]{\includegraphics[width=0.50\textwidth]{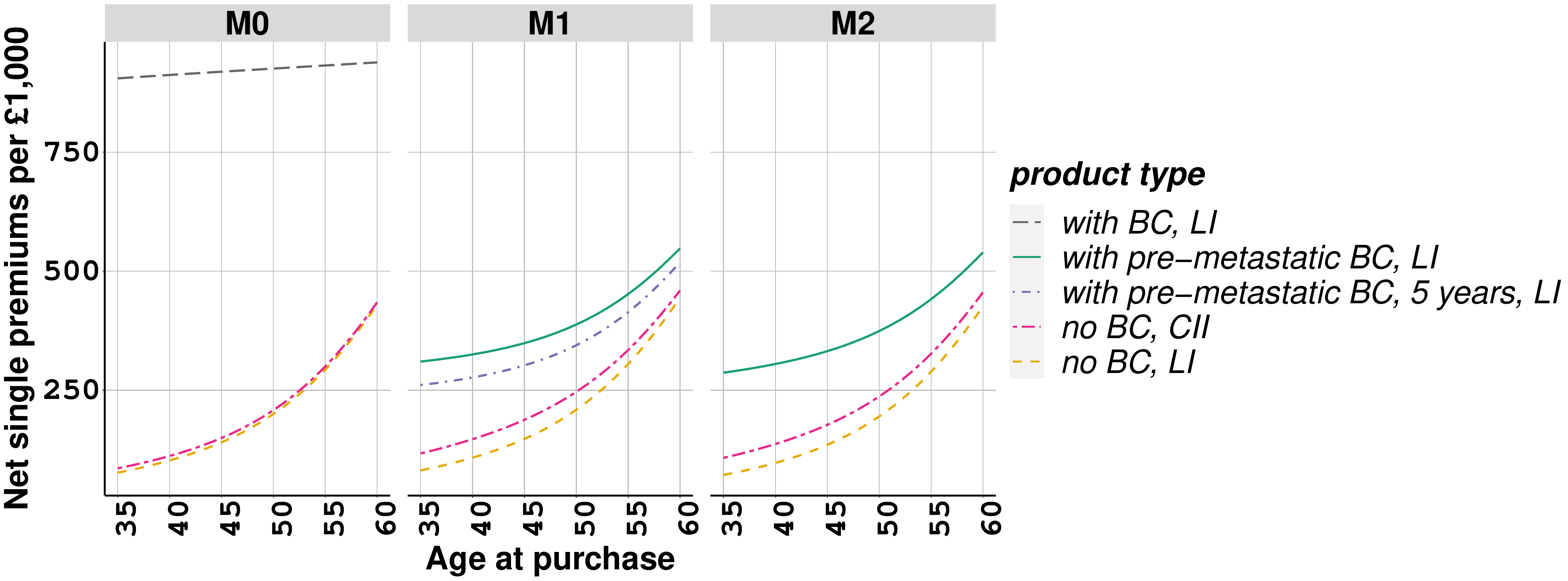}}
			\hfill
			\subfloat[25-year life insurance, $i=4\%$ \label{fig:25TIupto2019_LIv2_delta2}]{\includegraphics[width=0.50\textwidth]{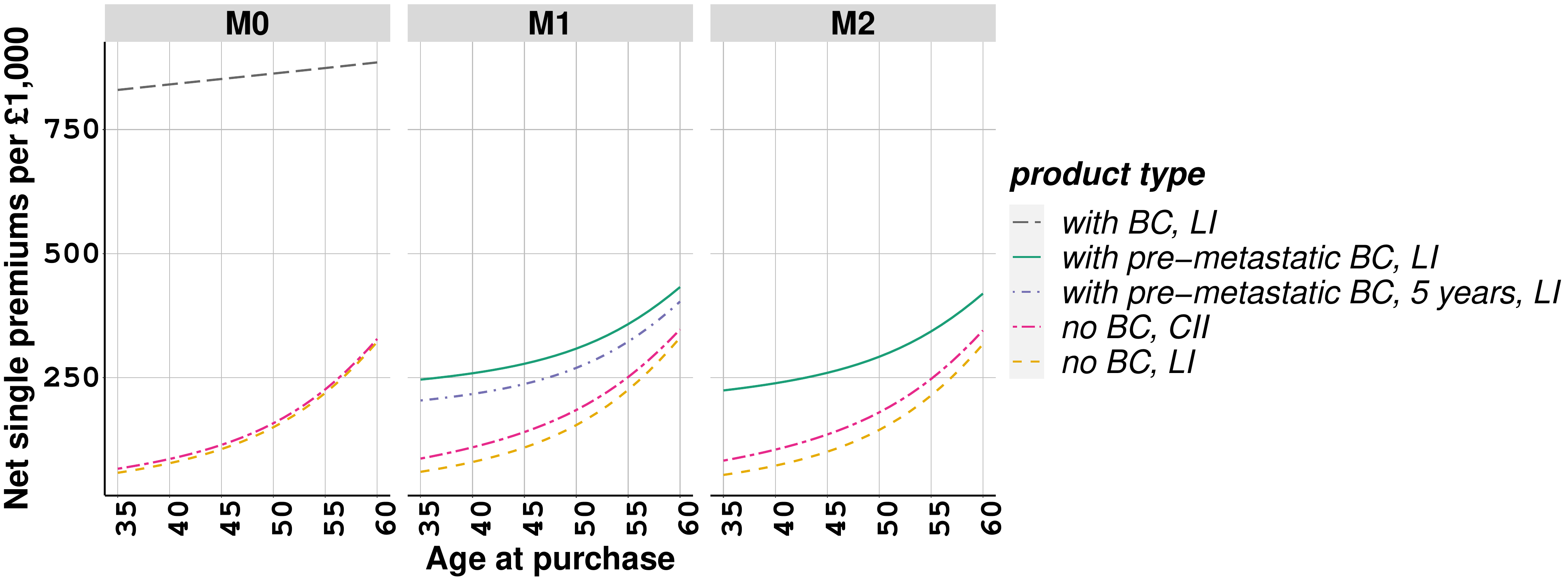}}
			\hfill
			\subfloat[10-year life insurance, $i=2\%$ \label{fig:10TIupto2019_LIv2_delta1}]{\includegraphics[width=0.50\textwidth]{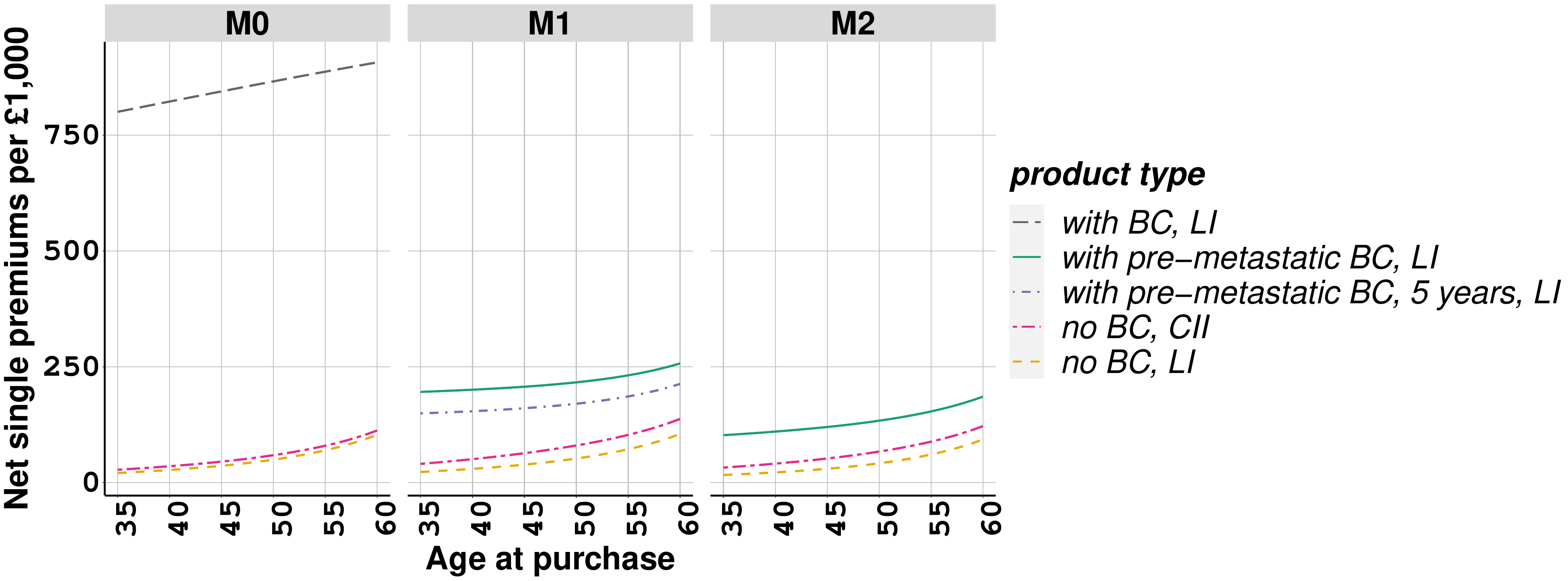}}
			\hfill
			\subfloat[10-year life insurance, $i=4\%$ \label{fig:10TIupto2019_LIv2_delta2}]{\includegraphics[width=0.50\textwidth]{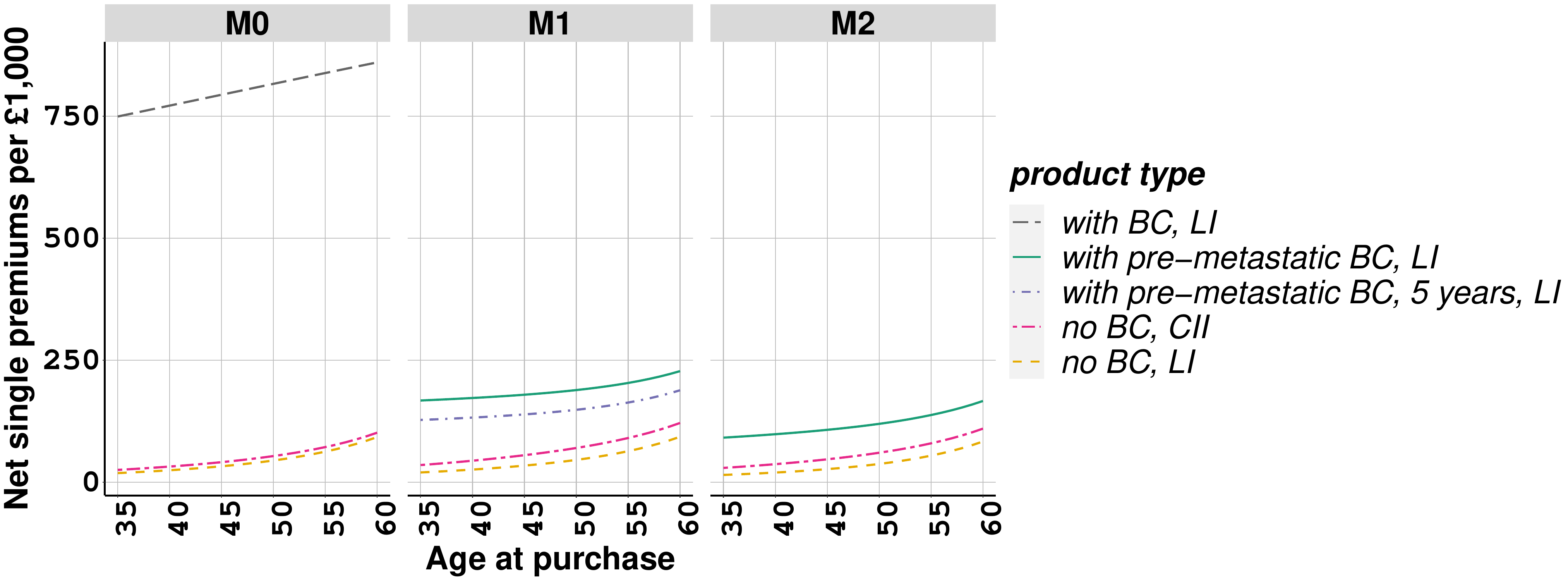}}
			\caption{Net single premium rates of a specialised life insurance contract, \eqref{eq:AxIndustry_death1}--\eqref{eq:AxSemiMarkov_death2} and \eqref{eq:AxSemiMarkov_death2_v2}, for
				policyholders with or without breast cancer at the time of purchase, \pounds1,000 benefit, {payable at the time of death}, based on M1 and M2, when $\alpha = 0.4$ and $\beta = 1/7$.}
			\label{fig:SinglePremiums_LI_DiffModels_v2}
		\end{figure}

	\begin{figure}[H]
		\subfloat[Whole life insurance, $i=2\%$ \label{fig:WLIupto2019_LIv3_delta1}]{\includegraphics[width=0.50\textwidth]{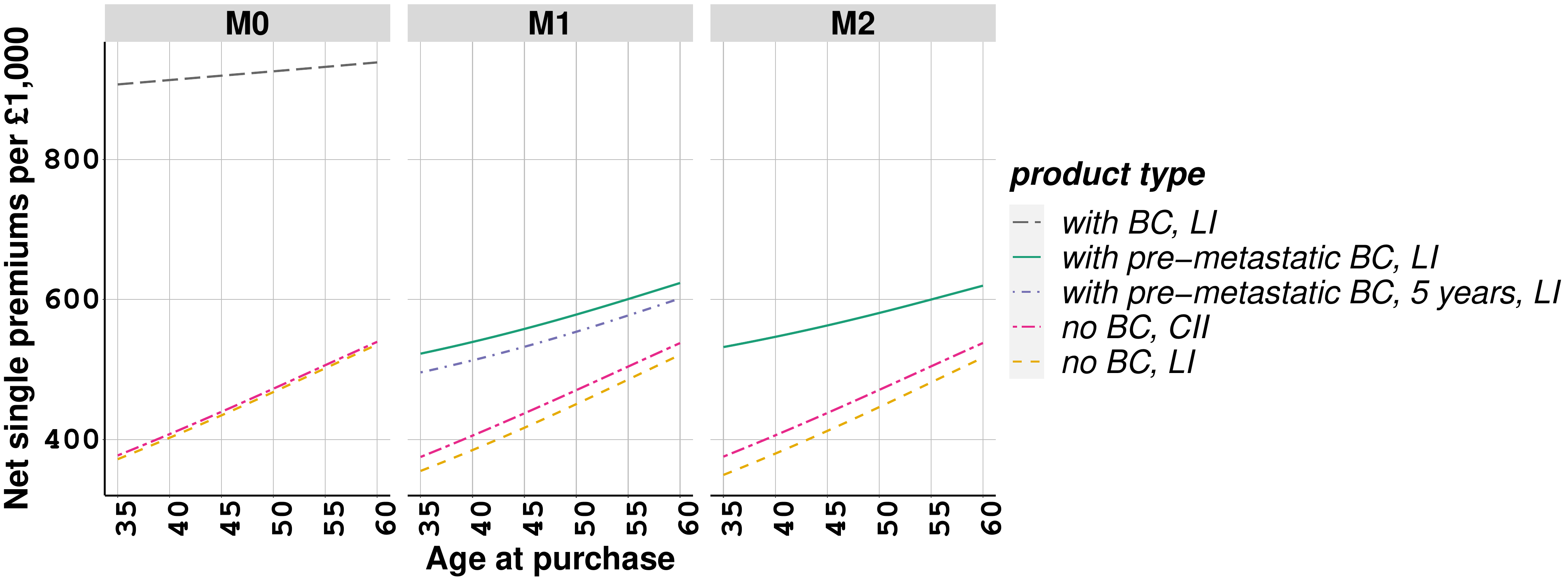}}
		\hfill
		\subfloat[Whole life insurance, $i=4\%$
		\label{fig:WLIupto2019_LIv3_delta2}]{\includegraphics[width=0.50\textwidth]{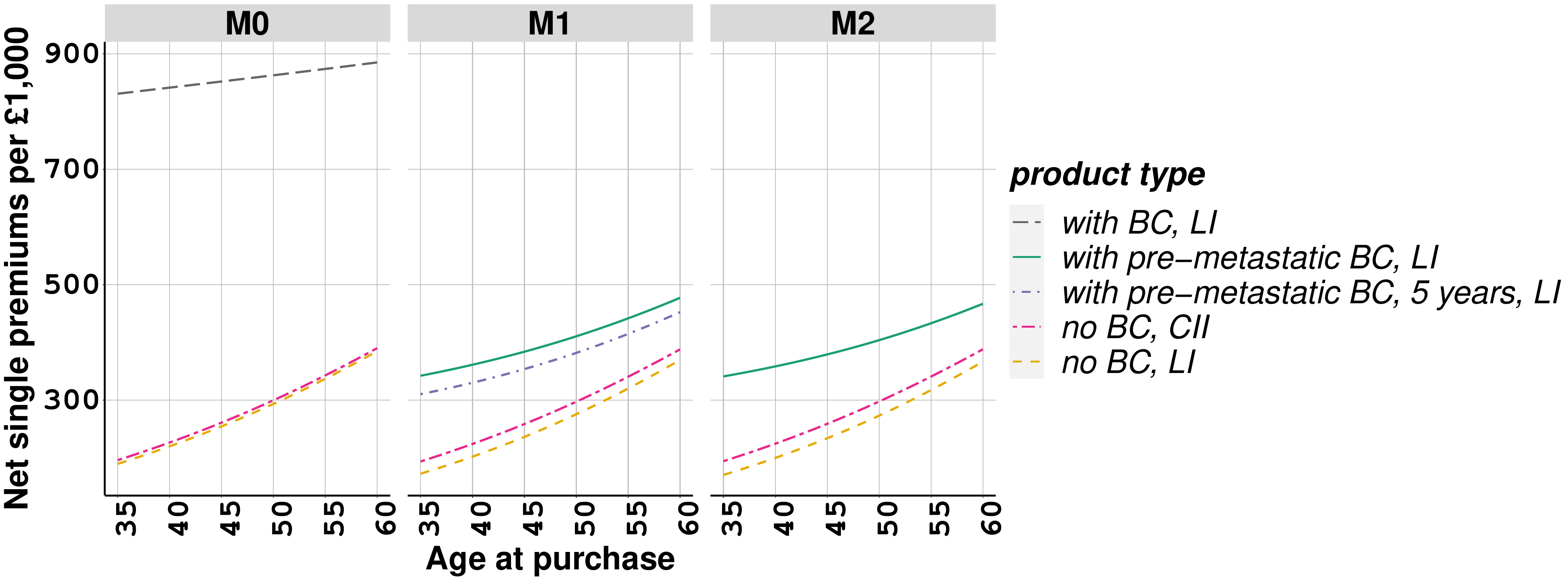}}
		\hfill
		\subfloat[25-year life insurance, $i=2\%$ \label{fig:25TIupto2019_LIv3_delta1}]{\includegraphics[width=0.50\textwidth]{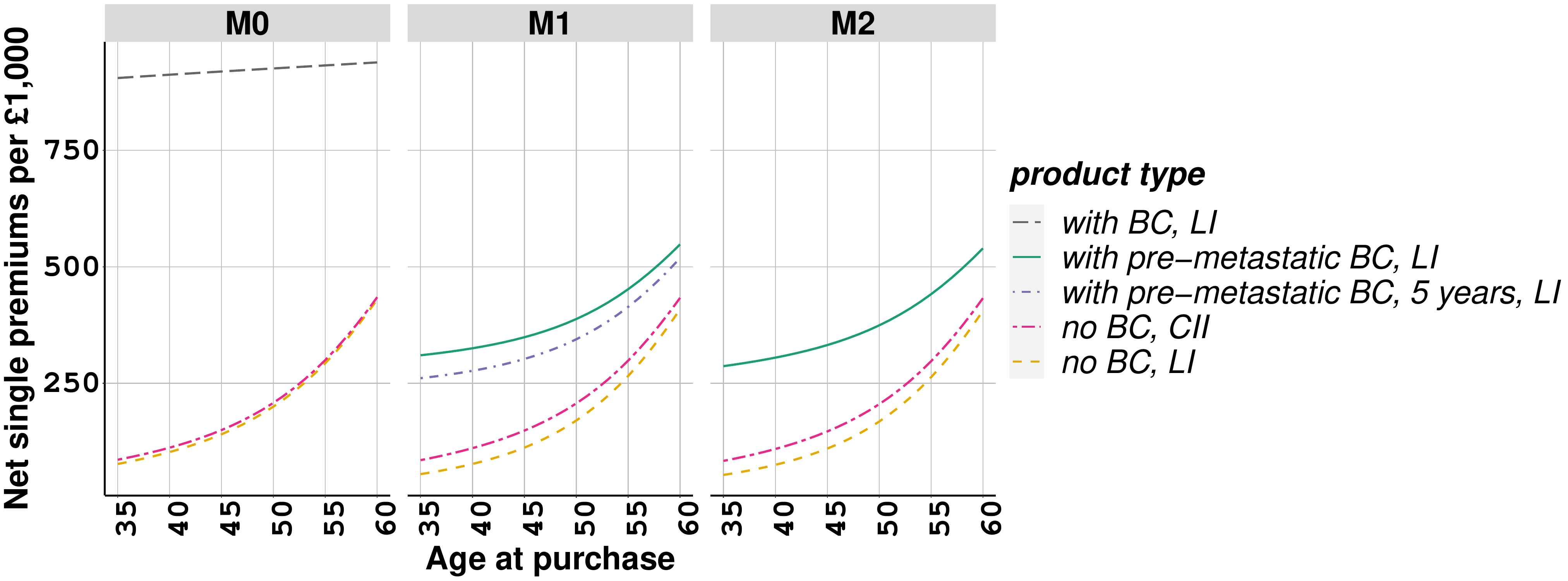}}
		\hfill
		\subfloat[25-year life insurance, $i=4\%$
		\label{fig:25TIupto2019_LIv3_delta2}]{\includegraphics[width=0.50\textwidth]{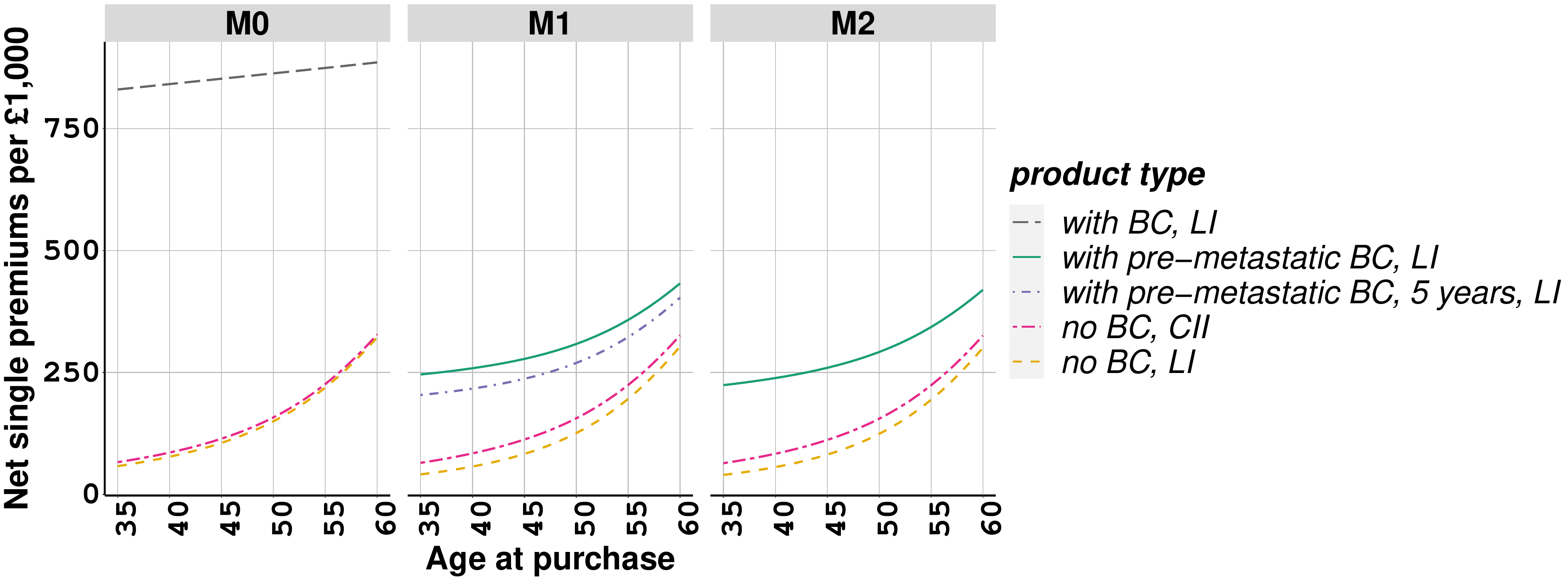}}
		\hfill
		\subfloat[10-year life insurance, $i=2\%$ \label{fig:10TIupto2019_LIv3_delta1}]{\includegraphics[width=0.50\textwidth]{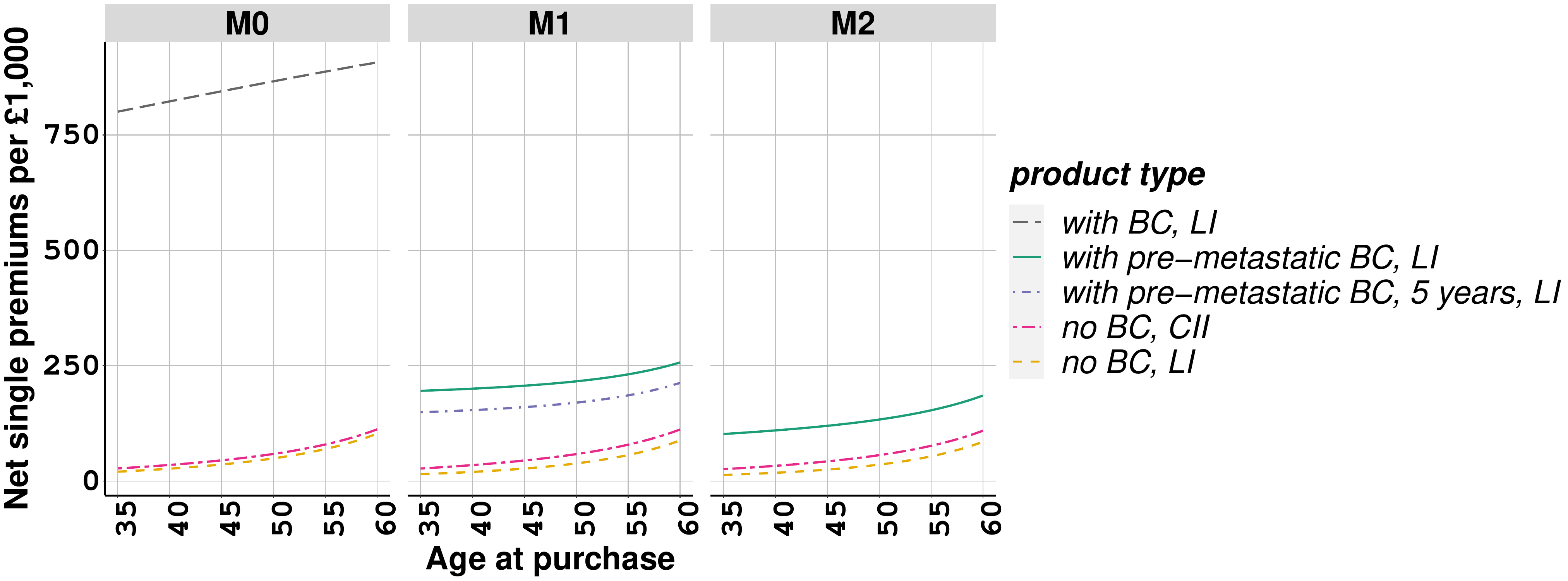}}
		\hfill
		\subfloat[10-year life insurance, $i=4\%$ \label{fig:10TIupto2019_LIv3_delta2}]{\includegraphics[width=0.50\textwidth]{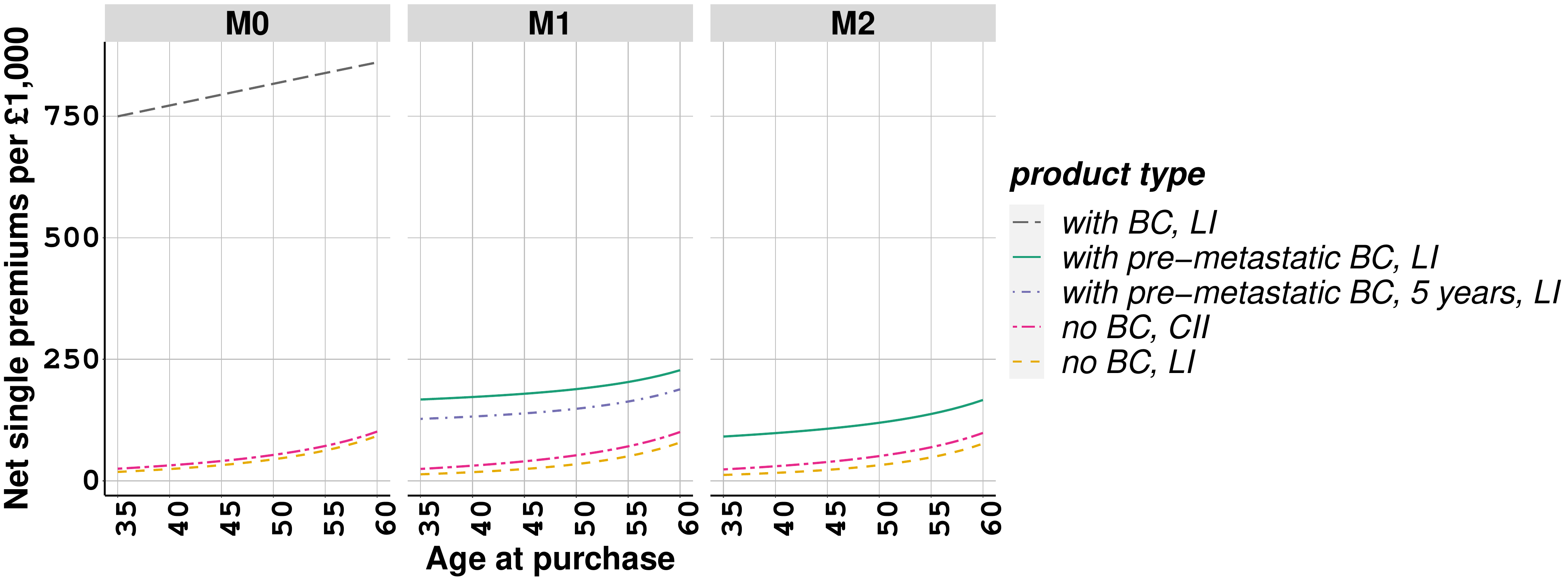}}
		\caption{Net single premium rates of a specialised life insurance contract, \eqref{eq:AxIndustry_death1}--\eqref{eq:AxSemiMarkov_death2} and \eqref{eq:AxSemiMarkov_death2_v2}, for
			policyholders with or without breast cancer at the time of purchase, \pounds1,000 benefit, {payable at the time of death}, based on M1 and M2, when $\alpha = 0.8$ and $\beta = 1/7$.}
		\label{fig:SinglePremiums_LI_DiffModels_v3}
	\end{figure}
	
	
}

\begin{figure}[H]
	\centering
		\subfloat[M1 \label{fig:OccupancyProb_Kx_DiffModels_Alpha1M0}]{\includegraphics[width=0.50\textwidth]{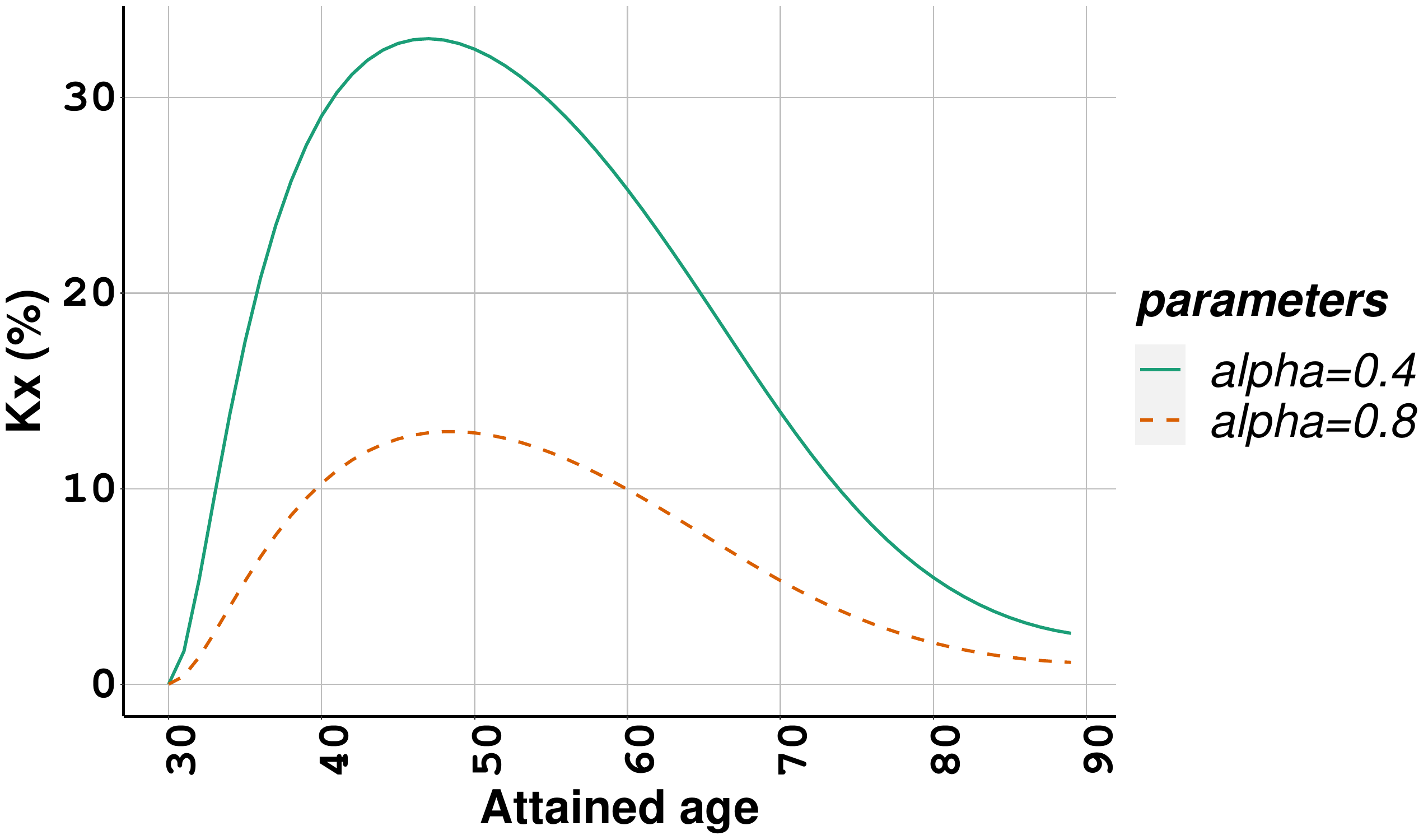}}
	\hfill
	\subfloat[M2 \label{fig:OccupancyProb_Kx_DiffModels_Alpha1M1}]{\includegraphics[width=0.50\textwidth]{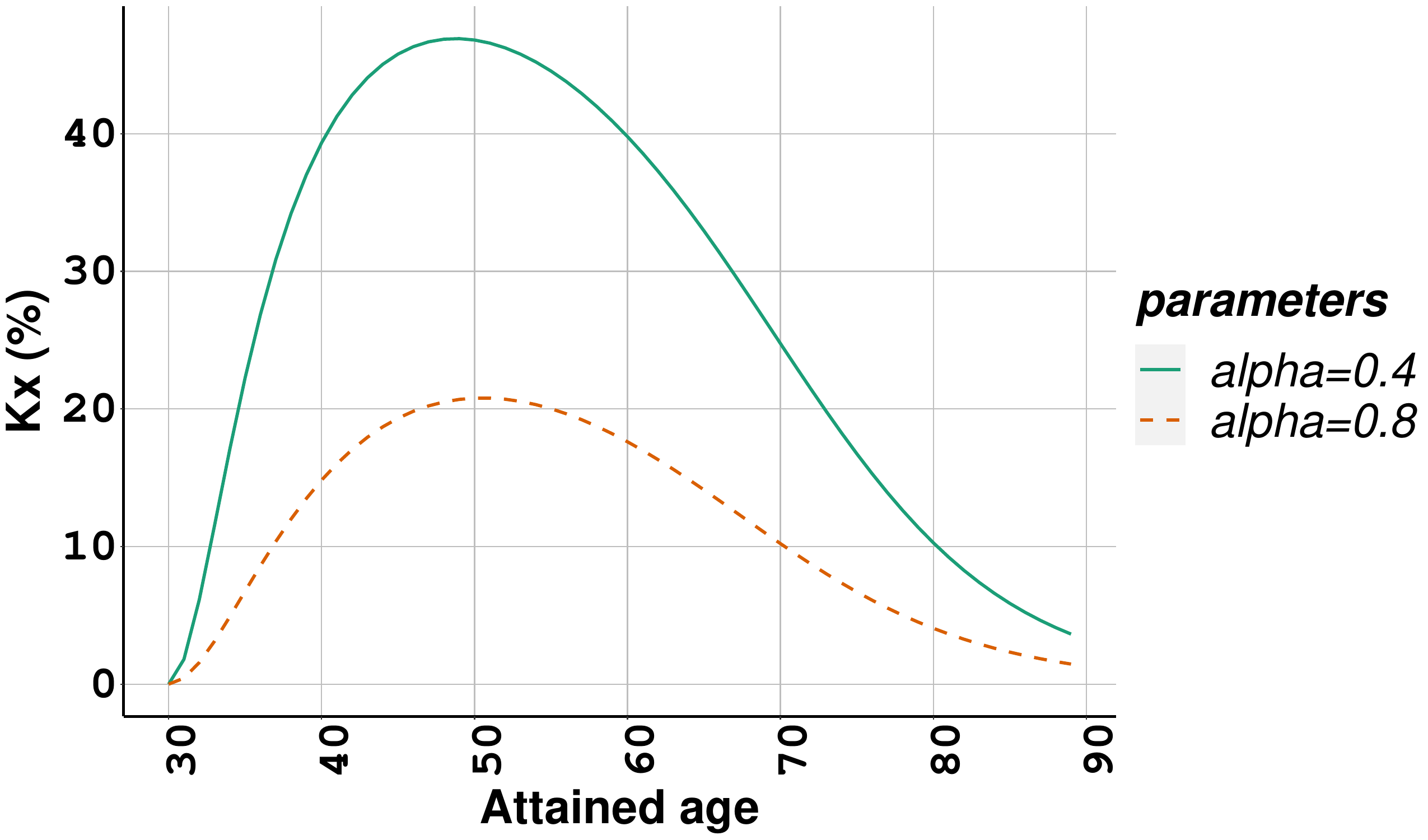}}
	\caption{Estimated $\hat{k}_{x}$ values for a policyholder aged 30, with no breast cancer, at time zero, based on M1 and M2, when $\alpha=0.4$ or  $\alpha=0.8$ and $\beta=1/7$.}
	\label{fig:OccupancyProb_Kx_DiffModels_Alpha1}
	\floatfoot{Note: $\text{Attained age} = \text{Age-at-entry} + \text{Time}$
	}
\end{figure}

\section{Main findings in Section \ref{sec:ImpactBeta}}

\begin{figure}[H]
	\subfloat[Whole life accelerated CII \label{fig:WLIupto2019v4}]{\includegraphics[width=0.5\textwidth]{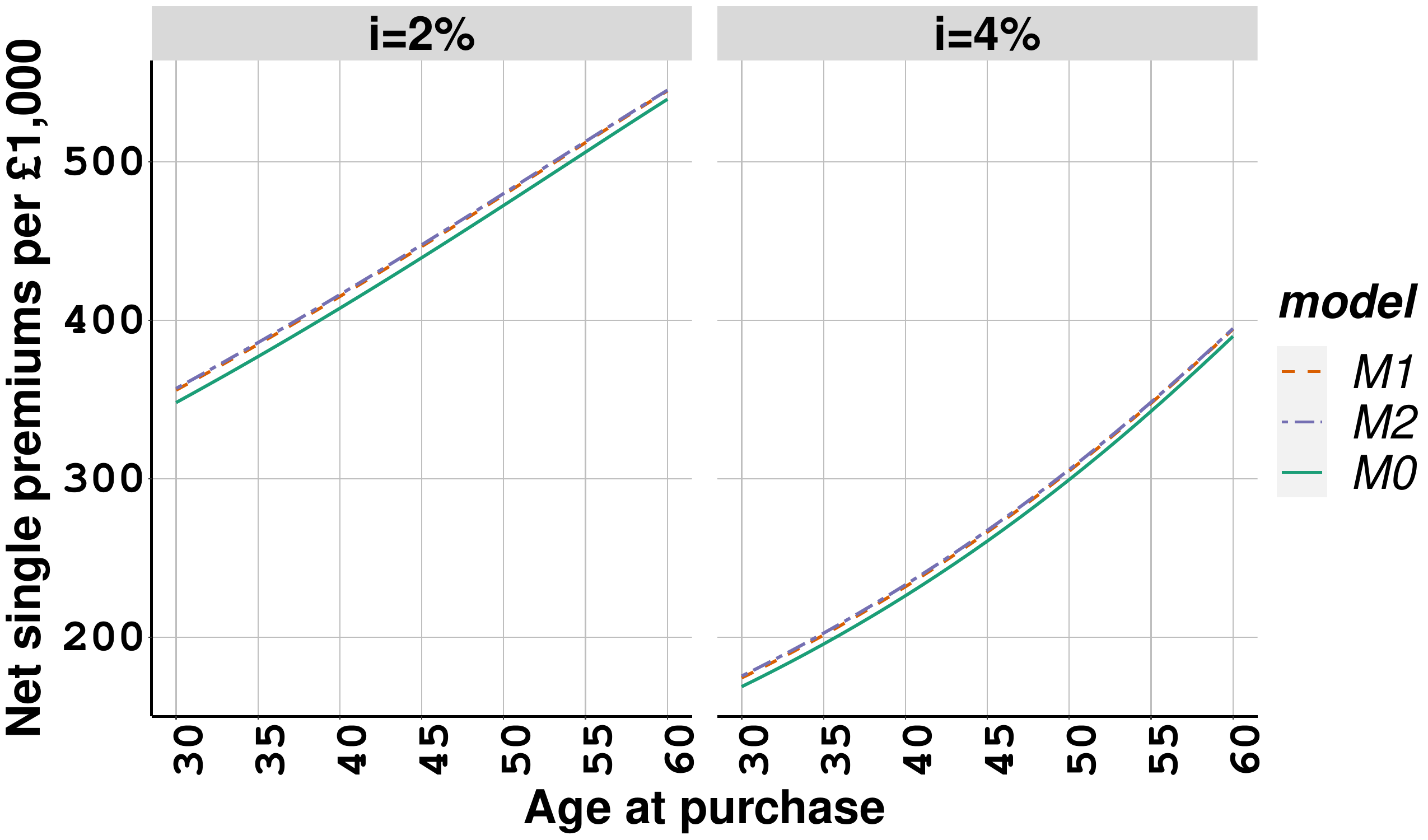}}
	\hfill
	\subfloat[25-year life accelerated CII \label{fig:25TIupto2019v4}]{\includegraphics[width=0.5\textwidth]{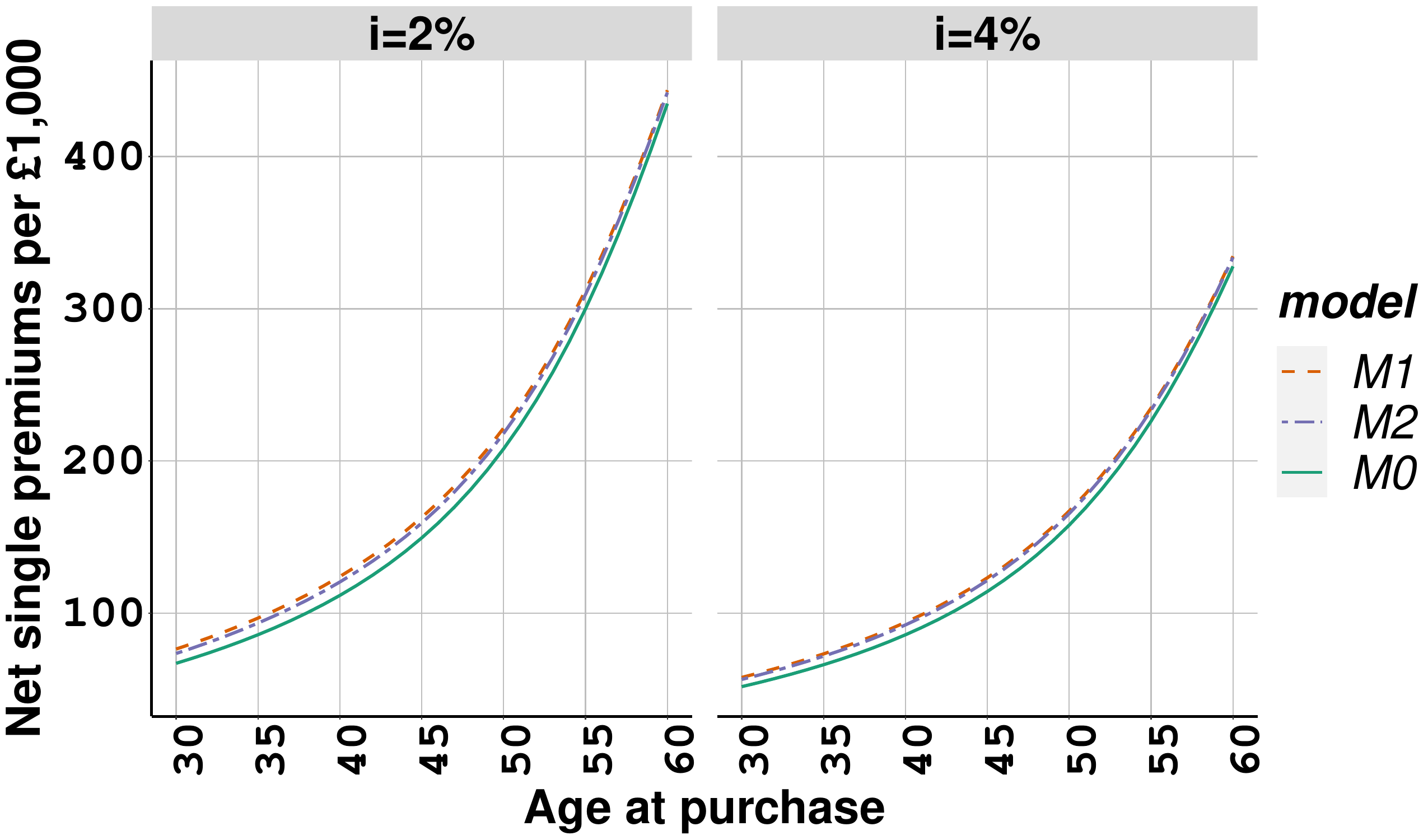}}
	\hfill
	\subfloat[10-year life accelerated CII \label{fig:10TIupto2019v4}]{\includegraphics[width=0.5\textwidth]{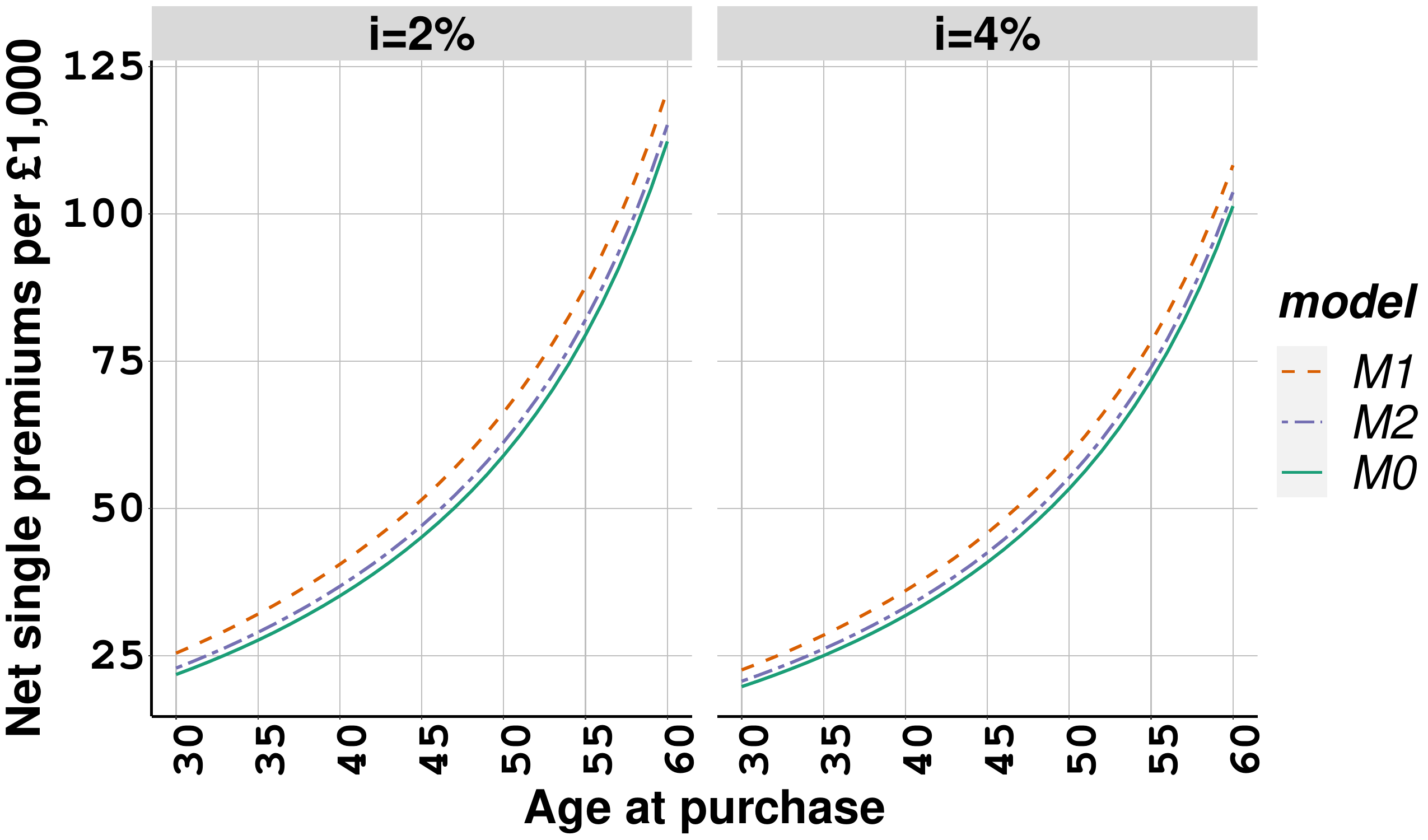}}
	\caption{Net single premium rates of a specialised CII contract, \eqref{eq:AxCIMarkov2} and \eqref{eq:AxSemiMarkov2}, for policyholders between ages 30--60 with \pounds1,000 benefit, {payable at the time of diagnosis of BC or at the time of death}, based on different model assumptions for $\alpha = 0.6$ and $\beta = 1/10$.}
	\label{fig:SinglePremiums_DiffModelsv4}
\end{figure}

\begin{figure}[H]
	\subfloat[Whole life accelerated CII \label{fig:WLIupto2019v5}]{\includegraphics[width=0.5\textwidth]{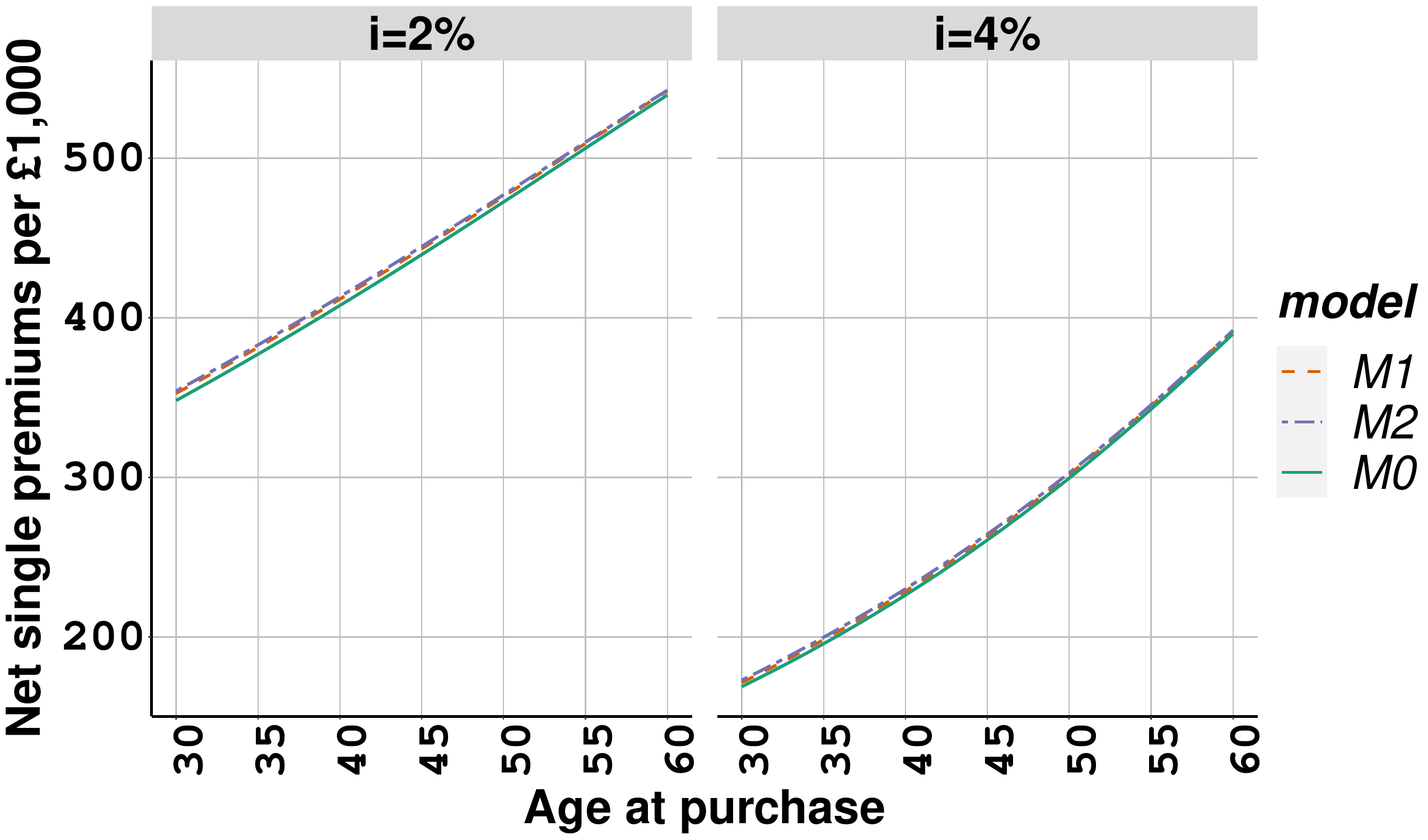}}
	\hfill
	\subfloat[25-year life accelerated CII \label{fig:25TIupto2019v5}]{\includegraphics[width=0.5\textwidth]{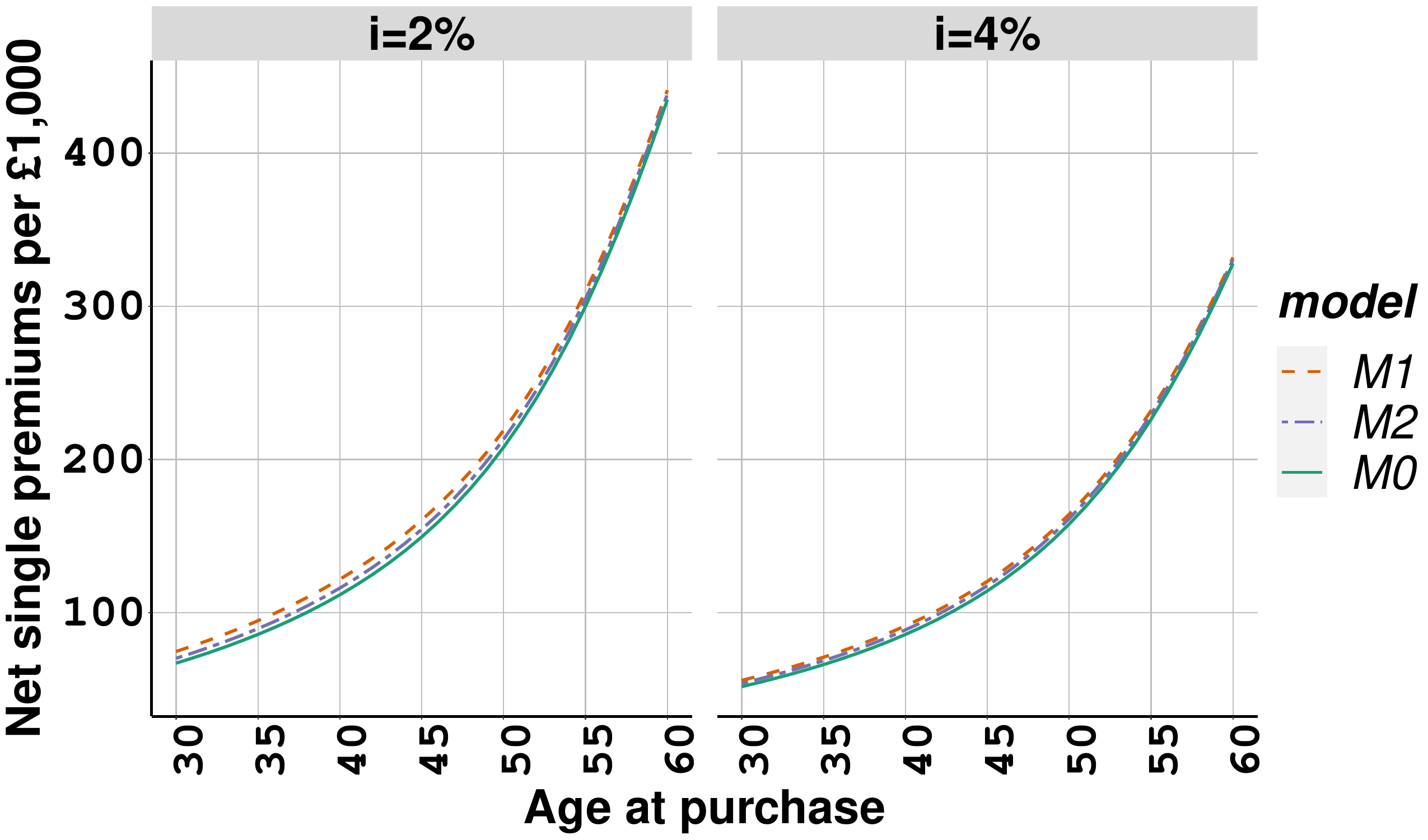}}
	\hfill
	\subfloat[10-year life accelerated CII \label{fig:10TIupto2019v5}]{\includegraphics[width=0.5\textwidth]{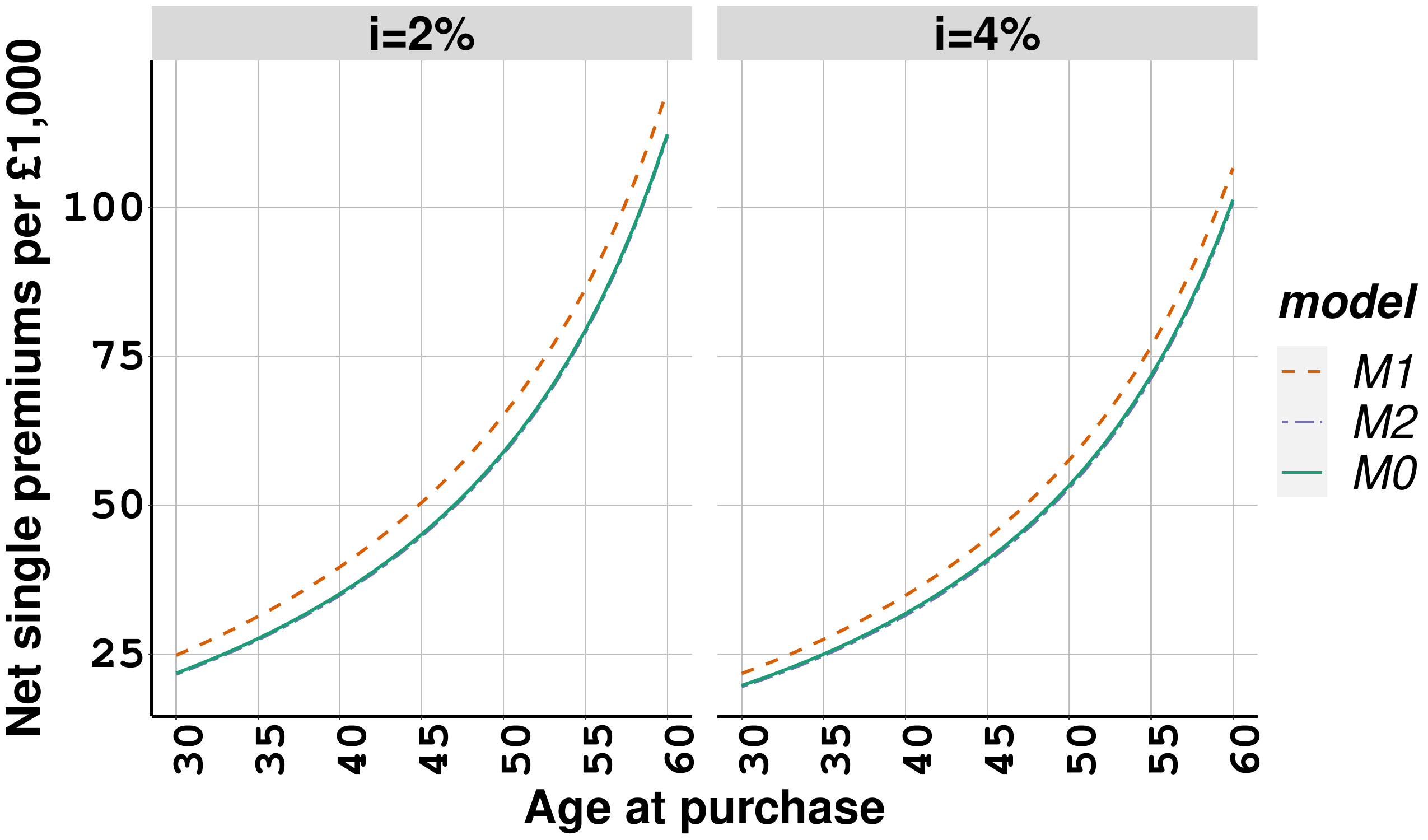}}
	\caption{Net single premium rates of a specialised CII contract, \eqref{eq:AxCIMarkov2} and \eqref{eq:AxSemiMarkov2}, for policyholders between ages 30--60 with \pounds1,000 benefit, {payable at the time of diagnosis of BC or at the time of death}, based on different model assumptions for $\alpha = 0.6$ and $\beta = 1/5$.}
	\label{fig:SinglePremiums_DiffModelsv5}
\end{figure}

{
	\begin{figure}[H]
	\subfloat[Whole life insurance, $i=2\%$ \label{fig:WLIupto2019_LIv4_delta1}]{\includegraphics[width=0.50\textwidth]{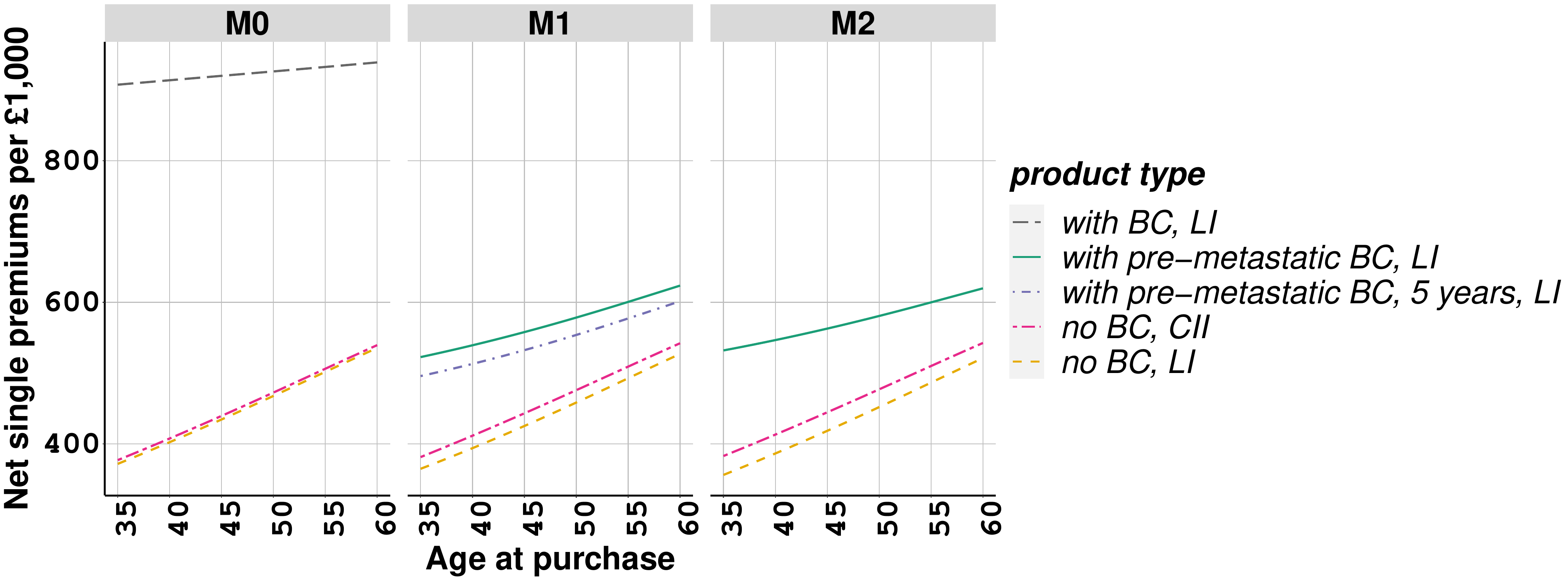}}
	\hfill
		\subfloat[Whole life insurance, $i=4\%$ \label{fig:WLIupto2019_LIv4_delta2}]{\includegraphics[width=0.50\textwidth]{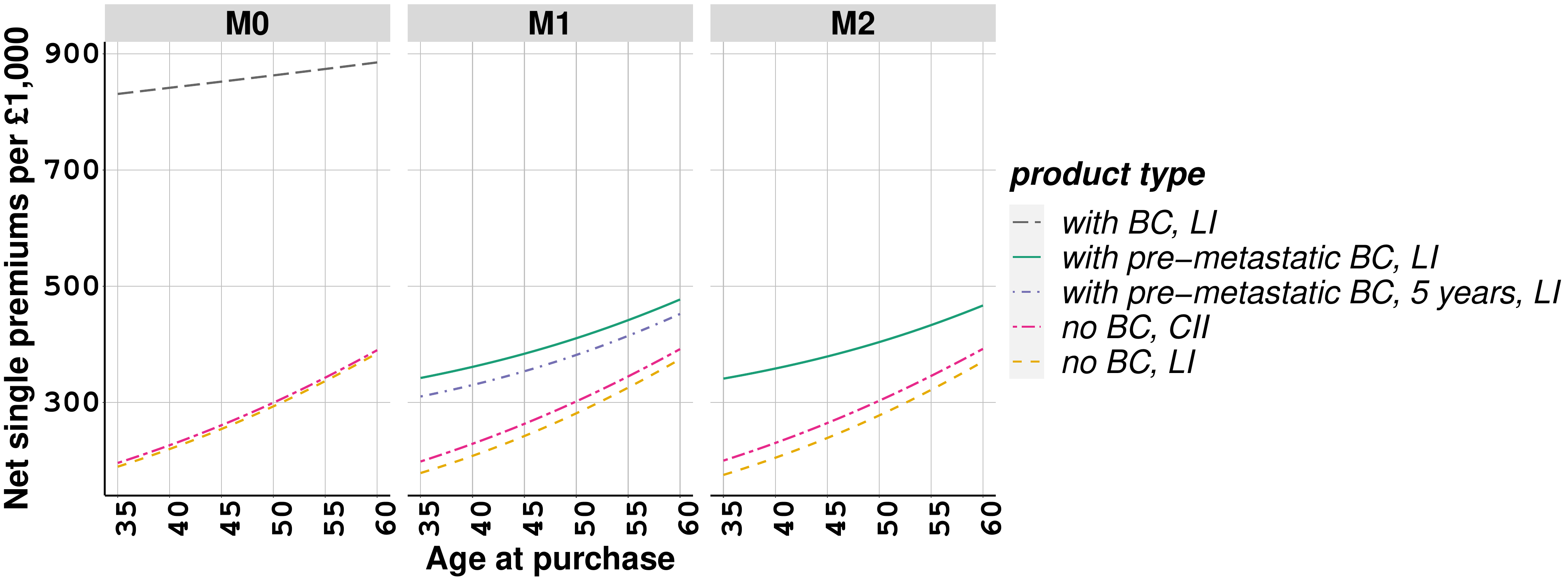}}
	\hfill	
	\subfloat[25-year life insurance, $i=2\%$ \label{fig:25TIupto2019_LIv4_delta1}]{\includegraphics[width=0.50\textwidth]{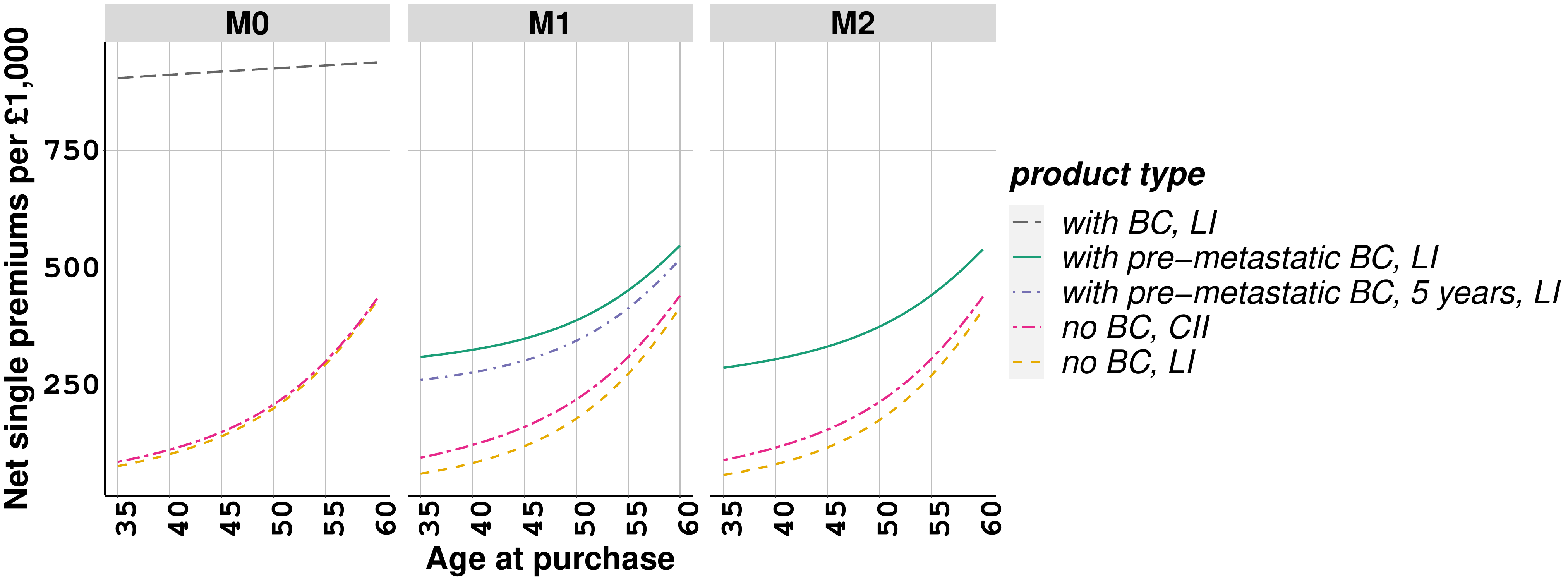}}
	\hfill
		\subfloat[25-year life insurance, $i=4\%$ \label{fig:25TIupto2019_LIv4_delta2}]{\includegraphics[width=0.50\textwidth]{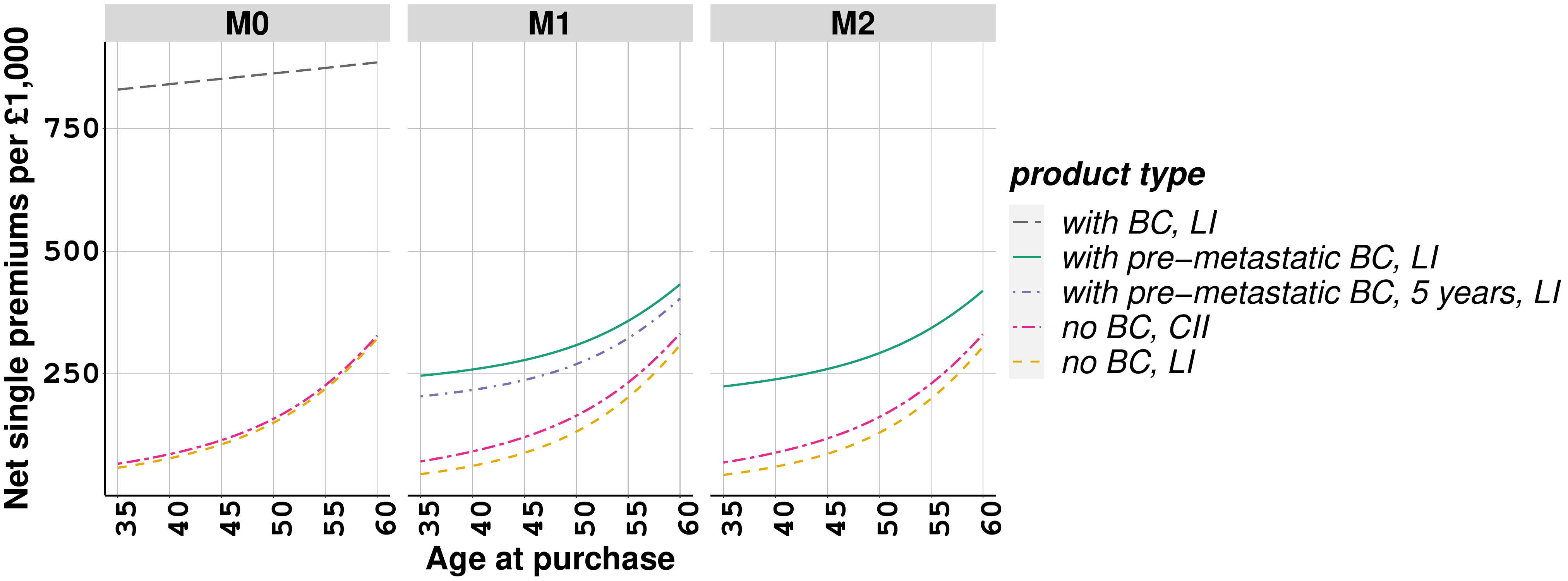}}
	\hfill	
	\subfloat[10-year life insurance, $i=2\%$ \label{fig:10TIupto2019_LIv4}]{\includegraphics[width=0.50\textwidth]{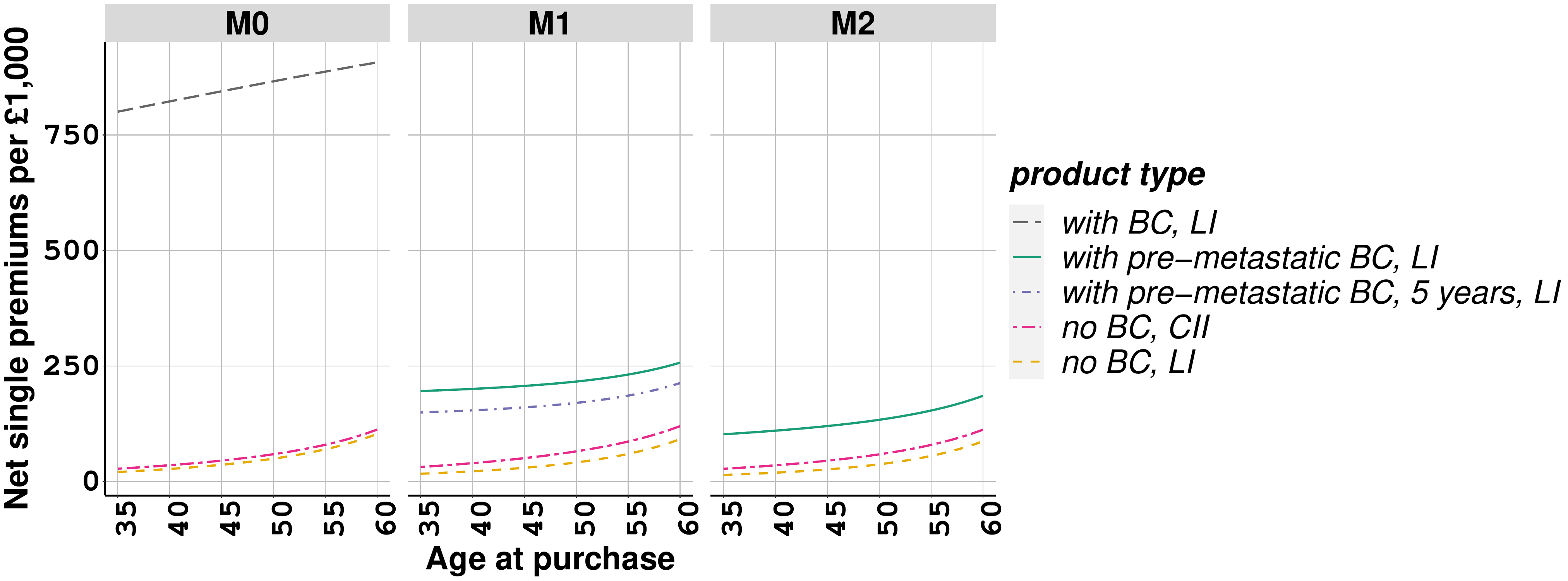}}
	\hfill
	\subfloat[10-year life insurance, $i=4\%$ \label{fig:10TIupto2019_LIv4_delta2}]{\includegraphics[width=0.50\textwidth]{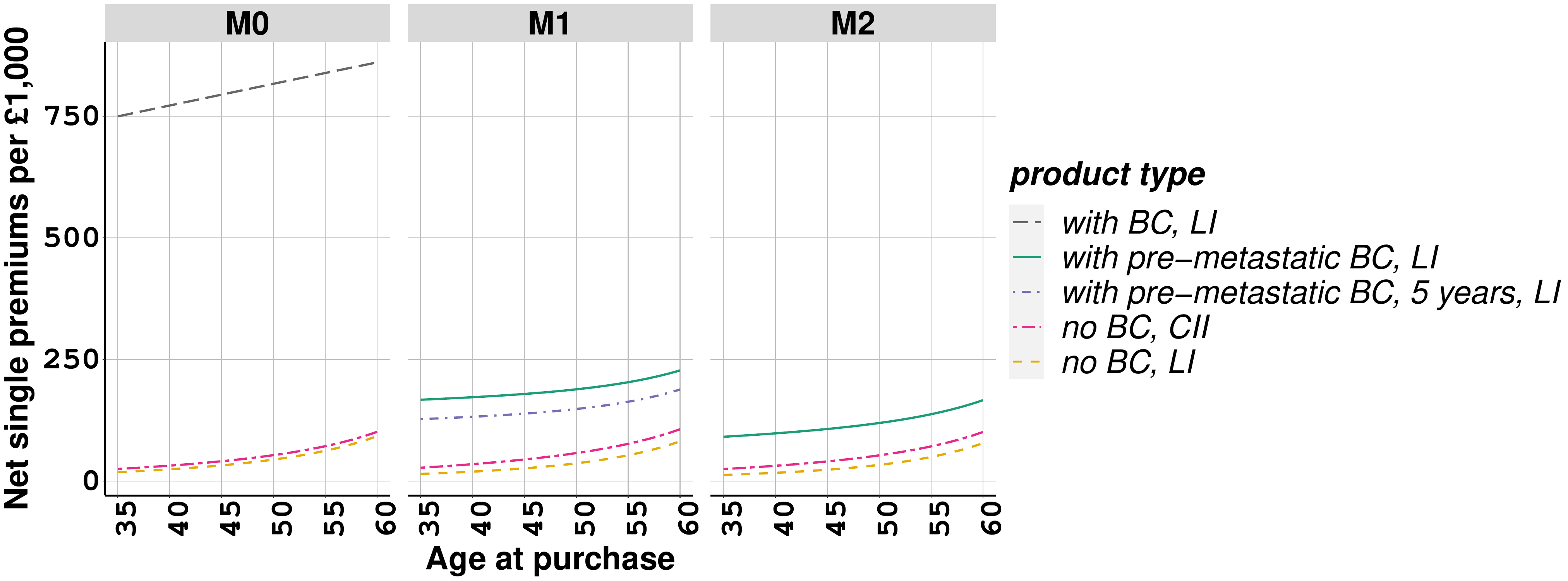}}
	\caption{Net single premium rates of a specialised life insurance contract, \eqref{eq:AxIndustry_death1}--\eqref{eq:AxSemiMarkov_death2}, and \eqref{eq:AxSemiMarkov_death2_v2}, for
		policyholders with or without breast cancer at the time of purchase, \pounds1,000 benefit, {payable at the time of death}, based on M1 and M2, when $\alpha = 0.6$ and $\beta = 1/5$.}
	\label{fig:SinglePremiums_LI_DiffModels_v4}
\end{figure}

\begin{figure}[H]
	\subfloat[Whole life insurance, $i=2\%$ \label{fig:WLIupto2019_LIv5_delta1}]{\includegraphics[width=0.50\textwidth]{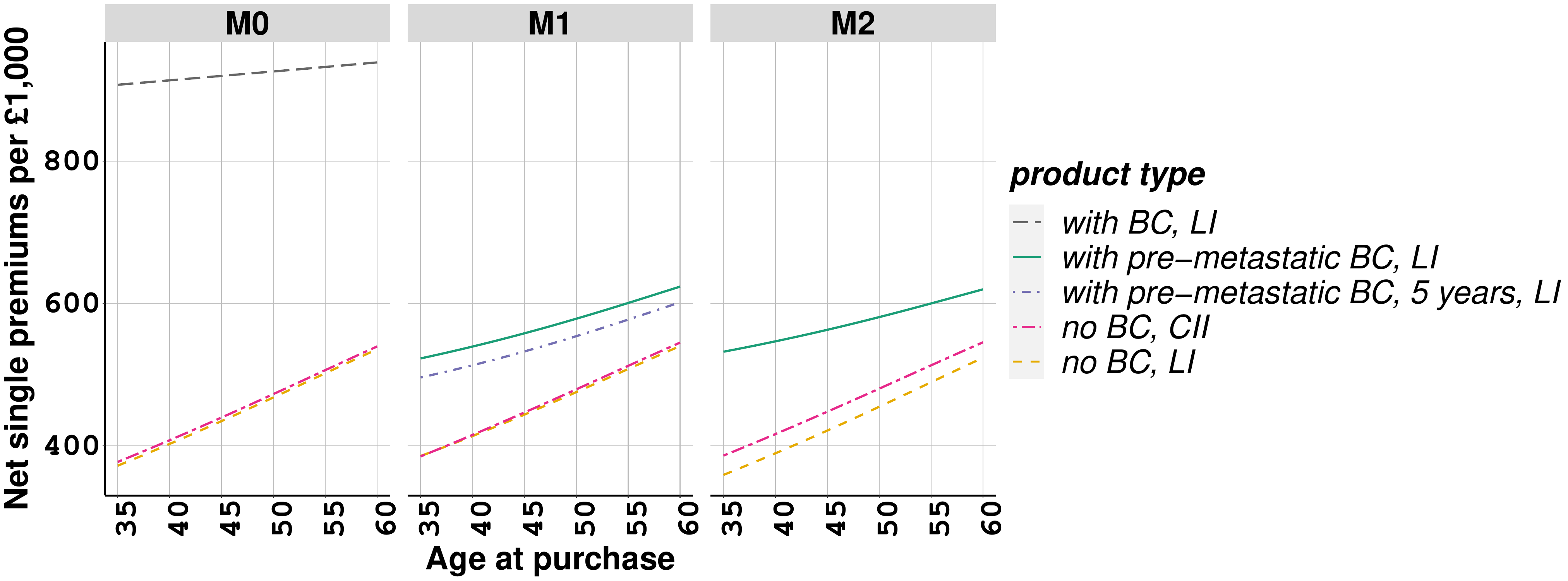}}
	\hfill
	\subfloat[Whole life insurance, $i=4\%$ \label{fig:WLIupto2019_LIv5_delta2}]{\includegraphics[width=0.50\textwidth]{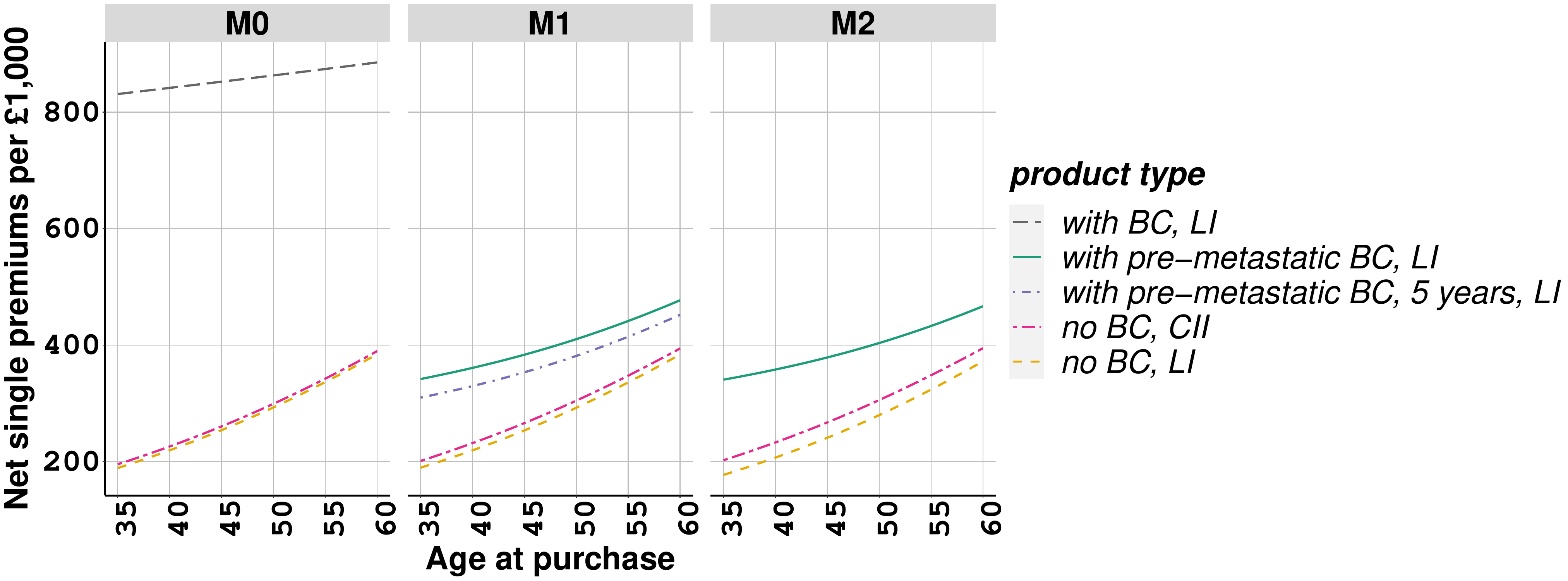}}
	\hfill	
	\subfloat[25-year life insurance, $i=2\%$ \label{fig:25TIupto2019_LIv5_delta1}]{\includegraphics[width=0.50\textwidth]{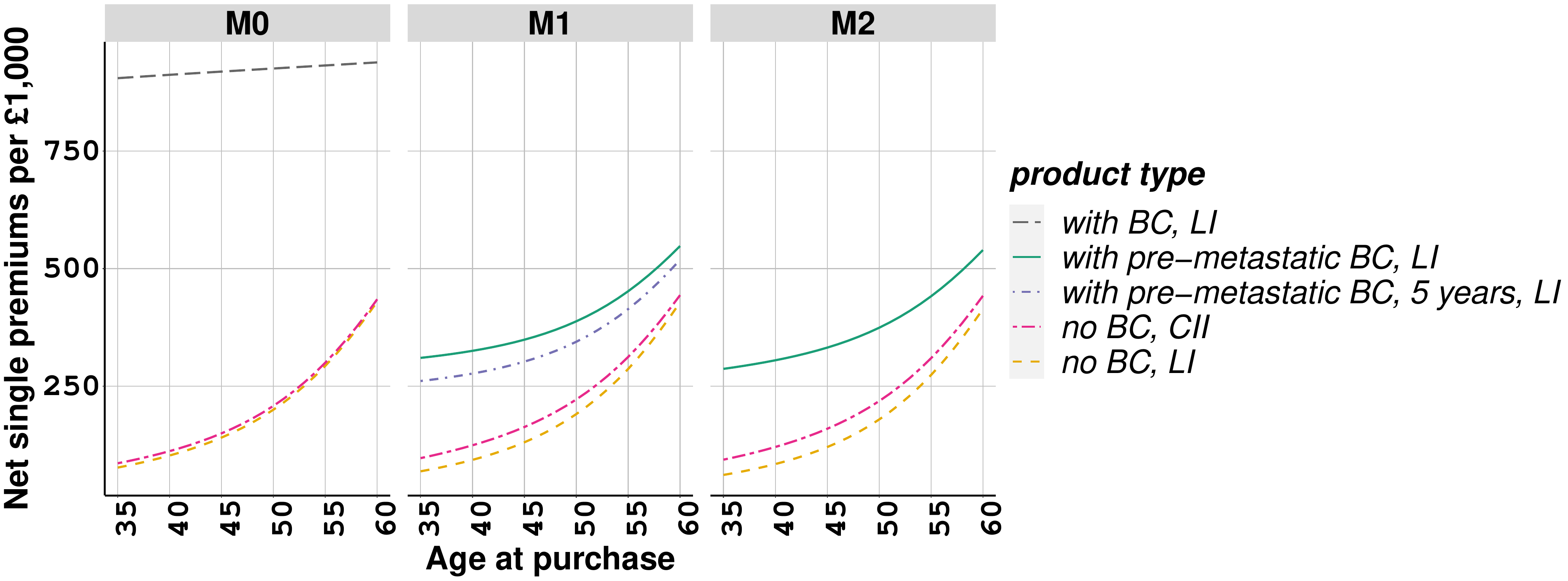}}
	\hfill
	\subfloat[25-year life insurance, $i=4\%$ \label{fig:25TIupto2019_LIv5_delta2}]{\includegraphics[width=0.50\textwidth]{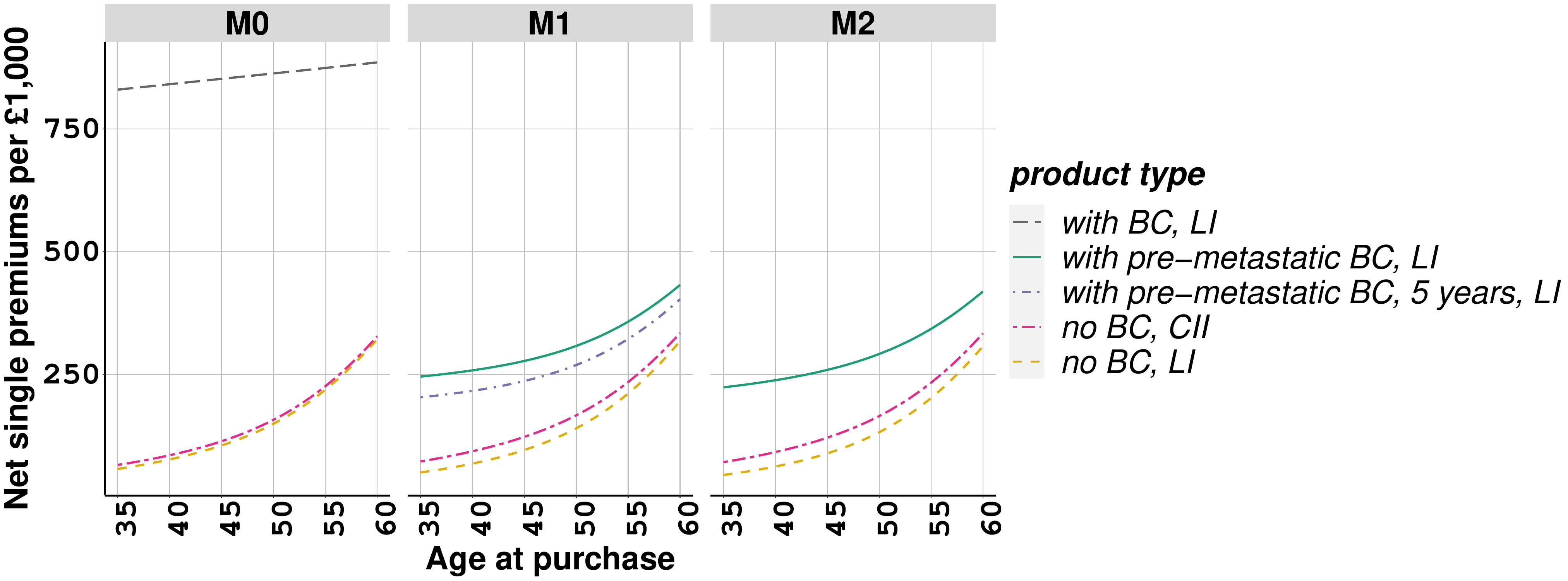}}
	\hfill
	\subfloat[10-year life insurance, $i=2\%$ \label{fig:10TIupto2019_LIv5_delta1}]{\includegraphics[width=0.50\textwidth]{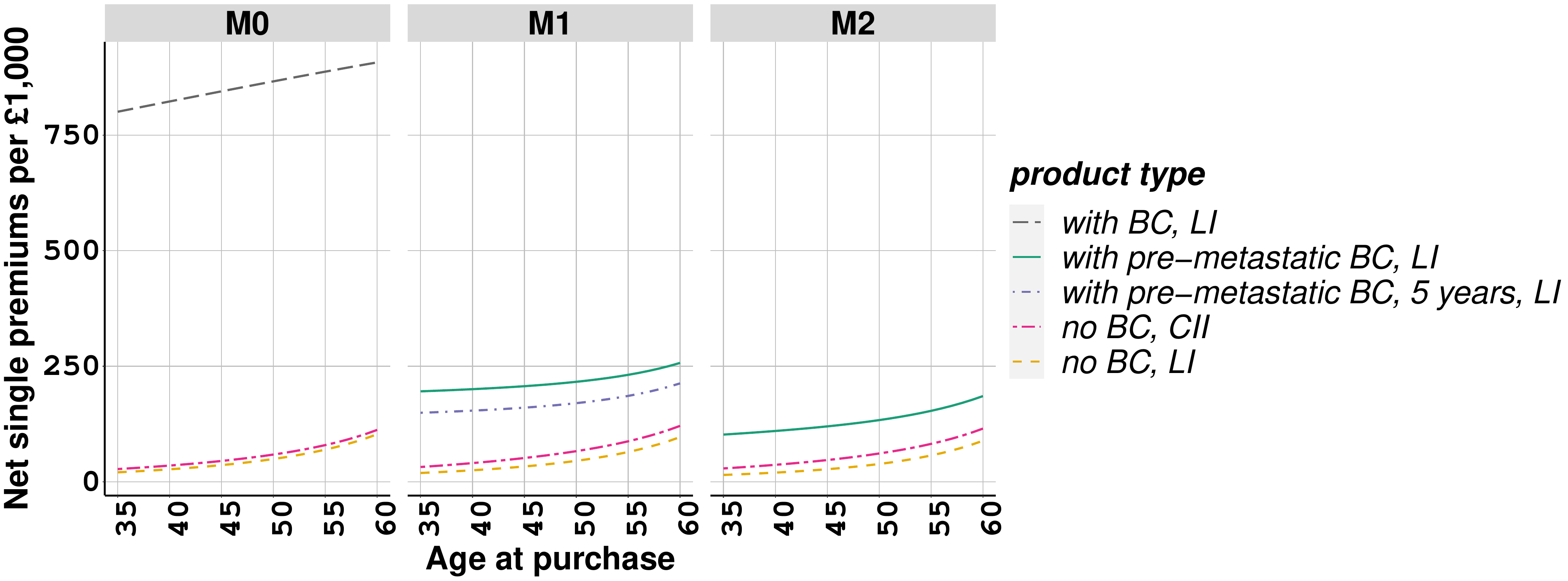}}
	\hfill
	\subfloat[10-year life insurance, $i=4\%$ \label{fig:10TIupto2019_LIv5_delta2}]{\includegraphics[width=0.50\textwidth]{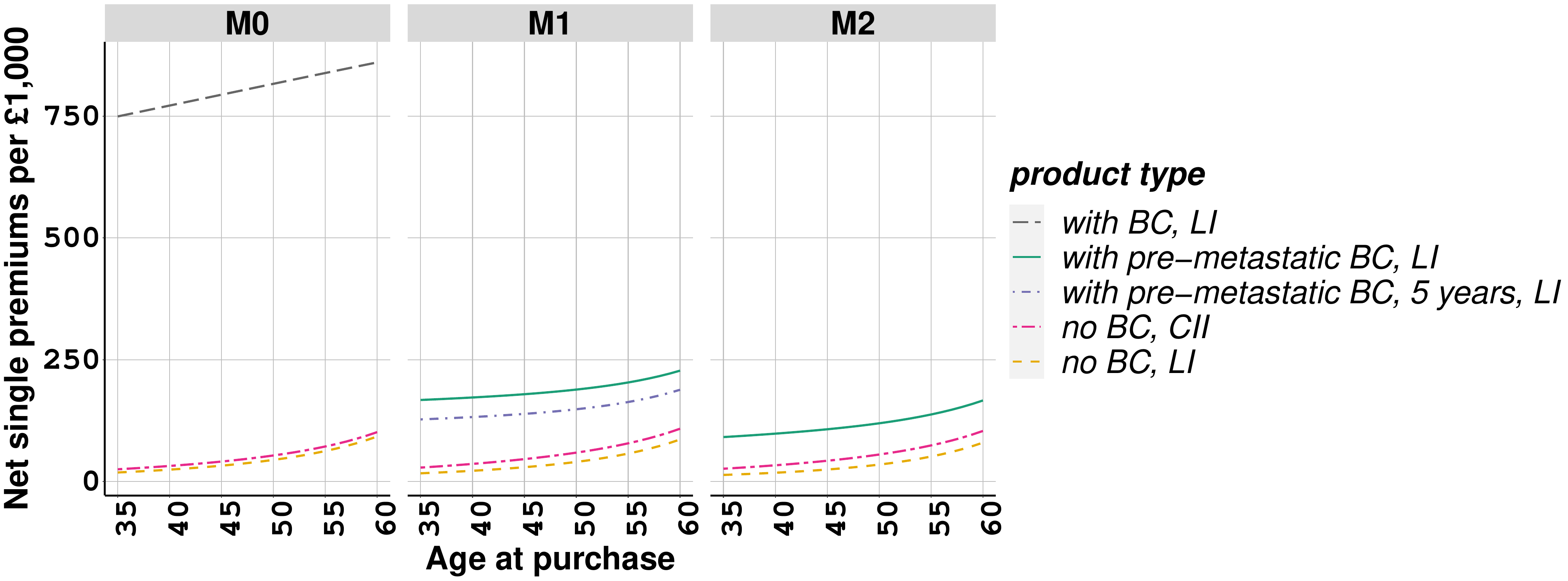}}
	\caption{Net single premium rates of a specialised life insurance contract, \eqref{eq:AxIndustry_death1}--\eqref{eq:AxSemiMarkov_death2}, and \eqref{eq:AxSemiMarkov_death2_v2}, for
			policyholders with or without breast cancer at the time of purchase, \pounds1,000 benefit, {payable at the time of death}, based on M1 and M2, when $\alpha = 0.6$ and $\beta = 1/10$.}
	\label{fig:SinglePremiums_LI_DiffModels_v5}
\end{figure}

}

\begin{figure}[H]
	\centering
	\subfloat[M1 \label{fig:OccupancyProb_Kx_DiffModels_Beta1M0}]{\includegraphics[width=0.50\textwidth]{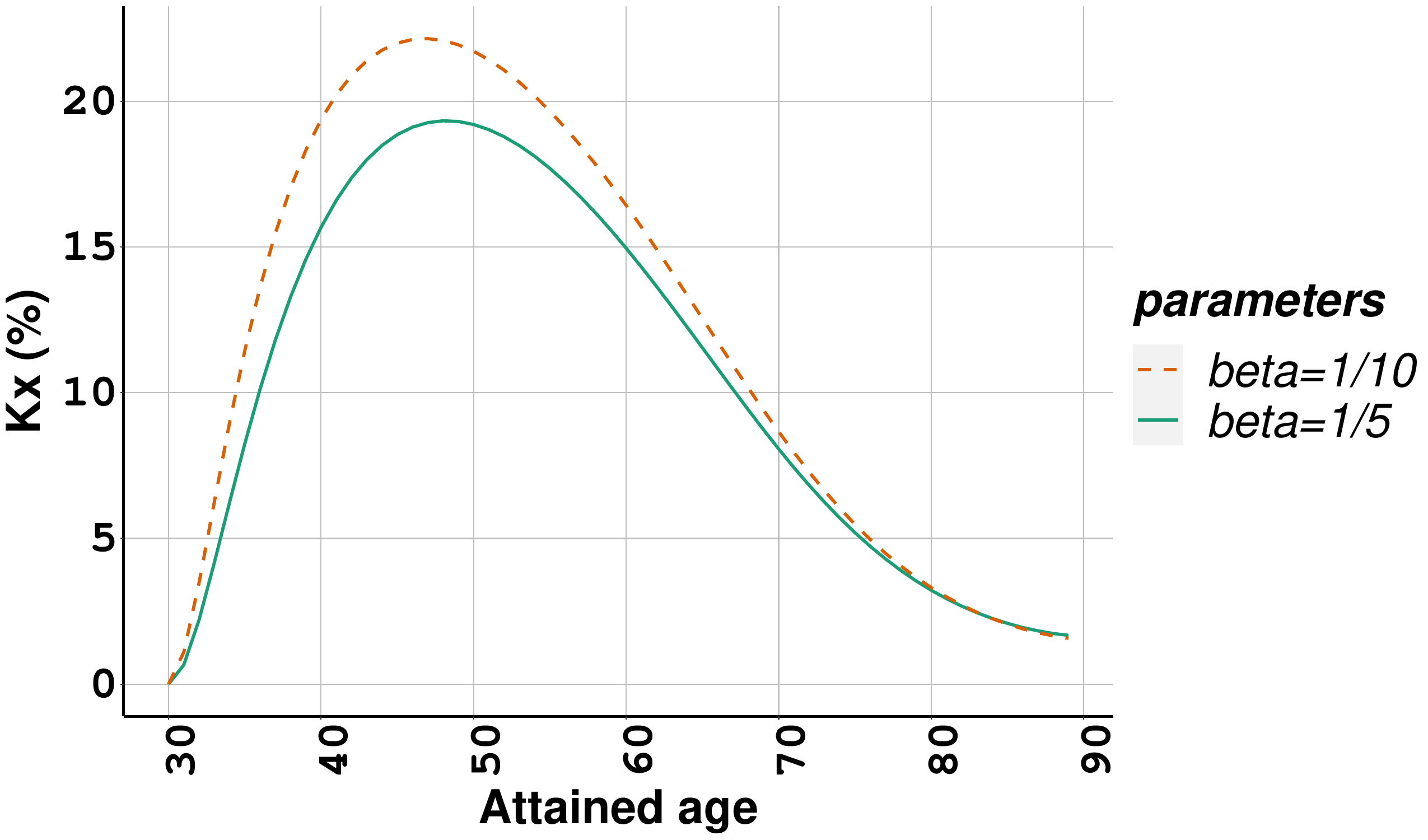}}
	\hfill
	\subfloat[M2 \label{fig:OccupancyProb_Kx_DiffModels_Beta1M1}]{\includegraphics[width=0.50\textwidth]{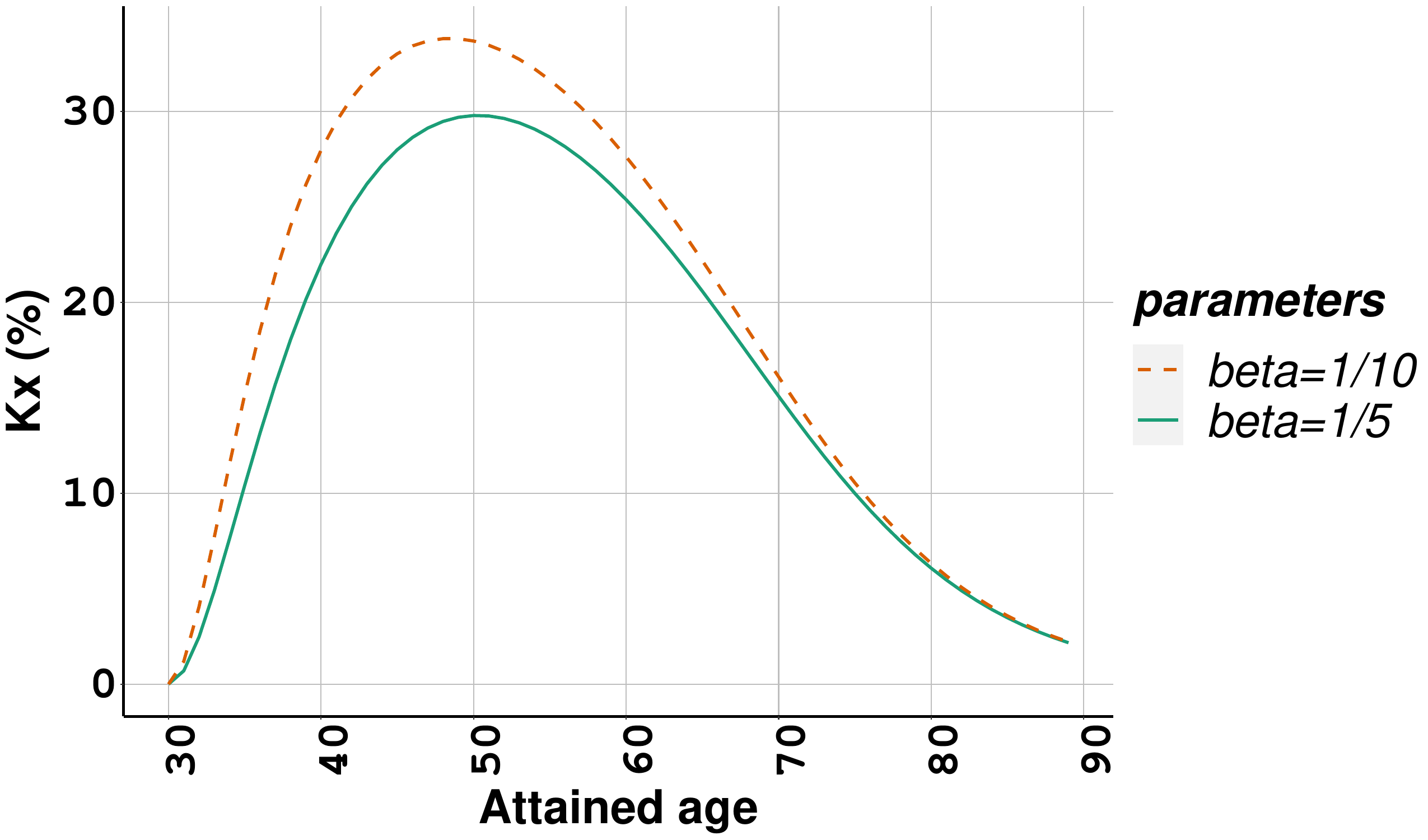}}
	\caption{Estimated $\hat{k}_{x}$ values for a policyholder aged 30, with no breast cancer, at time zero, based on M1 and M2, when $\beta=1/5$ or  $\beta=1/10$ and $\alpha=0.6$.}
	\label{fig:OccupancyProb_Kx_DiffModels_Beta1}
	\floatfoot{Note: $\text{Attained age} = \text{Age-at-entry} + \text{Time}$
	}
\end{figure}

\section{Main findings in Section \ref{sec:Impactmu13}}

\begin{figure}[H]
 \includegraphics[width=0.65\textwidth]{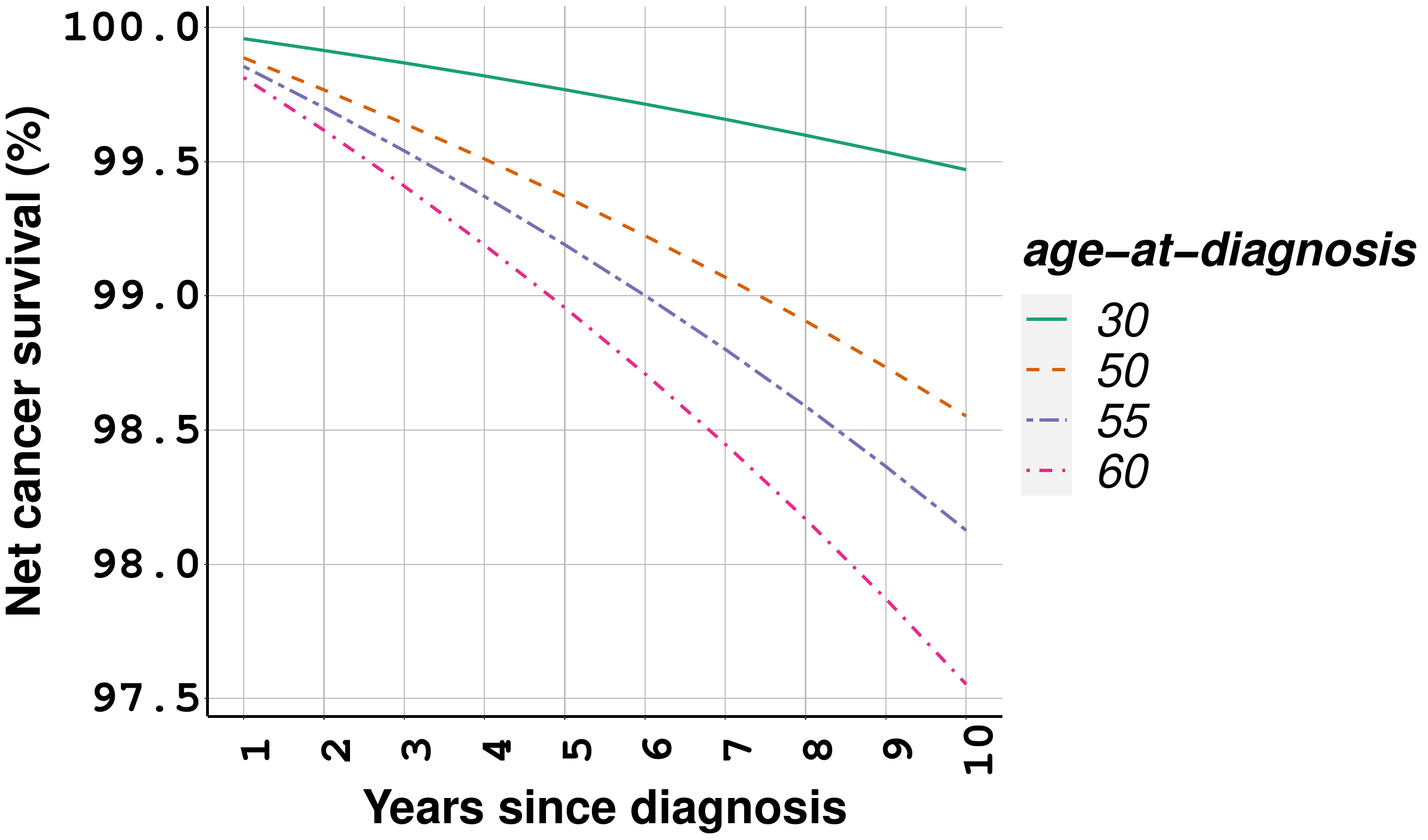}
	\caption{Estimated net cancer survival for a woman diagnosed with breast cancer	at different ages under M0.}
	\label{fig:NetCancerSurvivalMu13}
\end{figure}

\begin{figure}[H]
	\centering
	\includegraphics[width=0.65\textwidth]{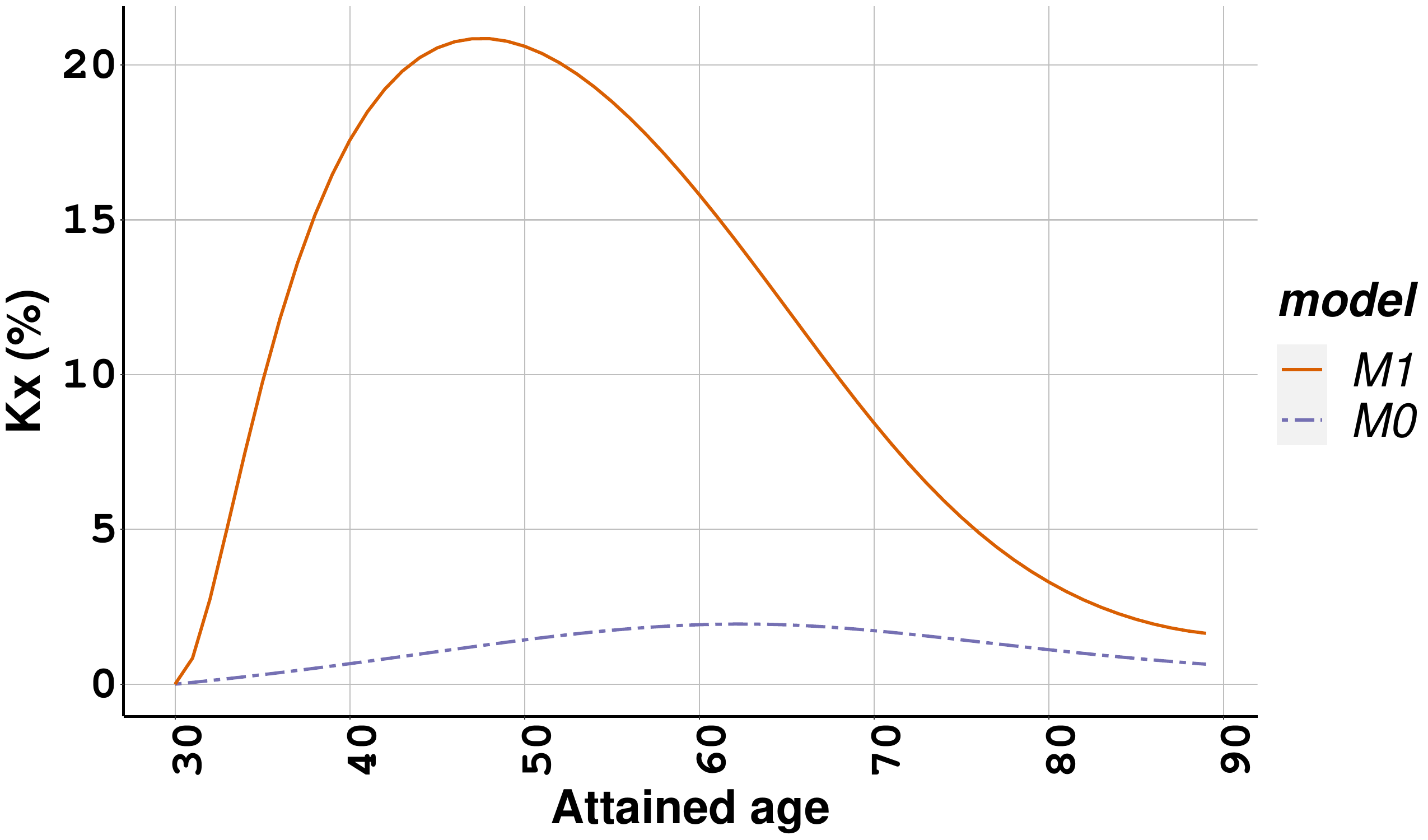}
		\caption{Estimated $\hat{k}_{x}$ values based on M0 and M1,  when $\alpha=0.6$ and $\beta=1/7$.}
	\label{fig:OccupancyProb_Kx_DiffModels_Mu13}
	\floatfoot{Note: $\text{Attained age} = \text{Age-at-entry} + \text{Time}$
		}
\end{figure}

		\begin{figure}[H]
	\subfloat[Whole life insurance, $i=2\%$ \label{fig:WLIupto2019_LIv6_delta1}]{\includegraphics[width=0.50\textwidth]{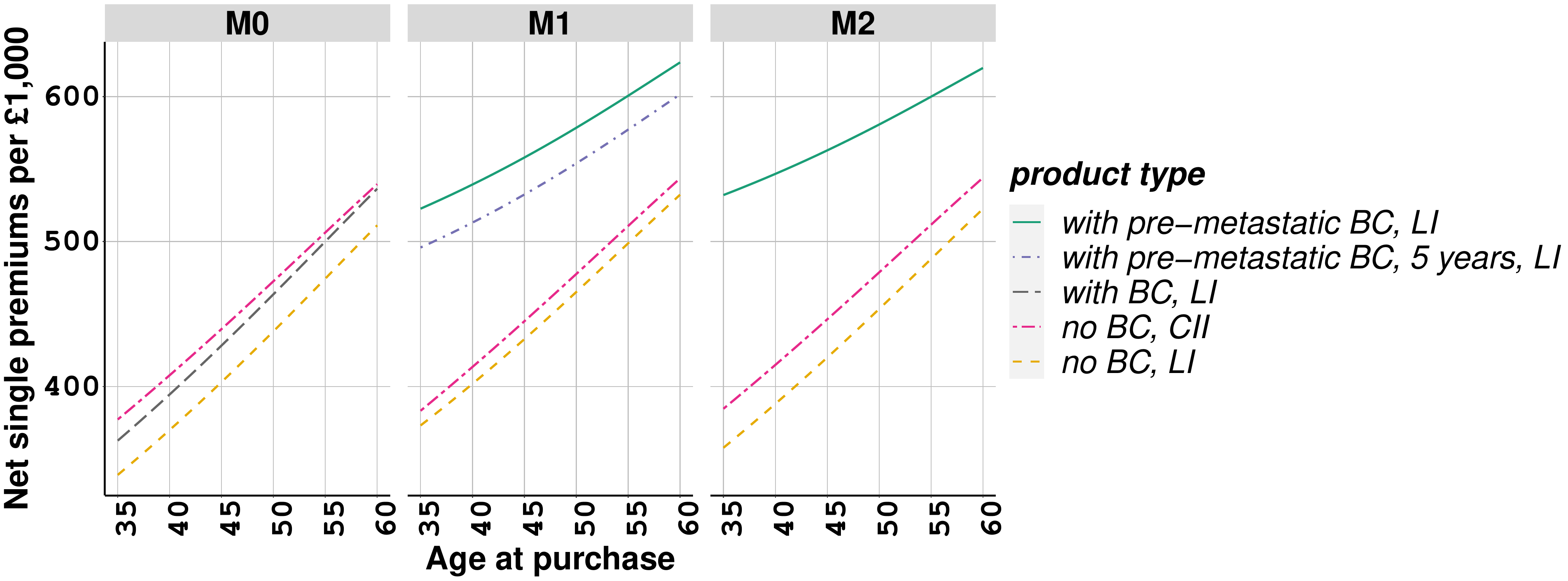}}
	\hfill
	\subfloat[Whole life insurance, $i=4\%$ \label{fig:WLIupto2019_LIv6_delta2}]{\includegraphics[width=0.50\textwidth]{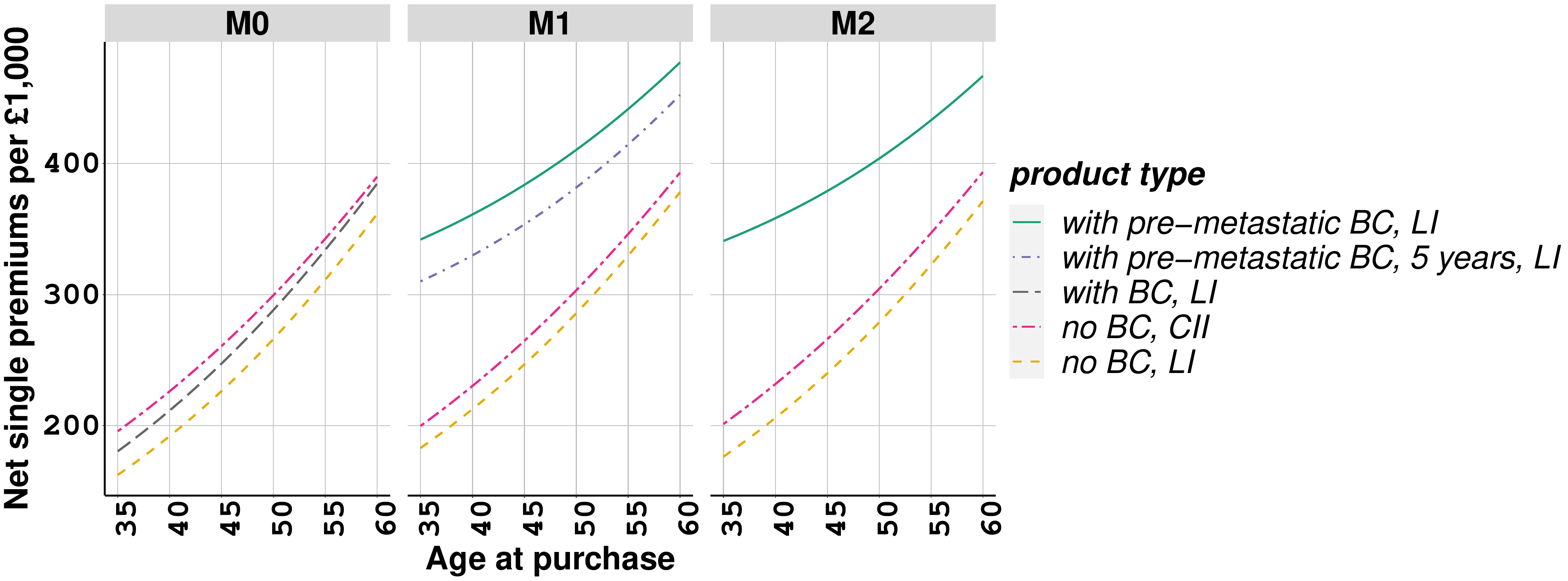}}
	\hfill
	\subfloat[25-year life insurance, $i=2\%$ \label{fig:25TIupto2019_LIv6_delta1}]{\includegraphics[width=0.50\textwidth]{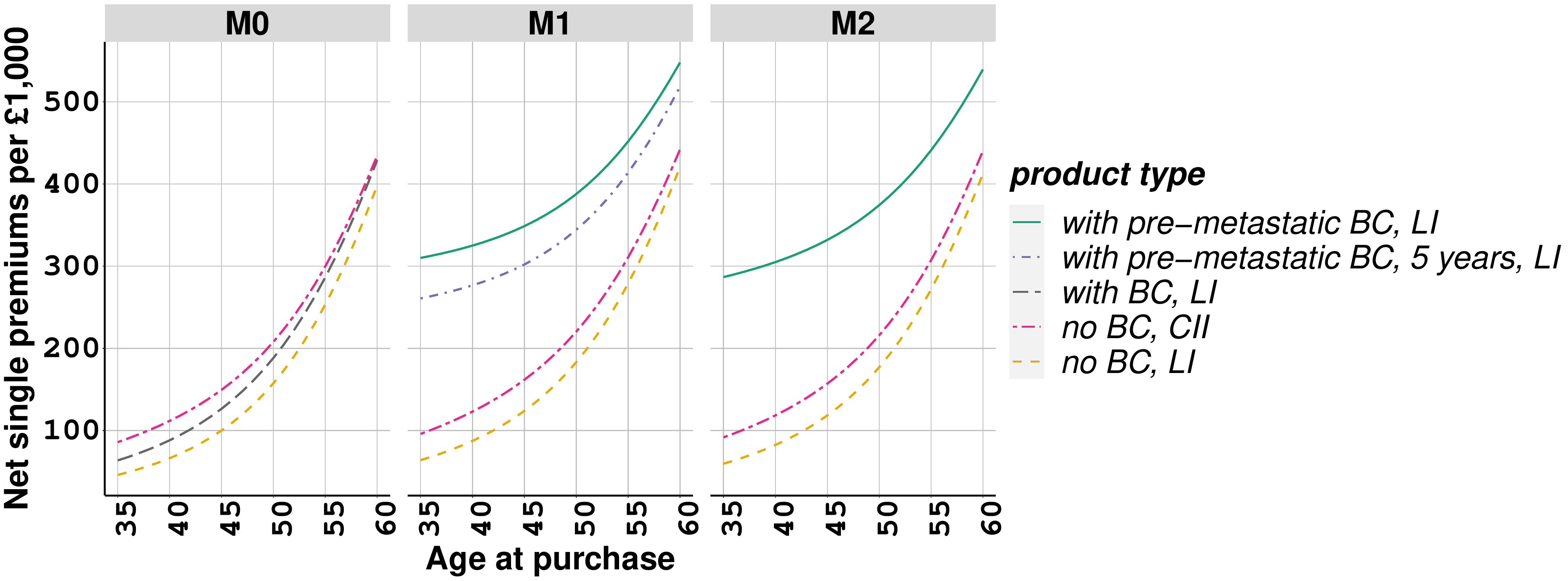}}
	\hfill
	\subfloat[25-year life insurance, $i=4\%$ \label{fig:25TIupto2019_LIv6_delta2}]{\includegraphics[width=0.50\textwidth]{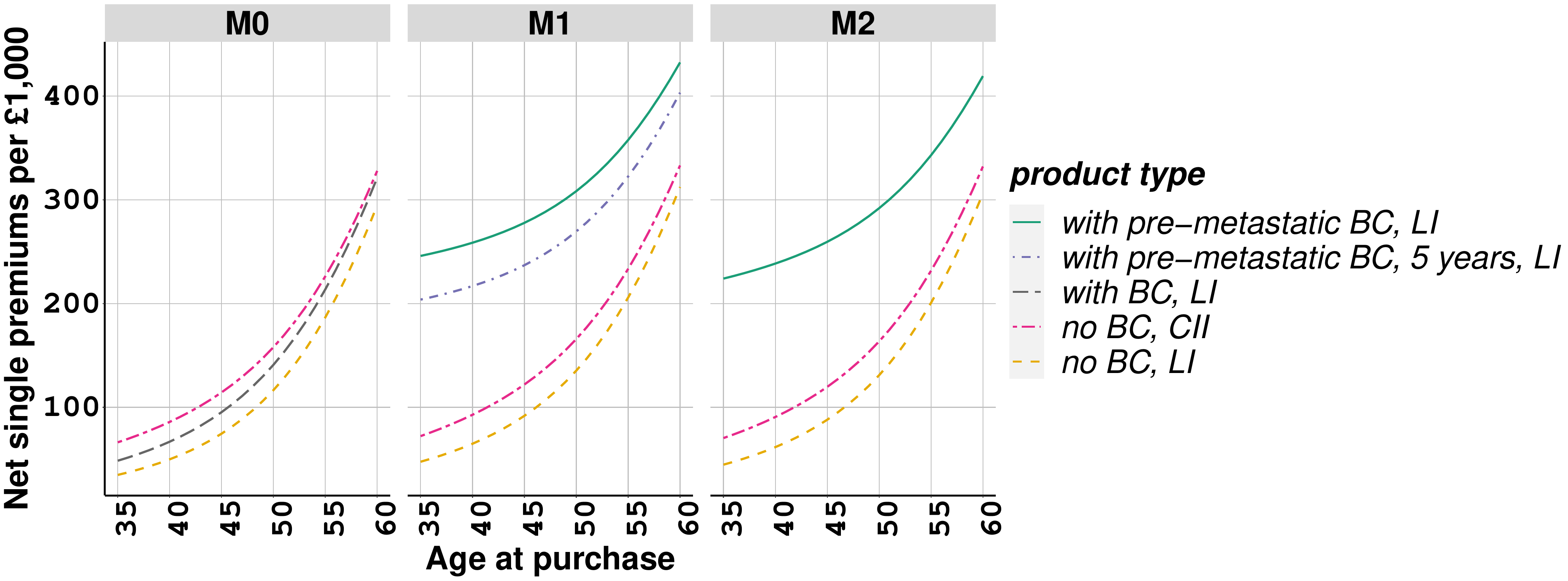}}
	\hfill
	\subfloat[10-year life insurance, $i=2\%$ \label{fig:10TIupto2019_LIv6_delta1}]{\includegraphics[width=0.50\textwidth]{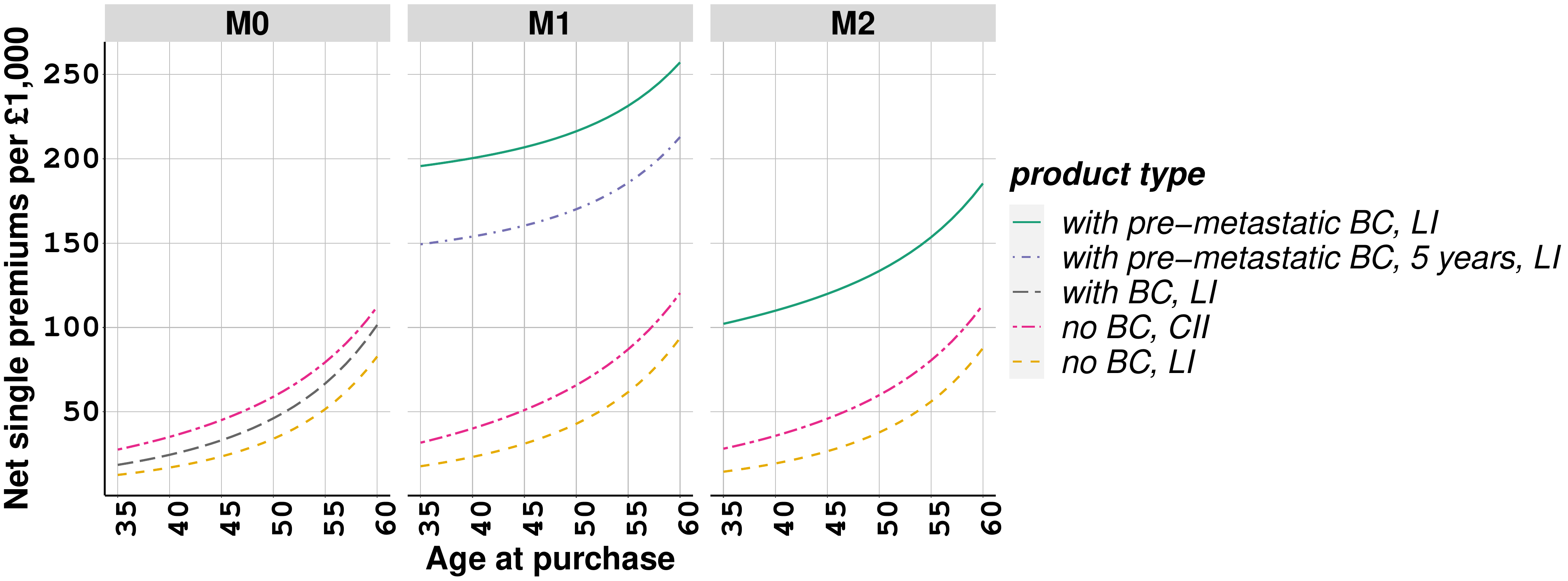}}
	\hfill
	\subfloat[10-year life insurance, $i=4\%$ \label{fig:10TIupto2019_LIv6_delta2}]{\includegraphics[width=0.50\textwidth]{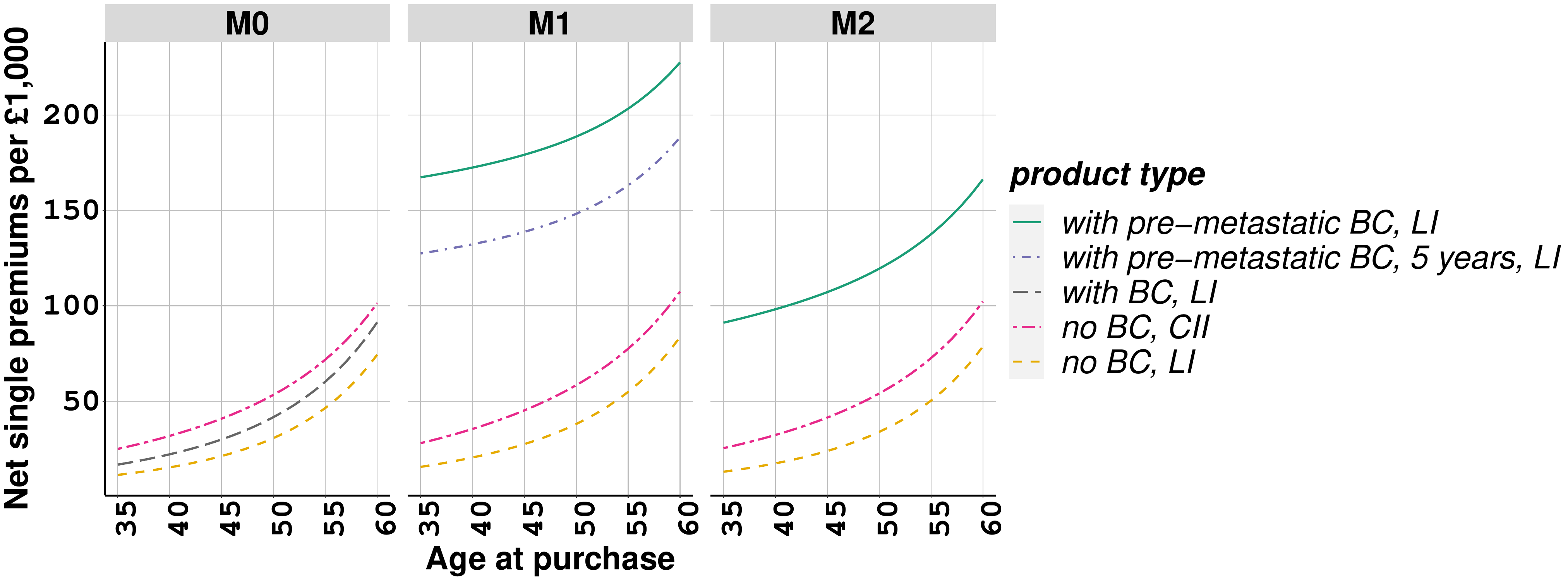}}
	\caption{Net single premium rates of a specialised life insurance contract, \eqref{eq:AxIndustry_death1}--\eqref{eq:AxSemiMarkov_death2} and \eqref{eq:AxSemiMarkov_death2_v2}, for
		policyholders with or without breast cancer at the time of purchase, \pounds1,000 benefit, {payable at the time of death}, based on M1 and M2, when $\alpha = 0.6$ and $\beta = 1/7$, and $\mu^{13}_x$ under M0 based on \tabref{tab:mu13M2}.}
	\label{fig:SinglePremiums_LI_DiffModels_Appendix}
\end{figure}

\end{appendices}

\end{document}